%% file: ZonedOut_arxivNov2025.tex
\documentclass[letterpaper,12pt]{article}

\usepackage{booktabs}
\usepackage{multirow}
\usepackage{xspace}

\usepackage{amsmath}
\usepackage{amsthm}
\usepackage{amssymb, mathtools}

\usepackage{fancyhdr}
\usepackage{wrapfig}
\usepackage{fullpage}
\usepackage{comment}
\usepackage{natbib}
\usepackage[dvipsnames]{xcolor}
\usepackage{hyperref}
\hypersetup{
    colorlinks,
    linkcolor={red!50!black},
    citecolor={blue!50!black},
    urlcolor={blue!80!black}
}
\usepackage{mathrsfs}
\usepackage[enable]{easy-todo}
\usepackage{centernot,cancel}
\usepackage[multiple]{footmisc}
\usepackage{setspace}
\usepackage{enumitem}

\usepackage{tikz}
\usetikzlibrary{calc,decorations.pathreplacing, arrows.meta}
\tikzset{vert/.style = {circle, fill, inner sep = 0, minimum size = 5}}
\newcommand{\circled}[2][inner sep=1pt]{\ifmmode
\tikz[baseline=(X.base),outer sep=0pt]{\node[circle,draw,#1](X){\ensuremath{#2}};}
\else
\tikz[baseline=(X.base),outer sep=0pt]{\node[circle,draw,#1](X){#2};}
\fi
}
\usetikzlibrary{patterns}
\usepackage{pgfplots}
\pgfplotsset{compat=1.18}
\usepackage{nicefrac}

\newtheorem{thm}{Theorem}
\newtheorem{prop}{Proposition}
\newtheorem{lm}{Lemma}

\newtheorem{cor}{Corollary}

\theoremstyle{definition}
\newtheorem{df}{Definition}

\newcommand{\argmax}{\mathop{\rm arg~max}\limits}

\newcommand{\1}{\mbox{1}\hspace{-0.25em}\mbox{l}}

\newcommand{\tblcng}[1]{{\footnotesize #1}}
\newcommand{\tblnmb}[1]{{\footnotesize #1}}

\def\e{\varepsilon}
\def\rich{{{\omega^R}}}

\def\poor{{{\omega^P}}}
\def\w{\mathbf{Y}}

\newenvironment{tabnotes}[2][1]{\begin{minipage}[t]{#1\textwidth}\vspace{0.1cm}\scriptsize{\emph{Notes:} #2}}{\end{minipage}}

\title{
Zoned Out: The Long-Term Consequences of School Choice for Wealth Segregation
}

\author{Georgy Artemov
\and Kentaro Tomoeda\footnote{Artemov: University of Melbourne, 111 Barry St., Parkville, VIC 3010, Australia; email: \url{georgy@gmail.com}; Tomoeda: UTS Business School, University of Technology Sydney, PO Box 123, Broadway, NSW 2007, Australia, email: \url{Kentaro.Tomoeda@uts.edu.au.} We thank 
Nageeb Ali, Ivan Balbuzanov, Christian Basteck, Jeff Borland, Aram Grigoryan, Daisuke Hirata, Byeonghyeon Jeong, Michihiro Kandori, Bettina Klaus, Fuhito Kojima, Simon Loertsher, Vikram Manjunath, Vincent Meisner, Alex Nichifor, Antonio Nicolo, Siqi Pan, Juan Pereyra, Andrzej Skrzypacz, Tayfun Sonmez, Alex Teytelboym, Jonas von Wangenheim, Benjamin Young, 
audiences at Australian National University, HKBU-NTU-Osaka-Kyoto-Sinica, Humboldt University, University of Melbourne, University of Queensland, University of Technology Sydney, and University of Tokyo theory seminars, 
Australian Conference of Economists, Deakin Economic Theory Workshop, Conference on Mechanism and Institution Design, East Asian Econometric Society conference, Conference on Economic Design, Matching Markets and Inequality Workshop, Stony Brook Game Theory conference. 
Georgy Artemov acknowledges financial support from the Faculty of Business and Economics at the University of Melbourne through its MatchLab funding.}}
\date{\today}

\usepackage{version}	

\begin{document}
\maketitle

\begin{abstract}
We study how school choice mechanisms shape wealth segregation in the long term by endogenizing residential choice. Families buy houses in school zones that determine admission priority, experience shocks to school preferences, and participate in one of three mechanisms: neighborhood assignment (N), Deferred Acceptance (DA), or Top Trading Cycles (TTC). Neighborhood segregation increases from N to DA to TTC. DA and TTC reduce school-level segregation relative to neighborhoods but typically not enough to reverse this ranking, and housing prices in oversubscribed zones rise in the same order. Two desegregation policies further illustrate how short- and long-term perspectives can differ. (JEL C78, D47, D63, I21, I28, R31)


\end{abstract}
{\small
{Keywords: wealth segregation, wealth sorting, school choice, housing, school admission priorities, Deferred Acceptance, Top Trading Cycles, market design} 
%
}


A substantial premium for housing in zones of desirable schools is well documented in the empirical literature.\footnote{Many studies estimate a premium of 2--4\% in housing prices per standard deviation increase in test scores (see \cite{SchoolHousingPrices:Black:1999} for a seminal contribution; \cite{ReviewSchoolHousingPrices:NHY:2011} for a review; \cite{AmenityHousingPrices2018Moon} for a recent study exploiting policy variation; and \cite{PricingNeighborhoods2023Eshaghnia} for a study employing an alternative methodology). \cite{SchoolHousingPricesBoston:La:2015} finds an even higher premium of about 7\% for family housing, where school considerations are most salient, though the aggregate effect is consistent with prior estimates.\label{foot:Empirics}} This premium reflects priority access to local school seats. The value of this access depends on the school assignment mechanism---particularly on the extent to which seats are available to out-of-zone applicants---and can influence residential location decisions. Yet most theoretical analyses of school assignment mechanisms treat residential locations---or priorities they confer---as fixed \citep[see, e.g., a review by][]{AbdulkadirogluAndersson2023SchoolChoiceReview}. Treating residential locations as fixed is appropriate for short-term analysis, but does not account for assignment mechanisms' impact on long-term residential wealth segregation (or sorting).

To provide a long-term analysis, we endogenize families' location choices in a two-stage model: families first purchase a house in one of the school zones and then participate in school assignment. In-zone housing gives residents priority at their local school, which we assume to be the sole reason families purchase in-zone housing. 

In our model, families' preferences over schools are heterogeneous---there is no uniformly best school---but there exists an undersubscribed school that is second best for most families. This assumption makes the matching problem tractable while preserving the key differences in school assignments under the mechanisms we study: Neighborhood Assignment (N), Deferred Acceptance (DA), and Top Trading Cycles (TTC). Assignment mechanisms matter in our model because families experience a school-preference shock after purchasing a house. The shock reflects the fact that, in practice, families choose when and where to buy housing based on factors outside our model, and may find it too costly to relocate once preferences change. Under the flexible mechanisms, DA and TTC, families who no longer wish to attend the in-zone school may be assigned elsewhere, freeing local seats for out-of-zone enrollments (Section \ref{sec:DAandTTC} explains how DA and TTC operate in our model).\footnote
{Heterogeneity of school preferences---the importance students place on different school characteristics---is well documented in the education policy literature \citep[see, e.g.,][]{GlazermanDotter2017NewOrleanPreferences,HarrisLarsen2023WhatFamiliesWant}. If instead there were a universally preferred school, our framework could readily accommodate it (see Section \ref{sec:Discussion}). Our model can also incorporate, at the cost of extra notation, families who do not experience a preference shock and instead purchase houses with complete information (see Section \ref{sec:Discussion}). Finally, while preferences for neighborhood amenities can also drive out-of-zone enrollment, we abstract from this channel to focus on uncertainty about school preferences.
}

Out-of-zone enrollments help DA and TTC reduce school-level segregation relative to N in the short term---when neighborhood segregation is held constant---by allowing students from poorer out-of-zone neighborhoods to access schools in richer ones.

In the long term, neighborhoods become segregated because families with different wealth levels resolve the tradeoff between the higher cost of in-zone housing and the expected benefit of a better school assignment by making different housing choices. Obtaining priority at an oversubscribed school improves the expected assignment quality under any mechanism with local priority. DA and TTC further improve the expected quality by providing out-of-zone assignment options, which, as we explain next, lead to different location choices across mechanisms: N produces the least, and TTC the most, segregated neighborhoods (Theorem \ref{thm:n_sorting}). As in the short term, DA and TTC continue to reduce school-level segregation relative to neighborhood-level one; yet, this reduction is usually insufficient to offset the increase in neighborhood segregation that arises under these mechanisms, resulting in more segregated schools than N (Theorem \ref{thm:s_sorting}).

At the time of housing decisions, families differ only in wealth and their ``signals:'' their expected values for schools. Because disutility from housing expenditure is lower for wealthier families, two families with identical signals but differing wealth may make different housing decisions. As wealthier families choose in-zone housing more often than poorer families, neighborhoods are segregated under any mechanism that grants in-zone priority. Yet, the level of segregation varies among the three mechanisms.

As DA and TTC provide an out-of-zone assignment option, signals play less of a role in housing decisions, compared to N. Consequently, under DA and TTC, wealthier families with weaker signals may outbid those poorer families with strong signals who buy in-zone housing under N. Thus, at the population level, as the importance of signals declines, wealth becomes the primary determinant of housing competition, increasing neighborhood-level segregation under flexible mechanisms compared to N. 

Between the two flexible mechanisms, TTC results in higher segregation because residents of oversubscribed schools' zones have a higher probability of obtaining an out-of-zone assignment under TTC than under DA. This probability is identical for all out-of-zone applicants under DA, implying no residential advantage. Under TTC, an out-of-zone assignment requires finding a partner willing to exchange seats, and such partners always exist for students residing in the zones of oversubscribed schools. Hence, out-of-zone assignments are more accessible under TTC, leading to greater segregation.

The argument connecting neighborhood segregation to the probability of obtaining an out-of-zone assignment ignores the housing price changes induced by the mechanisms, but we show that the price-driven effect only strengthens our argument. We also show that, under general conditions, housing prices increase from N to DA to TTC (Theorem \ref{thm:price}). This finding may be of independent interest, as prices capture externalities present in the school choice problem and are a persistent worry among residents of popular school zones in public debates.

Neighborhood segregation does not directly translate to segregation at the school level. Relative to N, two offsetting forces operate under DA: it increases residential segregation, as discussed above, but decreases segregation at each school relative to its own neighborhood. This decrease arises because residents of oversubscribed school zones are, on average, wealthier. When they forgo seats at their local school, these seats are filled by poorer students. Theorem \ref{thm:s_sorting} quantifies these forces and provides conditions under which one dominates the other. The shape of the signal distribution plays an important role: under a natural parameter restriction, we identify the shape that minimizes the forces driving segregation under DA (Lemma \ref{lm:F_max_pop}) and establish a condition under which DA yields higher neighborhood segregation than N for the segregation-minimizing---and hence any---distribution (Proposition \ref{prop:s_sorting_2types}). Finally, using this segregation-minimizing distribution, we numerically examine cases in which this condition is violated and show that, even then, DA typically produces more segregated schools. Overall, DA results in higher school-level segregation than N in most cases.

TTC results in the highest school-level segregation (Theorem \ref{thm:s_sorting} and Proposition \ref{prop:s_sorting_2types}). This follows from its high residential segregation, which is only weakly offset at the school level: residents of oversubscribed zones predominantly exchange seats with residents of other oversubscribed zones, making these exchanges effectively segregation-neutral. As a result, school segregation largely mirrors neighborhood segregation.

While the benefits of DA and TTC are well documented \citep[see, e.g.,][]{Pathak2011Review} and remain in our setting (see Section \ref{sec:benchmarks} for an example\footnote{The example also considers two hypothetical benchmarks without neighborhood priority. DA/TTC without neighborhood priority yields full integration but lower match quality than DA or TTC with priority. Auctioning school seats achieves higher match quality but, perhaps unexpectedly, produces segregation levels comparable to N and DA and well below TTC.}), our segregation results may help explain the hesitation of some policymakers to adopt flexible school-choice mechanisms. Policymakers should weigh these benefits against the potential for rising segregation and design desegregation policies to mitigate it. 
    
Desegregation policies, like assignment mechanisms, should be evaluated not only by their short-term outcomes but also by their long-term effects. To illustrate, we construct an environment where giving priority access to poor families from a less desirable school zone is highly effective in the short term but increases neighborhood segregation and, to a small extent, school segregation in the long term. By contrast, extending priority to all families from that zone---an intuitively inferior policy---achieves less desegregation when locations are fixed but becomes increasingly effective once families relocate in response to its incentives.

While early empirical contributions (see footnote \ref{foot:Empirics}) identified the housing premium using discontinuity at the school boundaries, recent papers are closer to our setting, as they use structural models that incorporate neighborhood priority, detailed housing supply, and residential sorting, taking the models closer to the one considered here  \citep{Caetano2019,Agostinelli2024,LocationSchool2023Park,Pietrabissa2024,SchoolSorting2024Greaves}. These papers do not allow for a preference shock after the neighborhood choice except for \cite{LocationSchool2023Park}. The latter estimates a structural model similar to ours in the context of New York school choice that uses DA. It does not compare DA to other assignment mechanisms and, naturally, focuses on quantitative rather than qualitative comparisons.

Recent theoretical contributions have significantly advanced our understanding of how school choice mechanisms interact with housing markets when access to schools is rationed, but, to our knowledge, only \cite{Grigoryan2021} models housing choices at the level of individual school zones. He allows for general preferences over neighborhoods and schools, but unlike our model, families do not experience preference shocks after housing decisions. He establishes that DA (with neighborhood priorities, as in our setting) generates higher welfare than N, but does not consider segregation. Thus, the two papers can be considered complementary, highlighting a tradeoff between welfare improvements of flexible school choice and negative consequences in terms of segregation.
    
\cite{Gonczarowski2024SeveralFields} analyze a school assignment problem that involves multiple school districts. They extend the analysis of \cite{PathakSonmez2008SincereSophisticated} of sincere and sophisticated players to introduce additional ``constrained''---those who must be assigned within the district---and ``unconstrained' types who can hop between districts. They show that multiple school districts may overturn classical results. Even though the focus and the models of that paper and our work are very different, the two share the main message that additional school enrollment constraints may play an important role in the analysis.

\cite{OpenEnrollment2018Jeong} studies a multi-district school system with multiple schools per district. He compares regimes with and without inter-district enrollment and analyzes how families sort by income and school preferences under Immediate Acceptance (IA) and DA mechanisms. He assumes that school quality is linked to housing prices. The main case in the paper is when preferences and wealth are correlated; when they are not, as in our model, schools become vertically differentiated, with complete segregation.

More generally, our paper contributes to the literature examining joint schooling and housing choices, often embedding education into broader models of residential sorting. Foundational work by \cite{LocalExpend1956Tiebout} and \cite{MobilityPrivateVouchers2000Nechyba} introduced the idea of families ``voting with their feet,'' embedding schools in general-equilibrium models with housing markets, vouchers, and taxation, but assumed simple school assignment rules. Later theoretical work, including \cite{NeigborhoodChoiceSchools2003Epple}, \cite{DeFrajaMartinezMora2014DesegragatinTracking}, \cite{BarseghyanClarkCoate2019PeerPublicSchool}, and \cite{AveryPathak2021PublicSchool}, explores stratification through mechanisms like peer effects, school effort, and competitive outside options, treating school assignments as neighborhood-based or open enrollment rather than those involving rationing. 

Other work on school choice studies segregation without modeling endogenous housing choices. \cite{Stratification2023Calsamiglia} define ``access to better schools'' as the number of students assigned to a school they prefer over their in-zone option, showing higher access under TTC than DA. Although this may seem related to segregation, the concepts differ: access can increase without any change in segregation---for example, when students from one desirable zone attend a school in another. Their analysis also assumes stratification, whereas in our model preferences over schools are highly heterogeneous (see Section \ref{sec:Discussion} for a discussion on introducing a universally top-ranked school). \cite{RiskSegregation2021Calsamiglia} propose the notion of ``cardinal segregation,'' arising when students with identical ordinal but different cardinal preferences submit different rank-order lists under IA. They show that IA creates more segregation than DA due to risk preferences.\footnote{They consider endogenous location choice in Section 5, but this only increases transportation costs rather than creating new sorting incentives.} Their framework differs from ours in both the mechanisms studied and the definition of segregation.

The rest of the paper is organized as follows. Section \ref{sec:Example} introduces an example that captures the main intuition behind our results. It focuses on signals drawn from a uniform distribution and provides a numerical illustration for other distributions. Section \ref{sec:model} introduces the model, describes the mechanisms, and derives equilibria. Section \ref{sec:main} presents the main results: the ranking of mechanisms by neighborhood segregation, school segregation, and housing prices. Section \ref{sec:Discussion} discusses possible extensions of the model. It then returns to the example from Section \ref{sec:Example} to examine a partial welfare measure---match quality---under N, DA, and TTC, and to compare them with two additional mechanisms that do not grant neighborhood priority, which serve as alternative benchmarks. The section also explores two possible desegregation policies. We conclude in Section \ref{sec:Concl}.

\section{Example}\label{sec:Example}

There are three schools, $c_{-1}, c_0,$ and $c_1$, each located in a neighborhood $n_{-1}, n_0,$ and $n_1$, and each granting priority to local residents. The total mass of agents is 2, split evenly between poor ($\poor=9/8$) and rich ($\rich=7/8$) families, so that the average wealth index is normalized to 1. 

First, each agent $i$ receives a signal $s_i \in [-1,1]$ representing their expected idiosyncratic value for schools. To draw a signal, Nature first selects either $[-1,0]$ or $[0,1]$ with equal probability, and then draws $s_i$ from the chosen interval. Let $F(\cdot)$ denote the cdf on $[0,1]$, extended symmetrically to $[-1,0]$. This setup facilitates generalization to more than three schools. After receiving a signal, each agent chooses a neighborhood $n_k$ and pays a housing price $p_k$, with $p_0$ normalized to zero. 

After choosing a neighborhood but before school assignment, agent $i$ experiences a preference shock $\e_i \in \{-1,0,1\}$, each with probability 1/3. After observing $\e_i$, agents submit preferences over schools $c_{-1}, c_0,$ and $c_1$ and are assigned via a mechanism $\varphi$. This timing abstracts from agents' full relocation decisions, which typically involve decisions on when and where to move based on multiple factors, including local school suitability. By condensing these multiple factors into a single housing decision under uncertainty, we isolate the effects of school choice.

When agent $i$ buys a house in neighborhood $n_k$ and attends school $c$, $i$'s utility is

\begin{align*}
	u_i(n_k,c,p_k|s_i+\e_i,\omega_i) =
	\begin{cases}
		s_i + \e_i - \omega_i p_k & \text{ if } c = c_1\\
		  - \omega_i p_k & \text{ if } c = c_0\\
		- (s_i + \e_i) - \omega_i p_k & \text{ if } c = c_{-1}
	\end{cases}
\end{align*}
Note that, ex ante, $c_1$ is the best school for any agent with $s_i > 0$, but ex post, with probability 1/3, $c_{-1}$ becomes $i$'s most preferred school.\footnote{In the main model, we allow $c_0$ to be the most preferred school for some agents.} Thus, allowing agents to attend schools outside their neighborhood increases efficiency. Poor agents (with $\omega_i=\poor=9/8$) incur greater disutility from housing costs than rich agents (with $\omega_i=\rich = 7/8$).

Schools $c_{-1}$ and $c_1$ have capacities $q=0.4$ each, and school $c_0$ has unlimited capacity. Neighborhood sizes are assumed to be equal to their respective school capacities. $n_{-1}$ and $n_1$ are overdemanded at zero price (we omit the qualifier ``at zero price'' hereafter), while $n_0$ is not, serving as an ``outside option.'' Because of symmetry in signals, shocks, and capacities, it is sufficient to focus on $n_1$ and $c_1$, and we write $p^{\varphi}$ as the price of overdemanded neighborhoods under mechanism $\varphi$ (i.e., $p^\varphi = p^\varphi_{-1} = p^\varphi_1$).

We study three assignment mechanisms $\varphi$: Neighborhood (N), Deferred Acceptance with neighborhood priority (DA), and Top Trading Cycles with neighborhood priority (TTC). Under N, agents in neighborhood $n_k$ are assigned to school $c_k$; there is no school choice. Under DA, agents in $n_1$ are guaranteed seats at $c_1$, but if $c_{-1}$ is their top choice, they face the same probability of assignment to $c_{-1}$ as any other non-local applicant due to a random tie-breaker. Under TTC, if agent $i$ in $n_1$ prefers $c_{-1}$, by symmetry, $i$ can always find an agent in $n_{-1}$ who prefers $c_1$, forming a cycle that guarantees $i$'s assignment to $c_{-1}$. Thus, a house in $n_1$ secures a seat at the school of the agent's choice. In turn, all seats in $c_1$ freed up by $n_1$ residents go to agents from $n_{-1}$ and all agents in $n_0$ are assigned to $c_0$ under TTC.\footnote{In the main model, some agents from $n_0$ are assigned to $c_{-1}$ or $c_1$ under TTC, which complicates the analysis but preserves the intuition.}

\subsection{Agent's optimal housing decision}\label{subsec:ex:agent}

We focus on a symmetric equilibrium strategy $\sigma$, which determines the probability $r^\varphi$ that an applicant from $n_0$ is rejected by $c_1$ due to a random tie-breaker. To simplify notation in the example, we omit $\sigma$. 
Agent $i$ chooses to live in $n_1$ if the expected utility gain relative to $n_0$ is non-negative: $\Delta u^\varphi(r^\varphi,p^\varphi|s_i,\omega_i) = E_{\epsilon_i}u_i^{\varphi}(n_1,r^\varphi,p^\varphi|s_i+\epsilon_i,\omega_i)-E_{\epsilon_i}u_i^{\varphi}(n_0,r^\varphi,p^\varphi|s_i+\epsilon_i,\omega_i) \geq 0$, where $u_i^\varphi$ denotes the expected utility under mechanism $\varphi$, given truth-telling and the induced $r^\varphi$. Under N, $r^N = 1$ because out-of-zone applicants are never assigned to $c_1$. In this example, $r^{TTC}=1$ as well, since all seats in $c_1$ vacated by $n_1$ residents are taken by $n_{-1}$ residents.

For $s_i > 0$, $\Delta u^\varphi(r^\varphi,p^\varphi|s_i,\omega_i)$ is increasing in $s_i$, as higher signals increase the expected benefit of living in $n_1$ (Lemma \ref{lm:monotone}). Therefore, for each mechanism $\varphi$ and wealth index $\omega_i$, there is a cutoff $s^\varphi_{\omega_i}$ solving $\Delta u^\varphi(r^\varphi,p^\varphi|s^\varphi_{\omega_i},\omega_i) = 0$, such that $i$ with $(s_i,\omega_i)$ buys in $n_1$ if and only if $s_i \geq s^\varphi_{\omega_i}$. We next find cutoffs under N, TTC, and DA.

Under N, all agents are assigned to their neighborhood schools. If agent $i$ resides in $n_1$, $i$'s expected value of the assignment is $s_i$, and it is 0 if $i$ resides in $n_0$. Hence, $\Delta u^N(r^N,p^N|s_i,\omega_i) = r^N s_i - {\omega_i}p^N$. The cutoff is $s^N_{\omega_i} = \omega_i  \frac{p^N}{r^N}$. 

Under TTC, $n_1$ residents are assigned to their most preferred school. If $\e_i \in \{0,1\}$, the most preferred school is $c_1$, with utility $s_i+\e_i$; if $\e_i =-1$, it is $c_{-1}$ with utility $1-s_i$. Thus, $\Delta u^{TTC}(r^{TTC},p^{TTC}|s_i,\omega_i) = r^{TTC}((s_i+1)/3+s_i/3+(1-s_i)/3) - \omega_i p^{TTC}$ and the cutoff is $s^{TTC}_{\omega_i} = 3 \omega_i \frac{p^{TTC}}{r^{TTC}} - 2$.

Under DA, agents in $n_0$ and $n_1$ have an equal chance to be assigned to $c_{-1}$, and $c_0$ accepts all rejected students. Hence, the only benefit of buying in $n_1$ is an increase in probability of getting $c_1$ when it is the top choice, from $(1-r^{DA})$ to 1. Hence, $\Delta u^{DA} (p^\varphi|s_i,\omega_i) = r^{DA} ((s_i+1)/3+s_i/3)-\omega_i p^{DA}$, and $s^{DA}_{\omega_i} = \frac{3}{2}\omega_i \frac{p^{DA}}{r^{DA}} - \frac{1}{2}$.

The cutoff values for each mechanism $\varphi$ and wealth index $\poor, \rich$ are summarized below.
\begin{align}
	&&\varphi=\text{N} && \varphi=\text{DA} && \varphi=\text{TTC}\nonumber\\
	s^\varphi_{\poor} = &&\poor\; \frac{p^\varphi}{r^\varphi} && \poor\; \frac{3}{2}\;\frac{p^{\varphi}}{r^{\varphi}} - \frac{1}{2} && \poor\; 3\; \frac{p^{\varphi}}{r^{\varphi}} - 2\label{ExEqCutoffsPoor}\\
	s^N_{\rich} = &&\rich\; \frac{p^\varphi}{r^\varphi} && \rich\; \frac{3}{2}\;\frac{p^{\varphi}}{r^{\varphi}} - \frac{1}{2} && \rich\; 3\; \frac{p^{\varphi}}{r^{\varphi}} - 2\label{ExEqCutoffsRich}
\end{align}

We map these cutoffs into neighborhood-level segregation in the next section.

\subsection{Neighborhood-level wealth segregation}

Using equations (\ref{ExEqCutoffsPoor}) and (\ref{ExEqCutoffsRich}), we quantify segregation under each mechanism. Since the example has only two wealth indices, we measure segregation by the share of poor agents in $n_1$. In the main model, we use the average wealth index as a more general metric. Both measures are linear transformations of the difference in masses of rich and poor agents in $n_1$: $(1-F(s^\varphi_{\rich}))-(1-F(s^\varphi_{\poor})) = F(s^\varphi_{\poor})-F(s^\varphi_{\rich})$. Although we focus on the uniform distribution here, the ranking of mechanisms in terms of neighborhood-level segregation is the same for other weakly concave distributions.

With uniform distribution, we have $F(s^\varphi_{\poor})-F(s^\varphi_{\rich}) = s^\varphi_{\poor} - s^\varphi_{\rich}$. This difference could be readily calculated from equations (\ref{ExEqCutoffsPoor}) and (\ref{ExEqCutoffsRich}), but it must also account for endogenous $\frac{p^\varphi}{r^\varphi}$. Equations (\ref{ExEqCutoffsPoor}) and (\ref{ExEqCutoffsRich}) and an observation that the total mass of agents in $n_1$, $(1-s^\varphi_{\rich})/2 + (1-s^\varphi_{\poor})/2$, must be equal to $n_1$'s capacity, $q$, means that $\frac{p^\varphi}{r^\varphi}$ must solve $(s^\varphi_{\poor} + s^\varphi_{\rich})/2 = 1-q = 0.6$. That, and equations (\ref{ExEqCutoffsPoor}) and (\ref{ExEqCutoffsRich}), with $\poor = 9/8, \rich=7/8$, give us:
\begin{align}
	&&\varphi=\text{N} && \varphi=\text{DA} && \varphi=\text{TTC}\nonumber\\
    \frac{(s^\varphi_{\poor} + s^\varphi_{\rich})}{2} = && \frac{p^{\varphi}}{r^{\varphi}} && \frac{3}{2}\;\frac{p^{\varphi}}{r^{\varphi}}-\frac{1}{2} && 3\;\frac{p^{\varphi}}{r^{\varphi}}-2\label{eq3}\\
    \frac{p^\varphi}{r^\varphi} = && \frac{9}{15} && \frac{11}{15} && \frac{13}{15}\label{eqPR}\\
    (s^\varphi_{\poor} - s^\varphi_{\rich}) = \frac{1}{4} \times && \frac{p^\varphi}{r^\varphi} && \frac{3}{2}\;\frac{p^{\varphi}}{r^{\varphi}} && 3\;\frac{p^{\varphi}}{r^{\varphi}}\label{eqDiffCutoffs}\\
    \text{\% poor agents in }n_1 = && 41\% && 33\% && 9\%,\label{eqPercPoor}
\end{align}
Line (\ref{eqPercPoor}) shows that segregation increases from N to DA to TTC. The multipliers in line (\ref{eqDiffCutoffs})---1, 3/2, and 3---reflect decreasing sensitivity of the utility from school assignment to the agent's signal, as the chance of being assigned to a non-local school increases across mechanisms. Even if $\frac{p^\varphi}{r^\varphi}$ were the same, these multipliers alone would lead to higher segregation.

The increases in $\frac{p^\varphi}{r^\varphi}$ from N to DA to TTC, as shown in line (\ref{eqPR}), further amplify segregation. This term captures the ``option value'' of buying a house in $n_1$: the utility gain of an $n_1$ resident relative to an $n_0$ resident rejected from $c_1$. 

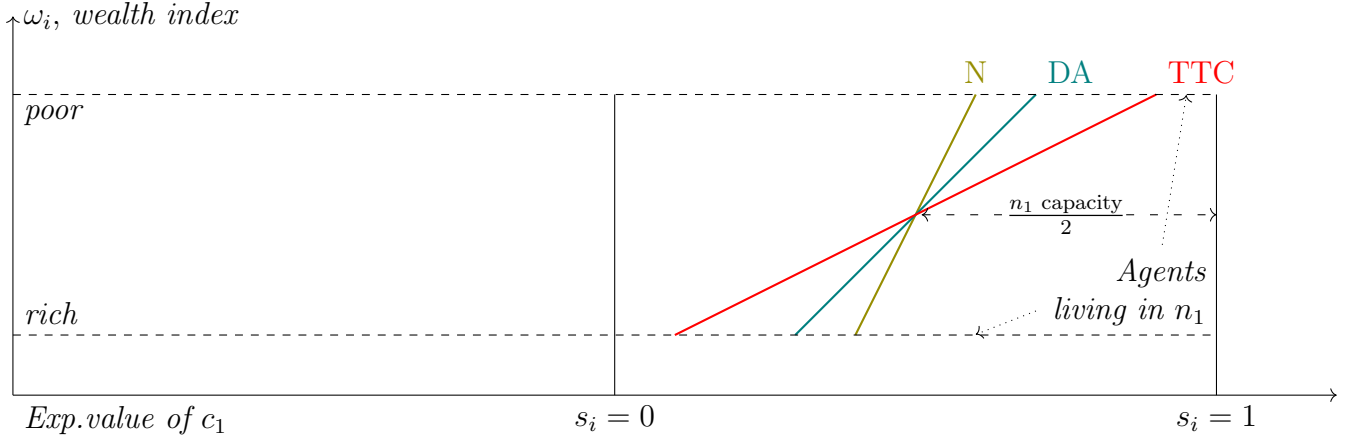
\begin{figure}
    \centering
    \begin{tikzpicture}[scale=0.8]
        \input{ex_pic_definitions3}
        \def\xLeft{-\xMax}
        \input{ex_pic_slopes}
        \input{ex_pic_mainbox3}
    \end{tikzpicture}
    \caption{Segregation in neighborhood $n_1$ increases across the school assignment mechanisms, from Neighborhood (N) to Deferred Acceptance (DA) to Top Trading Cycles (TTC).}
    \label{fig:slopeAll}
    \begin{tabnotes}
        Dashed lines labeled ``poor'' and ``rich'' represent two populations of mass 1, each with signals uniformly distributed on $[-1,1]$. Agent $i$'s signal, $s_i$, corresponds to $i$'s expected utility from attending school $c_1$ at the time of the housing decision. The intersections of the dashed line \textit{poor} with the solid lines labeled by assignment mechanisms---N, DA, and TTC---indicate the cutoffs for poor agents under each mechanism. The same applies to the \textit{rich} dashed line. Agents to the right of each cutoff, with higher signals, choose $n_1$. For each mechanism, the average of the poor and rich agents' cutoffs---the point of intersection of three solid lines---equals half of the total capacity of neighborhood $n_1$.
    \end{tabnotes}
\end{figure}

Our observations so far can be summarized in Figure \ref{fig:slopeAll}, which shows cutoffs for poor and rich families (which are intersections of solid and dashed lines), and the population living in $n_1$, which is to the right of these cutoffs. 
In perfectly unsegregated neighborhoods---such as those produced by a school assignment mechanism without neighborhood priorities\footnote{We provide an example of such mechanisms in Section \ref{sec:benchmarks}.}---the cutoffs would be at exactly one minus half of $n_1$'s capacity for each population. The lower the slope of the line connecting the cutoffs for rich and poor families, the higher the segregation. 
Theorem \ref{thm:n_sorting} shows that this result holds in our general model that allows for $m$ schools, preferences that allow $c_0$ to be the best school for some agents, and more than two wealth indices (in which case we use the average wealth index as a measure of segregation). 

\subsection{School-level wealth segregation}

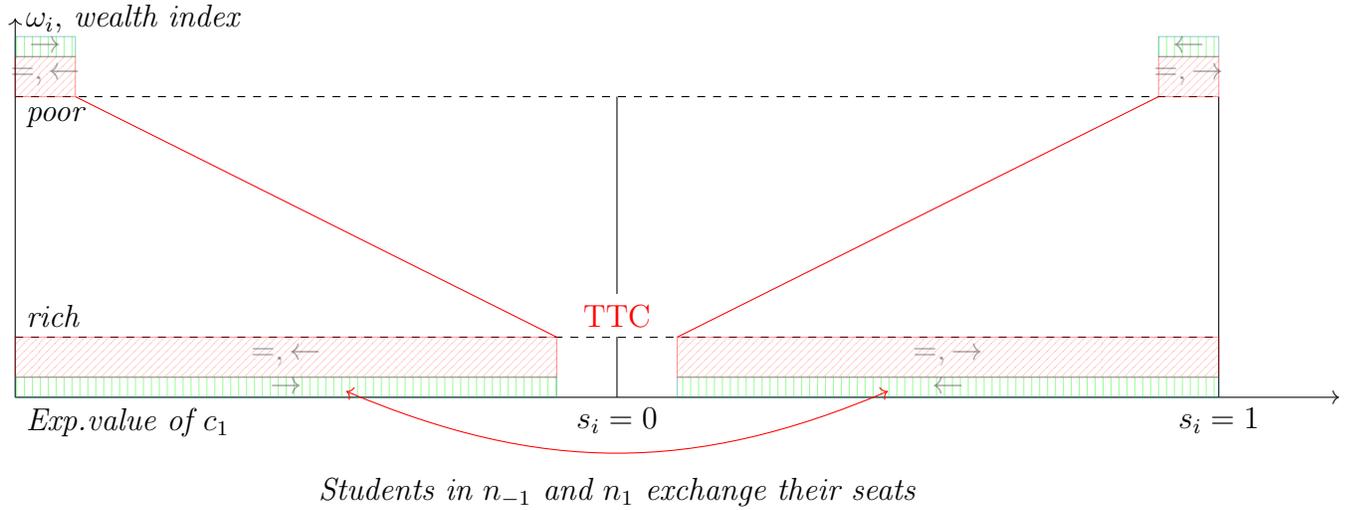
\begin{figure}
    \centering
    \begin{tikzpicture}[scale=0.8]
        \input{ex_pic_definitions3}
        \def\xLeft{-\xMax}
        \coordinate (O) at (0,0);
        \input{ex_pic_mainbox3}
        \input{ex_pic_TTC_reshufle3}
    \end{tikzpicture}
    \caption{School-level segregation is equal to neighborhood-level segregation under TTC.}
    \label{fig:TTCreshufle}
    \begin{tabnotes}
        Agents in the box labeled $=,\rightarrow$ have shocks of 0 or $+1$, while those in the box labeled $=,\leftarrow$ have shocks of 0 or $-1$. Red boxes with slanted fill show agents in $n_{-1}$ and $n_1$ attending their local schools, $c_{-1}$ and $c_1$, respectively. Green boxes labeled $\leftarrow$ contain agents in $n_1$ with a $-1$ shock whose most preferred school is $c_{-1}$. Conversely, green boxes labeled $\rightarrow$ contain agents in $n_{-1}$ with a $+1$ shock whose most preferred school is $c_1$. These agents trade seats: those in the left green box give up a seat in $c_{-1}$ in exchange for a seat in $c_1$, and vice versa. Intersections of solid red lines labeled TTC and the dashed lines indicate the cutoffs for rich and poor families and show the resulting neighborhood-level segregation, as shown in Figure \ref{fig:slopeAll}.
    \end{tabnotes}
\end{figure}

Under N, there is no school choice: all agents in $n_1$ attend $c_1$. Under TTC, reassignment does not reduce segregation in this example, as no agents from $n_0$ are ever assigned to $c_1$.\footnote{In the general model, $c_0$ is the most-preferred school for some agents from $n_1$, who exchange their seats with agents from $n_0$. Although this reduces school-level segregation, the reduction is small.} Figure \ref{fig:TTCreshufle} illustrates this: the green areas represent agents who prefer a school in the opposite neighborhood and exchange seats with their counterparts in the green area on the other side of $s_i=0$. As neither mechanism affects segregation that existed on the neighborhood level, schools are less segregated under N than under TTC.

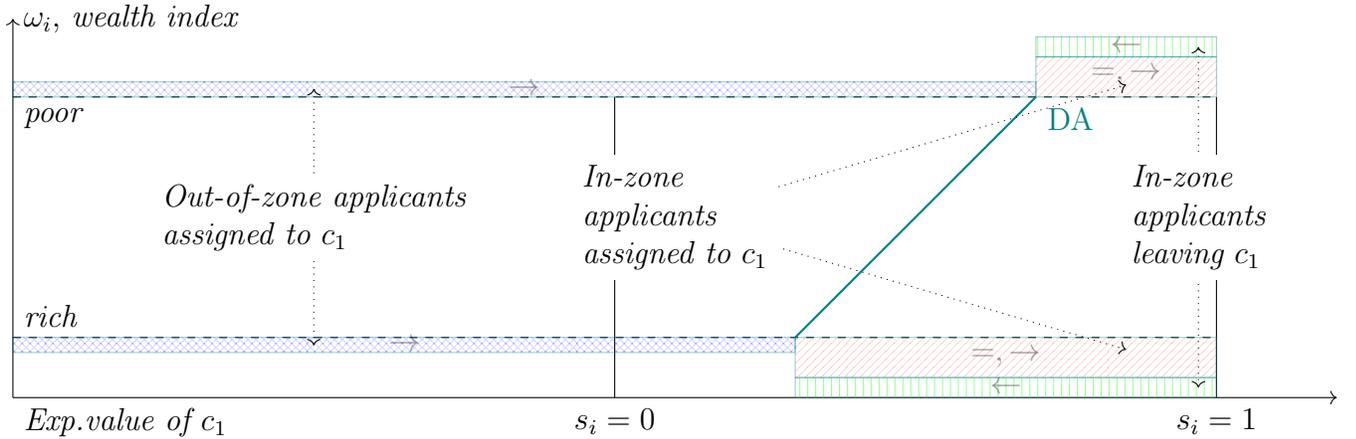
\begin{figure}
    \centering
    \begin{tikzpicture}[scale=0.8]
        \input{ex_pic_definitions3}
        \def\xLeft{-\xMax}
        \input{ex_pic_mainbox3}
        \input{ex_pic_DAreshufle3}
    \end{tikzpicture}
    \caption{Segregation on the school level is lower than on the neighborhood level under DA.}
    \label{fig:schoolsNDA}
    \begin{tabnotes}
        Red boxes with slanted fill labeled $=,\leftarrow$ show agents $n_1$ with shocks of 0 or $+1$ who attend their local school $c_1$. Green boxes labeled $\leftarrow$ contain agents in $n_1$ with a $-1$ shock whose preferences are $c_{-1}c_0c_1$. Blue boxes with crosshatch fill show agents outside of $n_1$ with a shock $+1$ whose preferences are $c_1c_0c_{-1}$ and who are assigned to $c_1$. The sizes of the two green boxes and the two blue boxes are equal. The agents in the green boxes are, on average, richer than agents in the blue boxes. A solid green line labeled DA indicates neighborhood-level segregation, as shown in Figure \ref{fig:slopeAll}.
    \end{tabnotes}
\end{figure}

Under DA, school choice unambiguously reduces segregation at the school level. Since $n_1$ comprises 33\% poor and 67\% rich agents, the population outside $n_1$ must be majority poor. Because agents outside $n_1$ have an equal chance to be assigned to $c_1$, DA replaces agents who are more likely to be rich with agents who are more likely to be poor, narrowing the wealth gap. Figure \ref{fig:schoolsNDA} provides an illustration analogous to the TTC case. Here, agents in the green area free up seats in $c_1$, and agents in the blue area to the left of the solid DA line are assigned to $c_1$. By comparing the relative lengths of blue lines capturing rich and poor families to the corresponding lengths of the green lines, we see that, on average, the green-area agents are richer than those in the blue area.

In general, whether DA results in lower school-level segregation than N depends on the signal distribution. Under a uniform distribution, both mechanisms yield identical segregation in $c_1$. For non-uniform cases, it suffices to consider single-kink cdfs: any weakly concave distribution has a linearized single-kink version that yields a weakly greater increase (or smaller decrease) in the share of poor agents at $c_1$ from N to DA (Lemma \ref{lm:F_max_pop}). That is, the distributions most favorable to DA are single-kink.

\begin{figure}
    \centering
    \includegraphics[width=0.7\textwidth]{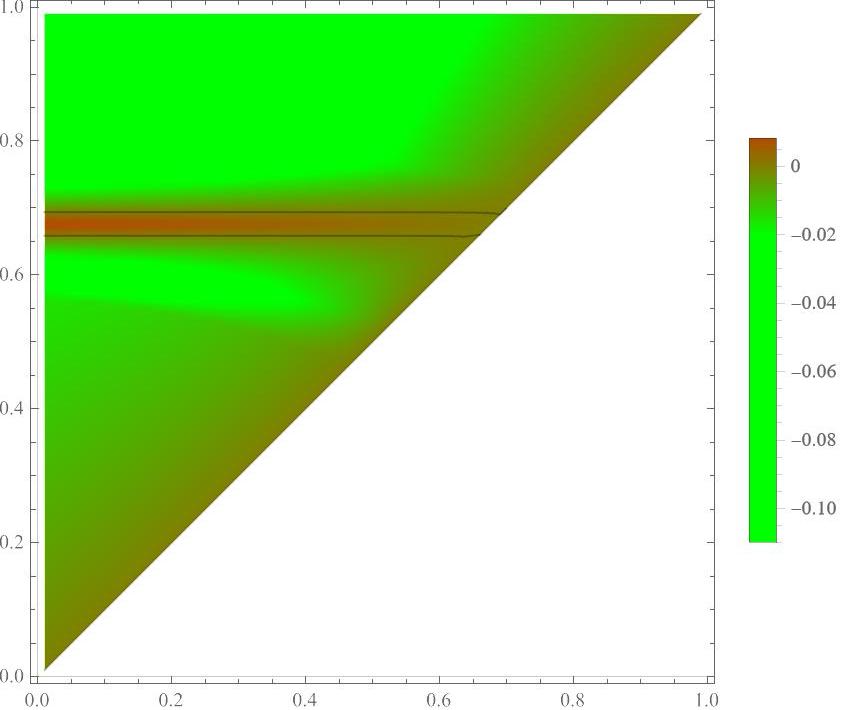}
    \caption{The difference in shares of poor agents in school $c_1$ under N and DA for all one-kink distribution of signals.}
    \label{fig:example_rich_poor}
    \begin{tabnotes}
        Each point $(x,y)$ above the 45-degree line corresponds to the kink of a one-kink signal distribution with $F(x)=y$. The color shading indicates which mechanism yields greater school segregation: green regions correspond to cases where DA generates higher segregation than N, while red regions correspond to the opposite.
    \end{tabnotes}
\end{figure}

To illustrate how often N results in lower segregation than DA, we compute the share of poor agents in $c_1$ across \emph{all} single-kink distributions for one set of parameters. Figure \ref{fig:example_rich_poor} shows the difference between these shares under N and DA. Each point $(x,y)$ corresponds to the kink of a one-kink signal distribution. Because we consider only weakly concave functions, the area below the 45-degree line is empty. The color scale shows the difference between school segregation under DA and N. Negative values, where DA results in higher segregation, are shown in green; positive values are shown in red. When the kink lies on the 45-degree line, the distribution is uniform, and segregation under N and DA is identical. The only region where DA yields lower segregation than N is a narrow horizontal red strip, where the neighborhood segregation gap between N and DA is small and the desegregating effect of DA is large.

In summary, TTC always produces more segregation than N, while the comparison between N and DA is more nuanced. Although N can be more segregated than DA for some distributions, most lead to the reverse ranking. These points are formally established in Theorem \ref{thm:s_sorting} for the general model.

\section{Model}\label{sec:model}

Consider $m+1$ schools $C=\{c_0, c_1, \ldots, c_m\}$, and corresponding neighborhoods (or zones) $N=\{n_0, n_1, \ldots, n_m\}$.
Each neighborhood $n_k \in N\setminus \{n_0\}$ has a fixed supply of housing $q \in [0,1]$, the same for each such $n_k$, whereas the housing supply of $n_0$ is sufficient to accommodate demand for $n_0$ (e.g., unlimited). 
In the main model, we assume that the capacity of each school $c_k \in C$ equals the housing supply of the corresponding neighborhood $n_k \in N$. 
We discuss an extension where $c_k$'s capacity is greater than the housing supply of $n_k$ in Section \ref{sec:delta_q}.
In either scenario, enrollment in the neighborhood school is guaranteed, which reflects a common feature in many school districts \citep[see, e.g.,][]{Musset2012SchoolChoiceEquity}. 

There is a mass $m$ of agents who choose their residential areas from $N$ and then participate in a school choice mechanism.\footnote{An ``agent'' refers to a family with one child. Typically, the parents decide their residential location, and the child attends one of the schools, but we treat the family as a single decision-maker.} 
An agent type is defined as $(t_i^1, t_i^2, s_i, \e_i, \omega_i) \in \{1,\ldots,m\}^2 \times [0, 1] \times \{-e, 0, e\} \times \Omega$, where $e\in \mathbb{R}_{>0}$, and $\Omega \subseteq \mathbb{R}_{>0}$ is a finite set.
$t_i^1$ and $t_i^2 (\neq t_i^1)$ are the indices of $i$'s primary and secondary ``fit schools,'' respectively, $s_i$ is a signal of $i$'s fit to these two schools, $\e_i$ is a random shock to the fit, and $\omega_i$ is $i$'s wealth type. The type
$(t_i^1, t_i^2, s_i, \e_i, \omega_i)$ determines the utility of an agent who chooses neighborhood $n_k$ and is assigned to school $c_l$ as follows: 
\begin{equation*} u_i(n_k,c_l,\boldsymbol{p}|t_i^1,t_i^2,s_i+\e_i,\omega_i) = \begin{cases}
 s_i+\e_i - \omega_i p_k & \text{ if } l=t_i^1 \\
 -(s_i+\e_i) - \omega_i p_k & \text{ if } l=t_i^2 \\
 g - \omega_i p_k & \text{ if } l=0 \\
 -\infty & \text{ otherwise, }
 \end{cases}
\end{equation*}
where $\boldsymbol{p} := (p_0,\ldots,p_m)$ is the vector of housing prices determined in equilibrium.
Agent $i$ receives utility of $s_i+\e_i$ from their primary-fit school $c_{t_i^1}$, whereas the secondary-fit school $c_{t_i^2}$ gives $i$ the negative of that value, $-(s_i+\e_i)$.
School $c_0$ gives fixed utility $g \ge 0$ irrespective of the student's type, and utility from all other schools is negative infinity.\footnote{This preference structure ensures that $c_0$ is always the first- or second-most preferred school, which keeps the model tractable.}
Schools $c_1, \ldots, c_m$ can be interpreted as schools with distinguished features, such as schools focusing on science, arts, or sports. 
$c_{t_i^1}$ (or $c_0$) is the most suitable school for agent $i$ \emph{ex ante}, but $c_{t_i^2}$ may become the most suitable school \emph{ex post} when $s_i+\e_i$ is small (negative) enough. 
$c_0$ is interpreted as a collection of schools that do not have such distinguished features and are underdemanded.

$\omega_i$ denotes agent $i$'s (constant) marginal value of money. A higher $\omega_i$ indicates a greater burden of housing payments at a given price, which can be interpreted as lower wealth or a tighter financial constraint.

The four components of $i$'s type, $(t_i^1, t_i^2)$, $s_i$, $\e_i$, and $\omega_i$, are mutually independent.\footnote{We take the Borel $\sigma$-algebra on the type space $\{1,\ldots,m\}^2 \times [0, 1] \times \{-e, 0, e\} \times \Omega$ to define the type distribution.}
$t_i^1$ is uniformly distributed over $\{1,\ldots,m\}$, and $t_i^2$ is uniformly distributed over $\{1,\ldots,m\} \setminus \{t_i^1\}$ conditional on $t_i^1$. 
This allows us to maintain symmetry among schools and neighborhoods indexed by $k \in \{1,\ldots,m\}$.
Let $F$ denote the cumulative distribution function of $s_i$ on $[0,1]$, which is weakly concave and continuous.
As we elaborate in Section \ref{sec:main}, the weak concavity of $F$ is a crucial assumption for our main results. This restriction is natural, as it implies that signals are not concentrated at the extremes. 
Let $\eta$ denote the probability distribution of $\e_i$ on $\{-e,0,e\}$. 
We assume symmetry: $\eta(e)=\eta(-e)=\pi \in [0,\frac{1}{2})$.\footnote{Throughout the analysis, we focus on $e>0$ and $\pi>0$ because otherwise all agents would attend their neighborhood school under any school choice mechanism, and school choice would play no role in the model.} 
Let $\rho$ denote the probability distribution of $\omega_i$ on $\Omega$, where $E[\omega_i]=1$ without loss of generality.
We assume $e+g\le 1$ without loss of generality, as we allow $F(s_i)=1$ for some $s_i < e+g$.

We consider a two-stage game of housing and school choice, where $\e_i$ is not realized in the first stage:
\begin{enumerate}
 \item Each agent $i$ observes their $(t_i^1,t_i^2,s_i,\omega_i)$ and decides on a neighborhood $n \in N$ to live in.
 \item Preference shocks $\e_i$ are realized.
 Given their chosen neighborhood and the realization of $\e_i$, each agent participates in a school choice mechanism, which determines their assignment to one of the schools in $C$. 
\end{enumerate}
In reality, housing choice and its timing depend on many other factors and vary across agents. 
However, our two-stage model is a good approximation of reality because we can interpret preference shocks $\epsilon_i$ as uncertainty about school fit that remains after the agents settle in their neighborhoods.

This paper studies three school choice mechanisms: the neighborhood assignment rule (N), Deferred Acceptance with neighborhood priority (DA), and Top Trading Cycles with neighborhood priority (TTC).
N assigns an agent to school $c_k$ if and only if they live in neighborhood $n_k$.
DA and TTC compute assignments by taking the agents' rank-order lists of schools and schools' priority rankings as inputs (see \cite{abdulkadirouglu2003school} for the algorithms defined for finite sets of students and schools).
In our model, we apply the DA algorithm of \cite{acy:15} and \cite{azevedo/leshno:16} and the TTC algorithm of \cite{leshno2021cutoff}, which are formally defined in a model with a continuum of students and a finite set of schools.
Under either mechanism, we consider random tie-breaking for those with the same priority level.
We illustrate how DA and TTC work in our setting in Section \ref{sec:DAandTTC}.

Since DA and TTC are strategy-proof for agents, we assume truth-telling in the school choice mechanisms.
Consequently, the outcomes of school choice mechanisms are computed using agents' true preferences and priorities. This allows us to focus on the static housing choice game.
Define a (pure) \emph{strategy} profile of agents (where all agents use the same Bayesian strategy) as a measurable function $\sigma: \{1,\ldots,m\}^2 \times [0,1] \times \Omega \to N$. 
Let
$\varphi(t_i^1,t_i^2,s_i,\e_i,\omega_i,\sigma) \in \Delta C$ be the stochastic school outcome of an agent with type $(t_i^1,t_i^2,s_i,\e_i,\omega_i)$ under a school choice mechanism $\varphi$ when all agents play strategy $\sigma$.
With a slight abuse of notation, we use $u_i$ to denote the expected utility derived from a stochastic school outcome.
$D_n(\sigma) \coloneqq \sum_{(t_i^1,t_i^2): t_i^1 \neq t_i^2}\frac{1}{m-1}\sum_{\omega_i \in \Omega}\rho(\omega_i)\int_0^1 \1_{\{\sigma(t_i^1,t_i^2,s_i,\omega_i)=n\}}dF(s_i)$ denotes the demand for neighborhood $n$ under strategy $\sigma$. Let $\Sigma$ denote the set of strategies $\sigma$ that satisfy $D_{n_k}(\sigma) \le q$ for each $n_k \in N \setminus \{n_0\}$; that is, the housing supply constraint is not violated. Combined with a condition on prices, given below, $\Sigma$ will form our housing market-clearing condition.
\begin{df}
 $(\sigma, \boldsymbol{p})$ is a \emph{symmetric equilibrium} of the neighborhood choice game induced by a school choice mechanism $\varphi$ if 
 \begin{enumerate}
  \item (Optimality) Given $\boldsymbol{p}$, for each $(t_i^1,t_i^2,s_i,\omega_i) \in \{1,\ldots,m\}^2 \times [0, 1] \times \Omega$, 
 \begin{equation*}
  \sigma(t_i^1,t_i^2,s_i,\omega_i) \in \argmax_{n \in N} E_{\epsilon_i}\Big[u_i(n,\varphi(t_i^1,t_i^2,s_i,\e_i,\omega_i,\sigma),\boldsymbol{p}|t_i^1,t_i^2,s_i+\e_i,\omega_i)\Big].
 \end{equation*}
 
 \item (Housing market clearing) $\sigma \in \Sigma$, and for each $n_k \in N \setminus \{n_0\}$, $D_{n_k}(\sigma)< q \Rightarrow p_k=0$.

 \item (Symmetry) $p_k=p_l$ for any $k,l \in \{1,\ldots,m\}$.\footnote{We normalize $p_0=0$ without loss of generality.} 
 \end{enumerate}
\end{df}
The optimality condition involves two types of uncertainty.
First, $u_i$ is the expected utility from a stochastic school assignment $\varphi(t_i^1,t_i^2,s_i,\e_i,\omega_i,\sigma)$, where uncertainty is generated by the tie-breaker of mechanism $\varphi$. Second, since each agent $i$ chooses their neighborhood before learning their own preference shock, each agent $i$ evaluates their expected utility with respect to $\e_i$. Note that the aggregate uncertainty about the preference shocks of other agents disappears because the agent set is a continuum. By the symmetry of all indices in $\{1,\ldots,m\}$, our analysis focuses on agents with $t_i^1=1$ without loss of generality.

We next define our key measure of interest: \emph{segregation by wealth} at both neighborhood and school levels.
A residential matching is defined as a measurable function $\nu: \{1,\ldots,m\}^2 \times [0,1] \times \Omega \to N$ that satisfies $\sum_{(t_i^1,t_i^2): t_i^1 \neq t_i^2}\frac{1}{m-1}\sum_{\omega_i \in \Omega}\rho(\omega_i) \int_0^1 \1_{\{\nu(t_i^1,t_i^2,s_i,\omega_i)=n\}}dF(s_i) \le q$ for all $n \in N\setminus\{n_0\}$.
Similarly, a school matching is defined as a measurable function $\mu: \{1,\ldots,m\}^2 \times [0,1] \times \{-e,0,e\} \times \Omega \to C$ that satisfies $\sum_{(t_i^1,t_i^2): t_i^1 \neq t_i^2}\frac{1}{m-1}\sum_{\omega_i \in \Omega}\rho(\omega_i)\sum_{\e_i \in \{-e,0,e\}}\eta(\e_i)\int_0^1 \1_{\{\mu(t_i^1,t_i^2,s_i,\e_i,\omega_i)=c\}}dF(s_i) \le q$ for all $c \in C\setminus\{c_0\}$. 
The average wealth parameters of a neighborhood $n \in N$ and a school $c \in C$, denoted $\bar{\omega}(\cdot,\cdot)$, are given respectively by: 
\begin{equation*}
 \begin{split}
  & \bar{\omega}(\nu,n) := \frac{\sum_{(t_i^1,t_i^2): t_i^1 \neq t_i^2}\frac{1}{m-1}\sum_{\omega_i \in \Omega}\rho(\omega_i)\omega_i \int_0^1 \1_{\{\nu(t_i^1,t_i^2,s_i,\omega_i)=n\}}dF(s_i)}{\sum_{(t_i^1,t_i^2): t_i^1 \neq t_i^2}\frac{1}{m-1}\sum_{\omega_i \in \Omega}\rho(\omega_i) \int_0^1 \1_{\{\nu(t_i^1,t_i^2,s_i,\omega_i)=n\}}dF(s_i)}, \text{ and } \\
  & \bar{\omega}(\mu,c) := \frac{\sum_{(t_i^1,t_i^2): t_i^1 \neq t_i^2}\frac{1}{m-1}\sum_{\omega_i \in \Omega}\rho(\omega_i)\omega_i\sum_{\e_i \in \{-e,0,e\}}\eta(\e_i)\int_0^1 \1_{\{\mu(t_i^1,t_i^2,s_i,\e_i,\omega_i)=c\}}dF(s_i)}{\sum_{(t_i^1,t_i^2): t_i^1 \neq t_i^2}\frac{1}{m-1}\sum_{\omega_i \in \Omega}\rho(\omega_i)\sum_{\e_i \in \{-e,0,e\}}\eta(\e_i)\int_0^1 \1_{\{\mu(t_i^1,t_i^2,s_i,\e_i,\omega_i)=c\}}dF(s_i)}.
 \end{split}
\end{equation*}

\begin{df}
 A residential matching $\nu$ has \emph{greater segregation by wealth} than another residential matching $\nu'$ if $|\bar{\omega}(\nu,n)-1| > |\bar{\omega}(\nu',n)-1|$ for all $n \in N$.
 Two residential matchings $\nu$ and $\nu'$ have \emph{the same level of segregation by wealth} if $|\bar{\omega}(\nu,n)-1|=|\bar{\omega}(\nu',n)-1|$ for all $n \in N$.
\end{df}
\begin{df}
 A school matching $\mu$ has \emph{greater segregation by wealth} than another school matching $\mu'$ if $|\bar{\omega}(\mu,c)-1| > |\bar{\omega}(\mu',c)-1|$ for all $c \in C$.
 Two school matchings $\mu$ and $\mu'$ have \emph{the same level of segregation by wealth} if $|\bar{\omega}(\mu,c)-1|=|\bar{\omega}(\mu',c)-1|$ for all $c \in C$.
\end{df}
These definitions measure across-neighborhood or across-school variations of the wealth types.
Intuitively, a neighborhood or school is more segregated in one matching than in another if its average wealth parameter deviates further from the population average, which is equal to one. When this holds for every neighborhood or school, we say that the matching is more segregated. When all neighborhoods in $N\setminus\{n_0\}$ are symmetric and overdemanded, it is sufficient to focus on one neighborhood, such as $n_0$ or $n_1$: one matching is more segregated than another if and only if the average wealth parameter at $n_0$ or $n_1$ deviates more from one. Similarly, when schools in $C\setminus\{c_0\}$ are symmetric and oversubscribed---have more applicants than they can accept---one can focus on a single school.

Lemma \ref{lm:uniqueness} establishes that in the class of symmetric equilibria we focus on, the equilibrium is unique under all three mechanisms. Then, we can naturally extend the definitions of segregation from matchings to mechanisms.\footnote{Although tie-breakers are embedded in the mechanisms, the outcome of each mechanism is deterministic almost surely because of the continuum of agents. Thus, we can apply our segregation measure defined for deterministic matchings to mechanisms.} For a school choice mechanism $\varphi \in \{N, DA, TTC\}$, let $(\sigma^{\varphi},p^{\varphi})$ denote its unique symmetric equilibrium.\footnote{Note that, due to symmetry, housing prices under mechanism $\varphi$ are equal across all $n \in N\setminus\{n_0\}$, so we use the scalar $p^{\varphi}$ to denote this common price.}
We say that a mechanism $\varphi$ results in \emph{greater neighborhood segregation} (resp., \emph{greater school segregation)} than $\varphi'$ if the residential (resp., school) matching achieved under $(\sigma^{\varphi}, p^{\varphi})$ has greater segregation by wealth than that under $(\sigma^{\varphi'}, p^{\varphi'})$. Similarly, we say that $\varphi$ and $\varphi'$ result in \emph{the same level of neighborhood segregation} (resp., \emph{school segregation)} if the respective matchings under $(\sigma^{\varphi},p^{\varphi})$ and $(\sigma^{\varphi'},p^{\varphi'})$ have the same level of segregation by wealth.

Throughout the paper, we assume the following condition: 

\medskip
\noindent
{\bf Assumption 1.} $F(g)<1-q<F(e-g)<1$.
\medskip

This essentially means that, given $q$, $g$ is small enough and $e$ is large enough.
This condition allows us to focus on cases where $n_k$ and $c_k$ with $k \in \{1,\ldots,m\}$ are oversubscribed, but some students residing in $n_k$ may prefer another oversubscribed school $c_l \in C\setminus \{c_0,c_k\}$. 
We elaborate on the equilibrium under this assumption in Section \ref{sec:cutoff_eq}.

\subsection{How DA and TTC work in our model}\label{sec:DAandTTC}

Neighborhood priority guarantees admission to a local school under both DA and TTC, but it has different implications for admission to other oversubscribed schools. Since school $c_0$, which can accommodate all students, is no lower than the second choice for any agent, only the following cases can arise under either mechanism: the agent is
\begin{enumerate}[noitemsep, nolistsep]
 \item assigned to their neighborhood school (as their first choice),
 \item assigned to $c_0$ (as their first choice), or
 \item ``applies'' (in DA) or ``points'' (in TTC) to an oversubscribed school outside their zone and is assigned either to that school (as their first choice), or to $c_0$ (as their second choice) if rejected by the first choice.
\end{enumerate}
The only difference between DA and TTC arises in the third case.
Under DA, an agent in $n_k \in N \setminus \{n_0\}$ who applies to $c_l \in C\setminus \{c_0,c_k\}$ is treated like any other out-of-zone applicant, including residents of $n_0$, via a tie-breaking lottery. As a result, they may be rejected from $c_l$; we denote the probability of rejection by $r^{DA}(\sigma)$ when all agents play a market-clearing strategy $\sigma \in \Sigma$. By contrast, under the symmetric equilibrium of TTC, an agent in $n_k$ whose top choice is $c_l$ always finds a cycle to be assigned to $c_l$ because the quotas of all oversubscribed schools are filled simultaneously. In this case, only agents in $n_0$ may be rejected from $c_l$; we denote the rejection probability by $r^{TTC}(\sigma)$, for $\sigma \in \Sigma$. Note that under both DA and TTC, these rejection probabilities are identical across all oversubscribed schools due to symmetry.

\subsection{Cutoff structure of the equilibrium}\label{sec:cutoff_eq}

Under any of the three mechanisms, an agent $i$ with $t_i^1=1$ never strictly prefers neighborhood $n_k \in N\setminus \{n_0,n_1\}$ to $n_1$ because housing prices are identical across these neighborhoods and the expected benefit of living in $n_1$ is not lower than $n_k$. Thus, we suppose that such agents choose between $n_1$ and $n_0$.\footnote{Under N and DA, agents with $t_i^1=1$ never choose $n_k$ with $k \in \{2,\ldots,m\}$ in equilibrium. Under TTC, agents are indifferent between all neighborhoods in $N\setminus \{n_0\}$ in a symmetric equilibrium. To simplify our analysis, we focus on an equilibrium in which agents with primary type $t_i^1$ do not choose $n_k$ with $k \in \{2,\ldots,m\}$.}
We first show that their incentive to live in $n_1$ increases with $s_i$ for all three mechanisms. 
Focusing on agents with $t_i^1=1$, let $Eu_i^{\varphi}(n,r,p|s_i,\omega_i)$ be the expected utility of an agent $i$ with $(s_i,\omega_i)$ when $i$ chooses $n$, the rejection probability under mechanism $\varphi$ is $r>0$, and the housing price in every $n \in N\setminus\{n_0\}$ is $p$.\footnote{Although there is no school choice under N, we use a general rejection probability $r>0$ for notational consistency. We model N as a school choice mechanism in which all $n_k$ ($k \in \{1,\ldots,m\}$) residents are assigned to $c_k$ for sure, and all $n_0$ residents are assigned to their primary-fit schools with probability $1-r$. In equilibrium, we always consider $r=1$.}
Define $\Delta u^\varphi(r^\varphi,p^\varphi|s_i,\omega_i) = E_{\epsilon_i}u_i^{\varphi}(n_1,r^\varphi,p^\varphi|s_i+\epsilon_i,\omega_i)-E_{\epsilon_i}u_i^{\varphi}(n_0,r^\varphi,p^\varphi|s_i+\epsilon_i,\omega_i) \geq 0$ to be the utility gain from living in $n_1$ relative to living in $n_0$. 

\begin{lm}\label{lm:monotone}
For any $\varphi \in \{N, DA, TTC\}$, $r>0$, $p\ge 0$, and $\omega_i \in \Omega$, $\Delta u_i^{\varphi}(r,p|s_i,\omega_i)$ is (i) increasing in $s_i$ weakly for $s_i \in [0,g]$, (ii) strictly for $s_i \in [g,1]$, and (iii) $\Delta u_i^{\varphi}(r,0|g,\omega_i) \geq 0$.
\end{lm}

Statements (i)-(ii) imply that any strategy $\sigma$ that is a best response to $r>0$ and $p\ge 0$\footnote{Although we slightly abuse the notion of best response when we use this wording, note that the strategies of agents other than $i$ affect that payoff of $i$ via aggregate variables $r$ and $p$. Thus, this wording should be taken as saying that other players use strategies that lead to rejection probability $r$ and price $p$.}
 under $\varphi$ has a semi-cutoff structure: $\sigma(s_i,\omega_i)=n_1$ for some $s_i \in (g,1]$ implies $\sigma(s_i',\omega_i)=n_1$ for any $s_i' \in [s_i,1]$.
Statements (ii)-(iii) and Assumption 1 imply that $n_1$ is overdemanded when $p=0$, and thus, any symmetric equilibrium $(\sigma, p)$ must satisfy $D_{n_1}(\sigma)=q$ and $p>0$.

Given the original problem, consider its ``reduced problem,'' in which $\Omega=\{1\}$ and all other parameters are unchanged.
While the reduced problem is uninteresting for our research questions, it is useful for clarifying our next assumption.
Lemma \ref{lm:monotone} immediately implies the existence of a unique (except for the strategies of agents with measure zero), symmetric equilibrium in the reduced problem. 
To see this, consider a strategy $\hat{\sigma}$ such that $\hat{\sigma}(s_i,1)=n_0$ for any agent $i$ with $s_i \in [0,F^{-1}(1-q))$ and $\hat{\sigma}(s_i,1)=n_1$ for any agent $i$ with $s_i \in (F^{-1}(1-q), 1]$.
Note that $F^{-1}(1-q)>g$ by Assumption 1.
Since there is only one wealth type, $\hat{\sigma}$ is the only candidate for the equilibrium strategy, as any other strategy would violate either housing market clearing condition or monotonicity with respect to $s_i$.
Consider the associated rejection probability $\hat{r}^{\varphi}$: 1 for $\varphi=N$, $r^{DA}(\hat{\sigma})$ for $\varphi=DA$, and $r^{TTC}(\hat{\sigma})$ for $\varphi=TTC$.
Since $\Delta u_i^{\varphi}(\hat{r}^{\varphi},p|s_i,1)$ is continuous and strictly decreasing in $p$, there exists a unique $\hat{p}^{\varphi}>0$ such that $\Delta u_i^{\varphi}(\hat{r}^{\varphi},p|s_i,1) = 0$, implying that $\hat\sigma$ is the best response to $(\hat{r}^{\varphi}, \hat{p}^{\varphi})$. 
This proves that $(\hat{\sigma},\hat{p}^{\varphi})$ is a unique equilibrium of the reduced problem under $\varphi$.

We next impose the key assumption on $\Omega$.
Let $\bar{p}^{\varphi}$ be the upper bound of the price range defined as follows: $\bar{p}^N \coloneqq \hat{r}^N(1-q-g)$, $\bar{p}^{DA} \coloneqq \hat{r}^{DA}[(1-\pi)(1-q-g)+\pi e]$, and $\bar{p}^{TTC} \coloneqq \hat{r}^{TTC}[(1-2\pi)(1-q)+2\pi e-g]$.\footnote{$\bar{p}^{\varphi}$ coincides with the equilibrium price of the reduced model for uniform $F$.}

\medskip
\noindent
{\bf Assumption 2.} 
For any $\varphi \in \{N,DA,TTC\}$, $\omega_i \in \Omega$, and $p \in [\hat{p}^{\varphi},\bar{p}^{\varphi}]$, $\Delta u_i^{\varphi}(\hat{r}^{\varphi},p|g,\omega_i) < 0 < \Delta u_i^{\varphi}(\hat{r}^{\varphi},p|e-g,\omega_i)$.
\medskip

\begin{lm}\label{lm:uniqueness}
For any $\varphi \in \{N,DA,TTC\}$, there exists a symmetric equilibrium $(\sigma^{\varphi},p^{\varphi})$ of the original problem, in which for any $\omega_i \in \Omega$, there is a cutoff signal type $s_{\omega_i}^{\varphi} \in (g,e-g)$ such that $\sigma^{\varphi}(s_i,\omega_i)=n_0$ for any $s_i \in [0,s_{\omega_i}^{\varphi})$ and $\sigma^{\varphi}(s_i,\omega_i)=n_{t_i^1}$ for any $s_i \in (s_{\omega_i}^{\varphi},1]$.
When all equilibrium cutoff signal types are in $(g,e-g)$, they are unique.
\end{lm}

By Assumption 1, the cutoff signal type $F^{-1}(1-q)$ of the reduced problem is in $(g, e-g)$.
We focus on such cases because some agents residing in $n_1$ prefer another oversubscribed school with a positive probability, which allows us to study the differentiated effects of the three mechanisms.
Assumption 2 restricts the upper and lower bounds of $\Omega$ in a way that we can still find the cutoff signal type within the same interval, $(g, e-g)$, for any $\omega_i \in \Omega$.
We leave the cases with larger wealth disparities outside the scope of our analysis because, in such cases, some wealth types may be entirely displaced from overdemanded neighborhoods or may fully occupy their housing supply.

Lemma \ref{lm:uniqueness} follows from Assumption 2, Lemma \ref{lm:monotone}, and the capacity constraint (i.e., the housing market clearing condition) 
\begin{equation}\label{eq:cap}
 \sum_{\omega_i \in \Omega} \rho(\omega_i)F(s_{\omega_i}^{\varphi})=1-q.
\end{equation}
The equilibrium price $p^{\varphi}$ of the original problem can be found in $[\hat{p}^{\varphi},\bar{p}^{\varphi}]$ to satisfy the capacity constraint.\footnote{The rejection probability of the original problem coincides with that of the reduced problem, $\hat{r}$, because only the weighted average of $F(s_{\omega_i}^{\varphi})$ matters. See \ref{sec:proof_s_sorting} for the derivation of the rejection probabilities under DA and TTC.}

Since $\{s_{\omega_i}^{\varphi}\}_{\omega_i\in \Omega}$ are characterized by $\Delta u_i^{\varphi}(r^{\varphi},p^{\varphi}|s_{\omega_i}^{\varphi},\omega_i)=0$ for each $\varphi \in \{N,DA,TTC\}$, where $r^N:=1$, $r^{DA} := r^{DA}(\sigma^{DA})$ and $r^{TTC} := r^{TTC}(\sigma^{TTC})$, the proof of Lemma \ref{lm:monotone} provides the closed-form solutions for the cutoff signal types $s_{\omega_i}^{\varphi} \in (g,e-g)$: 
\begin{equation}\label{eq:cutoffs}
 \begin{split}
 & s_{\omega_i}^N-g = \omega_i \frac{p^N}{r^N}, \\
 & (1-\pi)(s_{\omega_i}^{DA}-g)+\pi e = \omega_i \frac{p^{DA}}{r^{DA}}, \\
 & (1-2\pi)s_{\omega_i}^{TTC}+2\pi e-g = \omega_i \frac{p^{TTC}}{r^{TTC}}, \\
 \end{split}
 \end{equation}
for each $\omega_i \in \Omega$.
We will exploit these equations to derive our main results.

\section{Main results}\label{sec:main}
\subsection{Neighborhood-level wealth segregation}\label{sec:n_sorting}

First, we establish an unambiguous ranking of the three mechanisms by their levels of neighborhood segregation. A mechanism induces greater segregation when the differences in cutoffs across wealth levels are larger; this is summarized by dispersion $d^\varphi$ in the following expression:
\begin{equation}\label{eq:diff}
 s_{\omega_i}^{\varphi}-s_{\omega_j}^{\varphi} = d^{\varphi}(\omega_i-\omega_j).
 \end{equation}
For each mechanism $\varphi \in \{N, DA, TTC\}$, the corresponding value of dispersion follows from the equilibrium conditions in (\ref{eq:cutoffs}):
$d^N \coloneqq p^N$,
$d^{DA} \coloneqq \frac{1}{1 - \pi} \cdot \frac{p^{DA}}{r^{DA}}$, and
$d^{TTC} \coloneqq \frac{1}{1 - 2\pi} \cdot \frac{p^{TTC}}{r^{TTC}}$. To rank these values, we take the weighted average over wealth indices $\omega_i$ in equations (\ref{eq:cutoffs})---using the expectation symbol $E$ to denote this average---to obtain:
 \begin{equation}\label{eq:expected_cutoff}
 \begin{split}
 & E[s_{\omega_i}^N]-g = d^N, \\
 & E[s_{\omega_i}^{DA}]-g+\frac{\pi e}{1-\pi} = d^{DA}, \\
 & E[s_{\omega_i}^{TTC}]+\frac{2\pi e-g}{1-2\pi} = d^{TTC}. \\
 \end{split}
 \end{equation}

Equations (\ref{eq:cap}), (\ref{eq:diff}), and (\ref{eq:expected_cutoff}) jointly determine $d^{\varphi}$ and $E[s_{\omega_i}^{\varphi}]$ in equilibrium. 
The concavity of $F$, the capacity constraint (\ref{eq:cap}), and equations (\ref{eq:diff}) imply Lemma \ref{lmA}.

\begin{lm}\label{lmA}
 For each mechanism $\varphi \in \{N, DA, TTC\}$, $E[s_{\omega_i}^{\varphi}]$ is weakly increasing in $d^{\varphi}$. 
\end{lm}

Further, Lemma \ref{lmA} and equations (\ref{eq:expected_cutoff}) determine the unique pair of $E[s_{\omega_i}^{\varphi}]$ and $d^{\varphi}$ for each $\varphi$, which establishes the ranking of these values across the three mechanisms.

\begin{lm}\label{lmB}
 $E[s_{\omega_i}^N] \le E[s_{\omega_i}^{DA}] \le E[s_{\omega_i}^{TTC}] \le 1-q$ and $d^N<d^{DA}<d^{TTC}$.
\end{lm}

\begin{figure}[htbp]
    \centering
    \begin{tikzpicture}
        \begin{axis}[
            xscale=1.1,   
            yscale=1,   
            axis lines=middle,
            xlabel={$E[s_{\omega_i}]$},
            ylabel={$d^\varphi$},
            xmin=0.3, xmax=0.8,
            ymin=0, ymax=3,
            xticklabels={$F^{-1}(1-q)=0.36$,$1-q=0.6$},
            yticklabels={0.6,1.1,2.6},
            xtick={0.36,0.6},
            ytick={0.6,1.1,2.6},
            samples=100,
            clip=false
        ]
    
        \addplot[black, thick, domain=0.34:0.65] {(x + 2)} node[right] {$\frac{1}{1-2\pi}\frac{p^{TTC}}{r^{TTC}}$};
    
        \addplot[black, thick, domain=0.34:0.65] {(x + 1/2)} node[right] {$\frac{1}{1-\pi}\frac{p^{DA}}{r^{DA}}$};
    
        \addplot[black, thick, domain=0.34:0.65] {x} node[right]  {$p^N$};
        
        \node[below] at (axis cs:0.7, 2) {(\ref{eq:expected_cutoff})};
    
        \addplot[red, ultra thick, domain=0.36:0.45] {(48/25)*sqrt((25*x-9))};
        \node[below right] at (axis cs:0.423743, 2.423743) {(\ref{eq:cap}) and (\ref{eq:diff})};
    
        \draw[blue, thick] (axis cs:0.6, 0) -- (axis cs:0.6, 2.8) node[above,xshift=10] {uniform};
    
        \fill[red] (axis cs:0.361417, 0.361417) ellipse [x radius=2pt, y radius=2pt];
        \fill[red] (axis cs:0.368179, 0.868179) ellipse [x radius=2pt, y radius=2pt];
        \fill[red] (axis cs:0.423743, 2.423743) ellipse [x radius=2pt, y radius=2pt];
    
        \fill[blue] (axis cs:0.6, 0.6) ellipse [x radius=2pt, y radius=2pt];
        \fill[blue] (axis cs:0.6, 1.1) ellipse [x radius=2pt, y radius=2pt];
        \fill[blue] (axis cs:0.6, 2.6) ellipse [x radius=2pt, y radius=2pt];
    
        \end{axis}
    \end{tikzpicture}
    \caption{Equilibrium $E[s_{\omega_i}^{\varphi}]$ and $d^{\varphi}$ for each $\varphi$}
    \label{fig:lmB}
    \begin{tabnotes}
        The figure shows the equilibrium values of $d^\varphi$ for mechanisms $\varphi \in \{N, DA, TTC\}$. Three thin, slanted black lines represent the linear restrictions from equations (\ref{eq:expected_cutoff}) for each mechanism. They are 45-degree lines, showing with a lower slope in the figure due to scaling. The red curve represents the solutions to equations (\ref{eq:cap}) and (\ref{eq:diff}) for different values of $d^\varphi$. The vertical blue line indicates the position of the red curve in the special case where $F$ is uniform. The intersections of the black lines with the red curve show the equilibrium values of $d^\varphi$ for the three mechanisms for a general weakly concave $F$, while their intersections with the blue line show the corresponding values for the uniform distribution. The parameters in the picture are as in our leading numeric example with $F(x)=\sqrt{x}$.
    \end{tabnotes}
\end{figure}
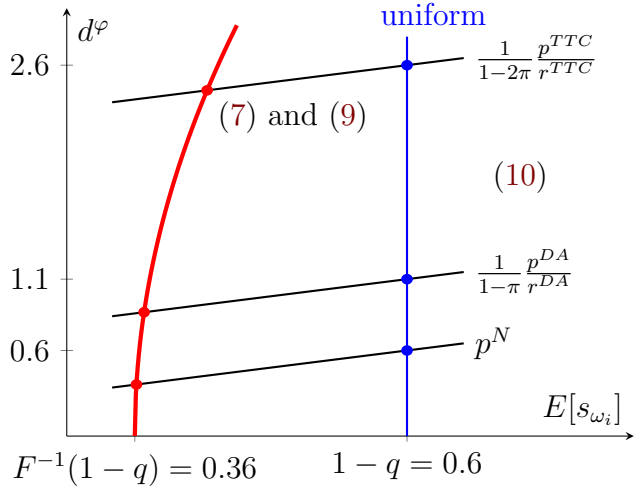

Figure \ref{fig:lmB} provides the intuition for Lemma \ref{lmB}.
Three thin, slanted black lines represent the linear restrictions from equations (\ref{eq:expected_cutoff}) and the thick red curve represents the restriction derived in Lemma \ref{lmA} (i.e., equations (\ref{eq:cap}) and (\ref{eq:diff})). When $d^\varphi = 0$, all cutoffs are the same (equation (\ref{eq:diff})) and $E[s_{\omega_i}^\varphi]$ solves $F\left(E[s_{\omega_i}^\varphi]\right)=1-q$. As $d^\varphi$ increases, the average cutoff $E[s_{\omega_i}^\varphi] = \sum_{\omega_i \in \Omega} \rho(\omega_i)s_{\omega_i}^{\varphi}$ must weakly increase to maintain $\sum_{\omega_i \in \Omega} \rho(\omega_i)F(s_{\omega_i}^{\varphi}) = 1-q$, due to the weak concavity of $F$. Under the uniform distribution $F$, the red curve becomes a vertical blue line, as would have been obtained in the example in Section \ref{sec:Example}. In that case, $\sum_{\omega_i \in \Omega} \rho(\omega_i)s_{\omega_i}^{\varphi} = \sum_{\omega_i \in \Omega} \rho(\omega_i)F(s_{\omega_i}^{\varphi}) = 1-q$ and equations (\ref{eq:expected_cutoff}) entirely determine $d^{\varphi}$ because the black slanted lines are parallel and the blue line is vertical. The equilibrium values of $E[s_{\omega_i}^{\varphi}]$ and $d^{\varphi}$ for each mechanism $\varphi$ correspond to the intersections of the black lines and the red curve (for a general weakly concave $F$) or the blue line (for uniform $F$).

Lemma \ref{lmB} implies that the dispersion $d^{\varphi}$ grows from N to DA to TTC, which proves our first main result.

\begin{thm}\label{thm:n_sorting}
The following ranking holds for neighborhood segregation by wealth: 
 
  1. DA results in greater neighborhood segregation by wealth than N. 
 
  2. TTC results in greater neighborhood segregation by wealth than DA.
\end{thm}

Theorem \ref{thm:n_sorting} establishes an unambiguous ranking of the three mechanisms by neighborhood-level segregation. We next discuss segregation at the school level.

\subsection{School-level wealth segregation}\label{sec:s_sorting}

School-level segregation is shaped by two forces: segregation ``inherited'' from neighborhoods and desegregation generated by flexible school choice mechanisms. 
Figure \ref{fig:lmB} from the previous section hints at the importance of the comparison of $F^{-1}(1-q)$ with the averaged cutoffs. Our school segregation results will depend on these related quantities, which we define below. Recall that, in a perfectly unsegregated neighborhood, the mass of agents with $\omega_i$ in $n_1$ would be $\rho(\omega_i)q$, while the actual mass under mechanism $\varphi$ is $\rho(\omega_i)\big(1-F(s_{\omega_i}^{\varphi})\big)$. The difference between these values measures the extent to which agents of type $\omega_i$ are ``over-'' or ``under-represented'' in $n_1$. For the richest agents, the difference $q-\big(1-F(s_{\omega_i}^{\varphi})\big)$ is negative for all mechanisms; for the poorest agents, it is positive. We define $\tilde{\Omega}(\varphi, \varphi') \subset \Omega$ as the set of wealth types for which the signs of $q - \big(1 - F(s_{\omega_i}^{\varphi})\big)$ and $q - \big(1 - F(s_{\omega_i}^{\varphi'})\big)$ are the same. This identifies the wealth types that are treated ``similarly'' by the two mechanisms $\varphi$ and $\varphi'$. 
For any $\omega_i \in \tilde{\Omega}(\varphi,\varphi')$, we define $\frac{|q - (1 - F(s_{\omega_i}^{\varphi'}))|}{|q - (1 - F(s_{\omega_i}^{\varphi}))|}$ as the ``neighborhood segregation expansion rate'' for $\omega_i$ when moving from mechanism $\varphi$ to $\varphi'$. Comparing school segregation between mechanisms thus reduces to comparing this expansion rate, which effectively measures how far neighborhood segregation is from the unsegregated benchmark, against how much each mechanism desegregates schools, a measure quantified in the theorem below. 

\begin{thm}\label{thm:s_sorting}
The following ranking holds for school segregation by wealth: 

 1. DA results in greater (resp., smaller) school segregation by wealth than N if the neighborhood segregation expansion rate from N to DA is greater (resp., smaller) than $\frac{1}{r^{DA}(1-\pi)}$ for every $\omega_i \in \tilde{\Omega}(N,DA)$.

 2. TTC results in greater (resp., smaller) school segregation by wealth than N if the neighborhood segregation expansion rate from N to TTC is greater (resp., smaller) than $\frac{1}{r^{TTC}}$ for every $\omega_i \in \tilde{\Omega}(N,TTC)$.

 3. TTC results in greater school segregation by wealth than DA if the neighborhood segregation expansion rate from DA to TTC is greater than $\frac{r^{DA}(1-\pi)}{r^{TTC}}$ for every $\omega_i \in \tilde{\Omega}(DA,TTC)$.
\end{thm}

We know from Theorem \ref{thm:n_sorting} that the expansion rate from N to TTC is higher than that from N to DA. Furthermore, the inequalities $\frac{1}{r^{DA}(1-\pi)} > \frac{1}{r^{TTC}} \geq 1$ always hold. Thus, while the condition in statement 2 for when N results in lower school segregation than TTC is weak, the comparison of N and DA (statement 1) is more nuanced, because the expansion rate is lower and the force reducing school segregation, captured by $\frac{1}{r^{DA}(1-\pi)}$, is stronger. The sufficient condition of statement 3 holds in most standard cases. For instance, when TTC moves the cutoffs ``outward'' from $F^{-1}(1-q)$ compared to DA, i.e., all cutoffs above $F^{-1}(1-q)$ increase while those below it decrease, Theorem \ref{thm:n_sorting} implies that the expansion rate is greater than 1 for every $\omega_i \in \tilde{\Omega}(DA,TTC)=\Omega$, which, in turn, is greater than $\frac{r^{DA}(1-\pi)}{r^{TTC}}$.

\begin{figure}
    \centering
    \def\practicaldata{
    23,16,14,9,9,7,6,3,3,2,1,0,0,0,
    29,17,19,12,12,8,8,4,4,2,1,1,0,0,
    21,16,14,9,8,5,4,2,1,1,0,0,0,0,
    29,17,19,11,11,7,6,3,2,1,0,0,0,0,
    19,14,12,8,7,5,2,1,0,0,0,0,0,0,
    27,16,16,9,9,5,3,1,0,0,0,0,0,0,
    19,14,9,6,5,3,0,0,0,0,0,0,0,0,
    25,16,14,8,7,4,0,0,0,0,0,0,0,0,
    18,12,6,5,0,0,0,0,0,0,0,0,0,0,
    23,14,9,5,0,0,0,0,0,0,0,0,0,0,
    12,7,0,0,0,0,0,0,0,0,0,0,0,0,
    20,9,0,0,0,0,0,0,0,0,0,0,0,0,
    0,0,0,0,0,0,0,0,0,0,0,0,0,0,
    0,0,0,0,0,0,0,0,0,0,0,0,0,0}
    
    \newcommand{\getpracticalvalue}[2]{%
      \pgfmathparse{{\practicaldata}[#1]}%
      \pgfmathtruncatemacro{#2}{\pgfmathresult}%
    }
    
    \newcommand{\getcolor}[1]{%
      \ifnum#1=0 green\else%
      \ifnum#1<5 green!25!yellow\else%
      \ifnum#1<10 yellow\else%
      \ifnum#1<15 yellow!50!red\else%
      \ifnum#1<20 yellow!25!red\else%
      red\fi\fi\fi\fi\fi%
    }
    
    \begin{tikzpicture}[scale=0.6]
        \foreach \i in {0,...,13} {
          \foreach \j in {0,...,13} {
            \pgfmathtruncatemacro{\index}{\i*14+\j}
            
            \getpracticalvalue{\index}{\cellval}
            
            \edef\mycolor{\getcolor{\cellval}}
            \fill[\mycolor] (\i,\j) rectangle ++(1,1);
            
            \node[font=\scriptsize] at (\i+0.5,\j+0.5) {\cellval};
          }
        }
        \node at (-0.25,-1.2) {$\rho^P \rightarrow$};
        \foreach \i [evaluate=\i as \val using \i*0.1+0.2] in {0,...,6} {
          \node at (\i*2+1,-0.45) {\pgfmathprintnumber[fixed,precision=1]{\val}};
        }
        \node at (-1.5,-0.25) {$q \uparrow$};
        \foreach \i [evaluate=\i as \val using \i*0.1+0.2] in {0,...,6} {
          \node[font=\normalsize] at (-0.65,\i*2+1) {\pgfmathprintnumber[fixed,precision=1]{\val}};
        }
        \draw[thin, gray] (0,0) grid (14,14);
        
        \foreach \x in {0,2,...,14} {
          \draw[very thick] (\x,0) -- (\x,14);
        }
        \foreach \y in {0,2,...,14} {
          \draw[very thick] (0,\y) -- (14,\y);
        }
    \end{tikzpicture}

    \begin{tikzpicture}[scale=0.8]
        \pgfmathsetmacro{\step}{2.5}
        \pgfmathsetmacro{\shift}{0.5}
        \pgfmathsetmacro{\yshift}{0.8}
        \coordinate (start) at (1.5, 0);
        \fill[green] (start) rectangle ++(0.5,0.3) node[right,black] at ($(start) + (\shift, 0.15)$) {0\%};
        \fill[green!25!yellow] ($(start) + (\step, 0)$) rectangle ++(0.5,0.3) node[right, black] at ($(start) + (\step, 0) + (\shift, 0.15)$) {1-4\%};
        \fill[yellow] ($(start) + (2*\step, 0)$) rectangle ++(0.5,0.3) node[right,black] at ($(start) + (2*\step, 0) + (\shift, 0.15)$) {5-9\%};
        \fill[yellow!50!red] ($(start) + (0, -\yshift)$) rectangle ++(0.5,0.3) node[right,black] at ($(start) + (\shift, 0.15) + (0, -\yshift)$) {10-14\%};
        \fill[yellow!25!red] ($(start) + (\step, 0) + (0, -\yshift)$) rectangle ++(0.5,0.3) node[right,black] at ($(start) + (\step, 0) + (\shift, 0.15) + (0, -\yshift)$) {15-20\%};
        \fill[yellow!25!red] ($(start) + (2*\step, 0) + (0, -\yshift)$) rectangle ++(0.5,0.3) node[right,black] at ($(start) + (2*\step, 0) + (\shift, 0.15) + (0, -\yshift)$) {$>$20\%};
        \node[black] at (9.25,-0.2) {$\pi=$};
        \draw[black] (10,-1) rectangle ++(.75,.75) node[midway] {0.3};
        \draw[black] (10.75,-1) rectangle ++(.75,.75) node[midway] {0.4};
        \draw[black] (10,-0.25) rectangle ++(.75,.75) node[midway] {0.1};
        \draw[black] (10.75,-0.25) rectangle ++(.75,.75) node[midway] {0.2};
    \end{tikzpicture}
    \caption{The percentage of single-kink cdfs that result in DA being less segregated at the school level than N for $(\pi,\rho^P,q) \in [0.1,0.4]\times[0.2,0.8]^2$}
    \label{fig:poor_q_pi_DA_N}
    \begin{tabnotes}
        There are $7 \times 7$ large boxes, with each large box containing $2 \times 2$ small boxes. Each large box show the results for $(\rho^P,q)$ parameters within $\{0.2,\dots,0.8\}^2$, and each small box show the results for $\pi \in \{0.1,0.2,0.3,0.4\}$. Each small box shows the percentage of times when a single-kink distribution results in DA being strictly less segregated than N, and each box is colored according to the legend above.
    \end{tabnotes}
\end{figure}
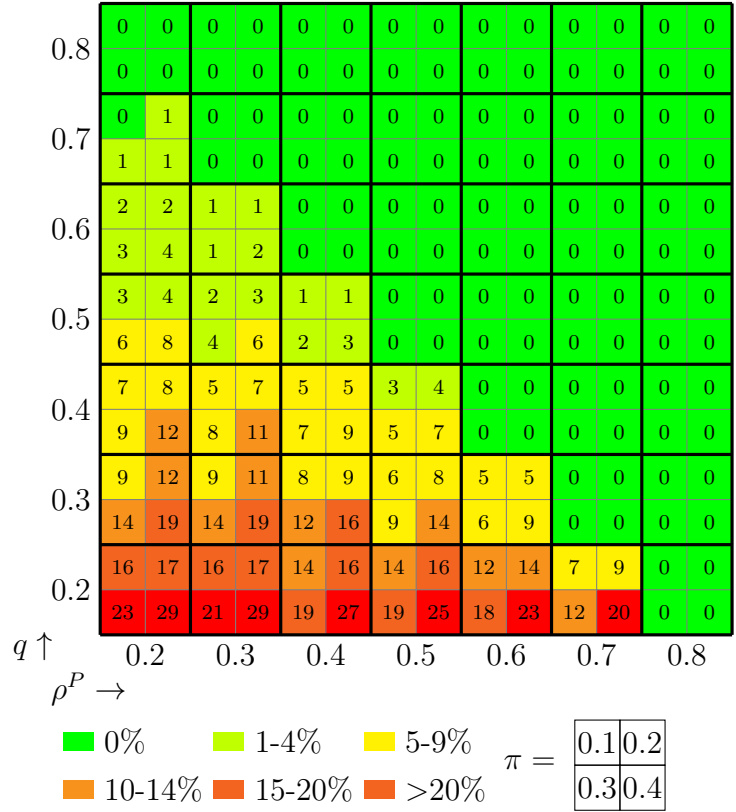

While Theorem \ref{thm:s_sorting} provides key conditions, they are expressed in terms of equilibrium cutoff types and are not straightforward to relate to the model primitives. To provide further results, we specialize the model to a uniform distribution for any number of wealth indices (Corollary \ref{cor:uniform}) and to an arbitrary distribution with two wealth indices, given by Proposition \ref{prop:s_sorting_2types} and Figure \ref{fig:poor_q_pi_DA_N}.

\begin{cor}\label{cor:uniform}
 Suppose $F(s_i)=s_i$. Then, 

 1. N and DA result in the same level of school segregation by wealth, and 

 2. TTC results in greater school segregation by wealth than N and DA.
\end{cor}

We next focus on cases with two wealth types, i.e., $\Omega=\{\omega^P, \omega^R\}$, where $\omega^P > \omega^R$. Two wealth types simplify the problem because any change for one wealth type has a one-to-one reflection in the change for the other, since there are no ``intermediate'' wealth types to absorb these changes. 
Let $\rho^P := \rho(\omega^P)$.
Given a parameter vector $\Pi$ of the model, let $\mathcal{F}(\Pi)$ denote the set of all concave cdfs of $s_i$ such that Assumption 2 is satisfied under $\Pi$.
Similarly, let $\mathcal{F}(\Pi,s_{\omega^P}^N,s_{\omega^R}^N)$ denote the set of all concave cdf's of $s_i$ such that Assumption 2 is satisfied under $\Pi$ and the cutoff signal types under N are $(s_{\omega^P}^N,s_{\omega^R}^N)$.
The next lemma provides a powerful result in finding the distribution that is ``most favorable to the school choice mechanisms (compared to N)'' in terms of school segregation.

\begin{lm}\label{lm:F_max_pop}
 Suppose $\Omega=\{\omega^P, \omega^R\}$, where $\omega^P > \omega^R$, $g=0$, and $e=1$.
 For any $(\Pi, s_{\omega^P}^N,s_{\omega^R}^N)$ such that $\mathcal{F}(\Pi,s_{\omega^P}^N,s_{\omega^R}^N) \neq \emptyset$, type $\omega^P$'s population change from N to $\varphi \in \{DA, TTC\}$ at school $c_1$ is maximized by $G \in \mathcal{F}(\Pi, s_{\omega^P}^N,s_{\omega^R}^N)$ which is piece-wise linear and has only one kink at $s_i=s_{\omega^P}^N$.
\end{lm}

By exploiting Lemma \ref{lm:F_max_pop}, we can characterize the school segregation ranking between N and the other two mechanisms under the binary wealth environments.

\begin{prop}\label{prop:s_sorting_2types}
 Suppose $\Omega=\{\omega^P, \omega^R\}$, where $\omega^P > \omega^R$, $g=0$, and $e=1$.
 Consider any $\Pi$ such that $\mathcal{F}(\Pi) \neq \emptyset$.

 1. DA results in greater school segregation by wealth than N for all $F \in \mathcal{F}(\Pi)$ if $1-q < \rho^P$.

 2. TTC results in greater school segregation by wealth than N for all $F \in \mathcal{F}(\Pi)$.

 3. TTC results in greater school segregation by wealth than DA for all $F \in \mathcal{F}(\Pi)$.
\end{prop}
Note that whenever $1 - q \geq \rho^P$, the comparison between N and DA depends on $F$, but we argue below that DA often leads to greater school segregation in this case as well. One example, with parameters $\pi = 1/3$, $q = 0.4$, and equal shares of poor and rich families, is shown in Figure \ref{fig:example_rich_poor} of Section \ref{sec:Example}, where a narrow red strip marks the cases in which DA yields lower school-level segregation than N. We extend this analysis across parameter configurations and compute the share of the red strip in the total area. Specifically, we compute school segregation under N and DA for all single-kink distributions---which, as argued in Lemma \ref{lm:F_max_pop}, represent the most favorable case for DA---and find that DA yields lower segregation only in a small minority of cases. These occur mainly when the share of rich agents is high, specialized schools have limited capacity, and uncertainty about school fit is large.

The calculations are presented in Figure \ref{fig:poor_q_pi_DA_N}. We partition the parameter space $(\rho^{P},q,\pi) \in [0.2,0.8] \times [0.2,0.8] \times [0.1,0.4]$ into 196 boxes. The $(\rho^{P},q)$-plane is divided into $7\times 7$ large boxes; within each large box, $\pi$ takes four values, yielding four small boxes. Thus, each large box corresponds to a $(\rho^{P},q)$ pair, and each small box to one of the four $\pi$ values. For the parameters of each element of the partition, we calculate school segregation under N and DA for all single-kink distributions defined by points $(x,y)$ on the 0.1 grid with $y \geq x$ (so that $F(x) = y$). We then report the fraction of such distributions that result in lower segregation under DA, out of all instances where the solutions to both problems exist within our restrictions.

\subsection{Housing prices}\label{sec:price}

Although the paper's main focus is on segregation by wealth induced by the mechanisms, policy debates on school choice often involve a much broader segment of the population, particularly those concerned about potential changes in housing prices.
However, the ranking of housing prices does not follow directly from the previous results, since Lemma \ref{lmB} only establishes the ranking of $d^{\varphi}$, namely $p^N < \frac{1}{1-\pi}\frac{p^{DA}}{r^{DA}} < \frac{1}{1-2\pi}\frac{p^{TTC}}{r^{TTC}}$. In this section, we derive the corresponding ranking for housing prices.

Let $r^{DA}_{\mathrm{uniform}}$ be the equilibrium rejection probability of DA under the uniform distribution of signals.

\begin{thm}\label{thm:price}
The housing prices satisfy the following inequalities:

1. $p^N \le p^{DA}$ if $r^{DA} \ge r^{DA}_{\mathrm{uniform}}$.

2. $p^{DA} < p^{TTC}$ if $e-g>1-q$.
\end{thm}

\begin{cor}
 Suppose $F(s_i)=s_i$. Then, $p^N = p^{DA} < p^{TTC}$.
\end{cor}

These rankings imply that, under mild sufficient conditions ($r^{DA} \ge r^{DA}_{\mathrm{uniform}}$ and $e-g>1-q$), the housing prices are unambiguously ranked: $p^N \le p^{DA} < p^{TTC}$. These conditions hold for a wide range of parameters of interest, particularly for sufficiently small $g$ and sufficiently large $e$.\footnote{More precisely, $r^{DA} \ge r^{DA}_{uniform}$ holds when $F(g)-g$ and $F(e+g)-(e+g)$ are small enough.}

\subsection{The role of $r^{\varphi}$ and school capacity greater than housing supply}\label{sec:delta_q}

The rejection probability $r^{\varphi}$ plays a central role in our analysis but affects neighborhood- and school-level segregation differently. Consider a situation where $r^{DA}$ and $r^{TTC}$ both decrease, while the equilibrium ratio $\frac{p^{\varphi}}{r^{\varphi}}$ remains constant across all mechanisms $\varphi$. In this case, neighborhood segregation would remain unchanged, as it depends solely on the equilibrium value of $\frac{p^{\varphi}}{r^{\varphi}}$. However, the ranking of prices and school segregation could reverse. Prices may be lower under DA and TTC than under N because $\frac{p^{\varphi}}{r^{\varphi}}$ is fixed, but $r^{DA}$ and $r^{TTC}$ are lower, implying lower prices. School segregation may be lower under DA and TTC because lower rejection probabilities allow oversubscribed schools to admit more out-of-zone applicants, thus reducing school-level segregation.

This difference helps explain the effects of expanding school capacity. While we previously assumed that each school's capacity equals the housing supply of its corresponding neighborhood, we now extend the model to allow a school $c_k \in C \setminus \{c_0\}$ to have capacity exceeding the housing supply of $n_k$ by an amount $\Delta q \ge 0$. As $\Delta q$ increases, the rejection probability $r^{\varphi}$ decreases for $\varphi \in \{DA, TTC\}$, yet the equilibrium ratio $\frac{p^{\varphi}}{r^{\varphi}}$ remains constant. Consequently, increases in $\Delta q$ generate the same qualitative effects discussed above.

\section{Discussions}\label{sec:Discussion}

In this section, we discuss several possible extensions of our model and their implications. We then turn to two numerical examples: the first introduces benchmarks for interpreting segregation levels, and the second highlights how the relative effectiveness of desegregation policies depends on the time horizon.

Our baseline model does not allow for a school that all families strictly prefer. In some contexts, this may be unrealistic. Suppose instead there exists a small-capacity ``universally preferred'' school, which every family prefers to any other school. Some agents would buy houses in its neighborhood and attend that school, leaving no seats available for others. The remaining agents would then play the game described in our main model. In this setting, wealth and signals would no longer be independent, but the logic of our results does not rely on that assumption.

Another possible extension would allow a fraction of families to choose their housing after observing their school fit. Our model can partially capture this case by lowering $\pi$, which increases the share of agents who receive a zero shock. For the analysis of school segregation, these families and zero-shock agents are indistinguishable. For the analysis of neighborhood segregation, however, they must be treated separately, as they no longer choose under uncertainty. The intuition for our main results would still apply.

Finally, many empirical models, following reality, incorporate preferences over peers. Our baseline model abstracts from such preferences, but we can readily accommodate two common versions: preference for homophily by wealth and uniform preference for wealthy peers, which may arise due to the associated resources they bring. Peer preference may expand the set of equilibria, but in the equilibrium we study, it would only reinforce our results: even without peer preference, $n_1$ and $c_1$ are wealthier. With peer preference, the willingness of families to pay for access would be even higher. In the homophily case, it comes only from rich families; under uniform preference for wealthy peers, it comes from all families. As wealthier families are less responsive to price changes, this would further increase segregation.

\subsection{Student-school match quality and other benchmarks} \label{sec:benchmarks}

In this paper we focus on segregation induced by the three widely used mechanisms. Much of the existing literature emphasizes the welfare gains of DA and TTC relative to neighborhood assignment (N), and policymakers also take these gains into account when deciding whether to adopt a mechanism. We do not attempt a full welfare comparison here, because we \textit{assume}, in line with widespread practice, that families residing near a school receive priority. Such priority may reflect positive externalities of local enrollment, the costs of long commutes, or political considerations. Because we do not model these factors, we cannot offer a full welfare analysis and instead focus on one of its components: the quality of student-school matches. While match quality is often taken in the literature as the sole measure of welfare, this interpretation is less suitable in our setting, since neighborhood priority is a policy goal in its own right.

Once we focus on match quality, two benchmarks that do not provide neighborhood priority become relevant. The first is DA (or TTC) without neighborhood priority. In this case the two mechanisms coincide: families whose top choice is $c_k$ are randomly assigned either to $c_k$ or to their second choice, $c_0$. This benchmark eliminates segregation but one may expect it to reduce match quality, since admission does not depend on how strongly families value the oversubscribed school. The second benchmark is an ex post auction, in which school seats are allocated to the highest bidders once preference shocks are realized. One may expect this mechanism to maximize match quality but also to produce highly segregated outcomes.

Table \ref{tab:bench} reports segregation and match quality under five mechanisms: N, DA, TTC, DA/TTC without neighborhood priority, and Auction for the example considered in Section \ref{sec:Example} with uniform distribution of signals. For N, DA, and TTC we previously reported segregation; here we add match quality for poor families, rich families, the total, and the poor families' share in match quality, as percentage. DA and TTC improve match quality relative to N in the short term (rows 2-3 vs. 1), when locations are fixed as under N, with both groups benefiting (cols. b-d). In the long term, relocation slightly reduces total match quality (within rounding) and shifts the gains decisively toward the rich (rows 2 vs. 4 and 3 vs. 5, cols. d-e).\footnote{Our assumption that $c_0$ is always ranked first or second favors DA relative to TTC. Under DA, allowing families to have orderings such as $c_2c_1c_0$ would instead keep families from $n_1$ in $c_1$ if they are unsuccessful at $c_2$, thus reducing DA match quality.}

We now turn to the two benchmark mechanisms without neighborhood priority. DA (or TTC) without neighborhood priority produces completely integrated schools and an equal division of match quality, but total match quality is lower than under DA and TTC with priority (col. d, row 6 vs. 2-3), since even families with low value for an oversubscribed school have an equal chance of admission. As expected, Auction delivers the highest total match quality (col. d, row 7). More surprising is that it generates school-level segregation comparable to N and DA (col. a) and gives the poor families a larger share of match quality than under those mechanisms (col. e, row 7 vs. 1 and 4). One way to understand this is to view N as an ex ante auction, conducted before preference shocks are realized, in contrast to the ex post auction described above. From this perspective, the ranking of segregation and the poor families' share is not obvious, and, in the example, Auction performs slightly better in both measures (within rounding for the percentage of poor families in $c_1$). In other words, neighborhood priority effectively turns school choice into an auction through the housing market, and mechanisms designed to offset preference uncertainty amplify segregation---even relative to an ex post auction. Finally, although Auction looks desirable in the table, it should be viewed only as a benchmark, not a policy recommendation, since our example abstracts from many other considerations.

\begin{table}[htbp]
    \centering
    \begin{tabular}{llccccc}
        \toprule
        && \multirow{2}{*}{\shortstack{poor\\families \\in $c_1$}} & \multicolumn{4}{c}{Student-school match quality $\times$ 100} \\
        \cmidrule(lr){4-7}
        && & poor & rich & total & $\frac{\text{poor}}{\text{total}}$ \\
        && \tblnmb{(a)} & \tblnmb{(b)} & \tblnmb{(c)} & \tblnmb{(d)} & \tblnmb{(e)} \\
        \midrule    
        \multirow{3}{*}{\shortstack{Short-term effect:\\ locations are fixed\\ as under N\\ (41\% poor families)}}
        &\tblnmb{(1)} N        & 41\% \tblcng{(=)} & 14 \tblcng{(=)} & 18 \tblcng{(=)} & 32 \tblcng{(=)} & 43\% \tblcng{(=)} \\
        &\tblnmb{(2)} DA       & 45\% \tblcng{(+4)} & 19 \tblcng{(+5)} & 24 \tblcng{(+6)} & 43 \tblcng{(+11)} & 45\% \tblcng{(+2)} \\
        &\tblnmb{(3)} TTC      & 41\% \tblcng{(=)}& 15 \tblcng{(+1)} & 22 \tblcng{(+4)} & 37 \tblcng{(+5)} & 41\% \tblcng{($-2$)} \\
        \midrule
        \multirow{2}{*}{\shortstack{Long-term effect:\\ endogenous locations}}
        &\tblnmb{(4)} DA      & 41\% \tblcng{(=)} & 17 \tblcng{(+3)} & 26 \tblcng{(+8)} & 43 \tblcng{(+11)} & 40\% \tblcng{($-3$)} \\
        &\tblnmb{(5)} TTC    & 9\% \tblcng{($-32$)} & 4 \tblcng{($-10$)} & 32 \tblcng{(+14)} & 36 \tblcng{(+4)} & 10\% \tblcng{($-33$)} \\
        \midrule
        \multirow{2}{*}{\shortstack{No neighborhood \\ priorities}}
        &\tblnmb{(6)} {\footnotesize DA/TTC}    & 50\% \tblcng{(+9)} & 17 \tblcng{(+3)} & 17 \tblcng{($-1$)} & 33 \tblcng{(+1)} & 50\% \tblcng{(+7)}\\
        &\tblnmb{(7)} Auction          & 41\% \tblcng{(+0)} & 25 \tblcng{(+11)}& 31 \tblcng{(+13)}& 56 \tblcng{(+24)}& 44\% \tblcng{(+1)}\\
        \bottomrule
    \end{tabular}
    \caption{Segregation and match quality under different mechanisms.}
    \label{tab:bench}
    \begin{tabnotes}
        The table reports the percentage of poor families in school $c_1$, match quality for poor, rich, and all families, and the poor families' share of the total. Match quality is the sum of ex post school fits for all students assigned to $c_1$. Rows correspond to: (1-3) N, DA, and TTC with locations fixed as under N (short-term effect of flexible school choice); (4-5) DA and TTC after relocation in response to the incentives created by these mechanisms; (6) DA and TTC---which are equivalent here---without neighborhood priority; and (7) Auction, which finds a market-clearing price for school seats.
    \end{tabnotes}    
\end{table}

\newcommand{\lAA}{\ifmmode\text{L}\else{DA+{L}}\fi\xspace}
\newcommand{\wlAA}{\ifmmode\text{{WL}}\else{DA+{WL}}\fi\xspace}

\subsection{Direct and indirect desegregation policies: short vs. long term}

Our main model shows that endogenizing location choices can reverse the segregation-based ranking of school choice mechanisms compared to models with fixed locations. In this subsection, we further contrast short- and long-term effects by considering a desegregation policy. Suppose a policymaker aims to desegregate neighborhoods in a DA-based school assignment system with neighborhood priorities, as in the main model, and considers two policies intended to increase the share of poor agents in the oversubscribed schools. These are schools $c_1$ and $c_{-1}$ in our example; by symmetry, we focus on $n_1$ and $c_1$. 
The first, a location-based (\lAA) policy, prioritizes residents of $n_0$ over other non-local applicants for seats in $c_1$ vacated by $n_1$ residents who opt out of their local school. Under the parameters we consider, demand from $n_0$ residents is sufficient to fill all such seats. Since agents in $n_0$ are, on average, poorer than those in $n_1$, this policy targets the poor families indirectly. The second, a wealth-and-location-based (\wlAA) policy, builds on the first by adding a direct wealth criterion: it prioritizes \emph{poor} residents of $n_0$, who again fill all the vacant seats in our example.

These two policies are motivated by---though not intended as direct analogues of---recent shifts from explicit to implicit race-based affirmative action in admissions \cite[see, e.g.,][]{DurKominersPathakSonmez2018ReserveDesign,CarlsonEtAl2020AA,EllisonPathak2021AA,Bleemer2023AA,PathakReesJonesSonmez2020ReversingReserves}. While our model focuses on segregation by wealth rather than race---and wealth is typically a legitimate basis for affirmative action---the contrast between short-term and long-term effects of direct and indirect targeting may remain relevant for a broader range of policies. In our example, the short-term effects align with existing findings: indirect targeting is much less effective at reducing segregation. In the long term, however, this conclusion may be reversed.

Our example uses the same setup as in {Section \ref{sec:Example}}, with two wealth levels, $\poor = 9/8$ and $\rich = 7/8$, uniformly distributed signals, capacities of $0.4$ for neighborhoods $n_{-1}$ and $n_1$ and schools $c_{-1}$ and $c_1$, unlimited capacity at $n_0$ and $c_0$, $g=0$, and preference shocks $\epsilon_i \in \{-1,0,1\}$ occurring with equal probability.

Under \lAA, the utility gain from living in $n_1$ instead of $n_0$ is:
\[
    \Delta u^{\lAA}(r^{\lAA},p^{\lAA}\mid s_i,\omega_i)
    = -(1-r^{\lAA})\frac{1-s_i}{3} + r^{\lAA}\left(\frac{s_i+1}{3}+\frac{s_i}{3}\right) - \omega_i p^{\lAA},
\]
where the first term reflects the fact that $n_0$ residents can be assigned to $c_{-1}$, but $n_1$ residents cannot.

Under \wlAA, expressions are similar to \lAA, except that rejection probabilities differ by wealth and are equal to $1$ for rich agents; thus, $r^{\wlAA}$ refers to the rejection probability for a poor agent:
\begin{align*}
    &\Delta u^{\wlAA}(r^{\wlAA}, p^{\wlAA}\mid s_i,\poor)
      = -(1-r^{\wlAA})\frac{1-s_i}{3} + r^{\wlAA}\left(\frac{s_i+1}{3} + \frac{s_i}{3}\right) - \poor p^{\wlAA},\\
    &\Delta u^{\wlAA}(r^{\wlAA}, p^{\wlAA}\mid s_i,\rich)
      = \frac{s_i+1}{3} + \frac{s_i}{3} - \rich p^{\wlAA}.
\end{align*}

Results are presented in Table \ref{tab:AAshares}. The first block, rows (1)-(3), shows the short-term effects of DA and the two desegregation policies, where residential locations are exogenously fixed as under DA. The second block, rows (4)-(5), reports the long-term effects of introducing each desegregation policy into a system previously operating under DA, with endogenous residential locations.

\begin{table}
    \centering
    \begin{tabular}{clcc}
        \toprule
        \multirow{2}{*}{} & \multirow{2}{*}{Policies} & \multicolumn{2}{c}{Share of poor agents in} \\
        \cmidrule(lr){3-4}
        & & $n_1$ & $c_1$ \\
        \midrule
        \multirow{3}{*}{\shortstack{Short-term effect:\\Locations fixed as in DA}}
        & \tblnmb{(1)} DA, no policy & 33\% & 41\% \\
        & \tblnmb{(2)} \lAA         & 33\% & 42\% \\
        & \tblnmb{(3)} \wlAA        & 33\% & 55\% \\
        \midrule
        \multirow{2}{*}{\shortstack{Long-term effect:\\Endogenous locations}}
        & \tblnmb{(4)} \lAA         & 36\% & 44\% \\
        & \tblnmb{(5)} \wlAA        & 10\% & 40\% \\
        \bottomrule
    \end{tabular}
    \caption{Share of poor agents under different desegregation policies, with fixed (short-term) or endogenous (long-term) locations.}
    \label{tab:AAshares}
    \begin{tabnotes}
        The table shows the share of poor agents in neighborhood $n_1$ and school $c_1$ under five scenarios: (1) DA with no policy; (2) location-based policy (\lAA), which restricts out-of-zone access to $c_1$ to $n_0$ residents, holding locations fixed as under DA; (3) wealth-and-location policy (\wlAA), which further restricts access to $c_1$ to poor $n_0$ residents, again with DA locations; and (4)-(5) the same \lAA and \wlAA policies with endogenous location choices.
    \end{tabnotes}
\end{table}

Comparing rows (2) and (3), we see that \wlAA is highly effective in the short term, increasing the share of poor agents in $c_1$ from $41\%$ to $55\%$. By contrast, \lAA is barely effective, increasing the poor agents' share by just $1$ percentage point. However, in the example, these conclusions reverse in the long term, once agents adjust their locations in response to the new policies: \wlAA becomes counterproductive, slightly reducing the share of poor agents in $c_1$ (row 5), while \lAA delivers a modest increase (row 4).

The intuition for \lAA mirrors that in {Section \ref{sec:Example}}. By removing $n_1$ residents' access to $c_{-1}$, the policy slightly worsens their out-of-zone options. At the same time, it improves $n_0$ residents' access to $c_{-1}$ by eliminating competition from $n_1$ residents. Together these effects reduce segregation in $n_1$, which in turn reduces segregation in $c_1$.

By contrast, under \wlAA, rich agents with positive signals face strong incentives to live in $n_1$, since this becomes their only route to $c_1$. Poor agents, however, have weaker incentives to be in $n_1$, because their chances of admission to $c_1$ from $n_0$ improve relative to DA. As a result, $n_1$ becomes heavily segregated, and even the strong correction at the school level---allocating all out-of-zone seats in $c_1$ to poor agents---cannot offset the increase in residential segregation, leaving $c_1$ more segregated than under DA.

\section{Concluding remarks}\label{sec:Concl}

Priorities shaped through neighborhood location are an important but often overlooked aspect of evaluating school assignment mechanisms. By ignoring this channel, one adopts a short-term perspective that leaves out the longer-term forces linking housing and school assignment. Our analysis suggests that while flexible school choice can reduce segregation in the short term, endogenous location choices increase segregation in neighborhoods and, in some cases, in schools. Our examples shed light on the importance of long-term segregation outcomes in evaluating school assignment policies.

This paper is among the first to connect two strands of literature: contests and matching (see, e.g., \cite{BodohHickman2018MatchingContest} for an alternative approach focusing on investment and market structure). Contest models allow for the endogenous formation of priority but make restrictive assumptions on preferences, whereas matching models take priorities as fixed but allow for rich preferences. Such settings, where contests determine priority in a matching problem, extend well beyond housing; investing effort, or money spent on tutoring, to improve school admission chances is one example. Segregation by wealth is a particularly salient concern in education, labor, and housing markets, yet it remains underexplored at the intersection of matching and contest theory. Future theoretical, empirical, and experimental work can further illuminate how assignment mechanisms shape segregation and inequality through the endogenous acquisition of priority.

\clearpage
\bibliographystyle{ecta}
\bibliography{merged_refs}    

\newpage
\appendix
\def\thesection{Appendix \Alph{section}}
\section{Omitted proofs}

\subsection{Proof of Lemma \ref{lm:monotone}}

Take arbitrary $\omega_i \in \Omega$, $r>0$, $p \ge 0$, and fix them.

For notational consistency, we model N as a school choice mechanism in which all $n_k$ ($k \in \{1,\ldots,m\}$) residents are assigned to $c_k$ for sure, and all $n_0$ residents are assigned to their primary-fit schools with probability $1-r$. In equilibrium, we always consider $r=1$.
Then, 
$\Delta u_i^N(r,p|s_i,\omega_i) = r(s_i-g)-\omega_i p$ is strictly increasing in $s_i \in [0, 1]$ and equal to zero when $s_i=g$ and $p=0$.

\begin{equation*}
\begin{split}
 & E_{\epsilon_i}u_i^{DA}(n_0,r,p|s_i+\epsilon_i,\omega_i) \\
 &= \begin{cases}
  \pi[(1-r)(s_i+e)+rg]+(1-2\pi)g+\pi[(1-r)(-s_i+e)+rg] \\
  \text{ for } s_i \in [0,g] \\
  \pi[(1-r)(s_i+e)+rg]+(1-2\pi)[(1-r)s_i+rg]+\pi[(1-r)(-s_i+e)+rg] \\
  \text{ for } s_i \in [g,e-g] \\
  \pi[(1-r)(s_i+e)+rg]+(1-2\pi)[(1-r)s_i+rg]+\pi g \\
  \text{ for } s_i \in [e-g,e+g] \\
  \pi[(1-r)(s_i+e)+rg]+(1-2\pi)[(1-r)s_i+rg]+\pi[(1-r)(s_i-e)+rg] \\
  \text{ for } s_i \in [e+g,1] \\  
 \end{cases} \\
\end{split}
\end{equation*}
and 
\begin{equation*}
\begin{split}
 & E_{\epsilon_i}u_i^{DA}(n_1,r,p|s_i+\epsilon_i,\omega_i) \\
 &= \begin{cases}
  \pi(s_i+e)+(1-2\pi)g+\pi[(1-r)(-s_i+e)+rg]-\omega_i p & \text{ for } s_i \in [0,g] \\
  \pi(s_i+e)+(1-2\pi)s_i+\pi[(1-r)(-s_i+e)+rg]-\omega_i p & \text{ for } s_i \in [g,e-g] \\
  \pi(s_i+e)+(1-2\pi)s_i+\pi g-\omega_i p & \text{ for } s_i \in [e-g,e+g] \\
  \pi(s_i+e)+(1-2\pi)s_i+\pi(s_i-e)-\omega_i p & \text{ for } s_i \in [e+g,1] \\  
 \end{cases} \\
\end{split}
\end{equation*}
lead to 
\begin{equation*}
\begin{split}
 & \Delta u_i^{DA}(r,p|s_i,\omega_i) = \begin{cases}
  r\pi(s_i+e-g)-\omega_i p & \text{ for } s_i \in [0,g] \\
  r\Big(\pi(s_i+e-g)+(1-2\pi)(s_i-g)\Big)-\omega_i p & \text{ for } s_i \in [g,e+g] \\
  r(s_i-g)-\omega_i p & \text{ for } s_i \in [e+g,1]. \\  
 \end{cases} \\
\end{split}
\end{equation*}
Because the utility difference between living in $n_1$ instead of $n_0$ only arises in the event the $n_0$ resident is rejected by $c_1$, the expected gain is multiplied by rejection probability $r$; the rest of that term captures different submitted rank-order lists, and, correspondingly, different assignment probabilities for agents with signals in intervals $[0,g]$, $[g,e+g]$, and $[e+g,1]$. 
Since $r>0$, $\Delta u_i^{DA}(r,p|s_i,\omega_i)$ is strictly increasing in $s_i \in [0, 1]$. When $s_i=g$ and $p=0$, $\Delta u_i^{DA}(r,0|g,\omega_i) > 0$ as $\pi e>0$.
\begin{equation*}
\begin{split}
 & E_{\epsilon_i}u_i^{TTC}(n_0,r,p|s_i+\epsilon_i,\omega_i) \\
 &= \begin{cases}
  \pi[(1-r)(s_i+e)+rg]+(1-2\pi)g+\pi[(1-r)(-s_i+e)+rg] \\
  \text{ for } s_i \in [0,g] \\
  \pi[(1-r)(s_i+e)+rg]+(1-2\pi)[(1-r)s_i+rg]+\pi[(1-r)(-s_i+e)+rg] \\
  \text{ for } s_i \in [g,e-g] \\
  \pi[(1-r)(s_i+e)+rg]+(1-2\pi)[(1-r)s_i+rg]+\pi g \\
  \text{ for } s_i \in [e-g,e+g] \\
  \pi[(1-r)(s_i+e)+rg]+(1-2\pi)[(1-r)s_i+rg]+\pi[(1-r)(s_i-e)+rg] \\
  \text{ for } s_i \in [e+g,1] \\  
 \end{cases} \\
\end{split}
\end{equation*}
and 
\begin{equation*}
\begin{split}
 & E_{\epsilon_i}u_i^{TTC}(n_1,r,p|s_i+\epsilon_i,\omega_i) \\
 &= \begin{cases}
  \pi(s_i+e)+(1-2\pi)g+\pi(-s_i+e)-\omega_i p & \text{ for } s_i \in [0,g] \\
  \pi(s_i+e)+(1-2\pi)s_i+\pi(-s_i+e)-\omega_i p & \text{ for } s_i \in [g,e-g] \\
  \pi(s_i+e)+(1-2\pi)s_i+\pi g-\omega_i p & \text{ for } s_i \in [e-g,e+g] \\
  \pi(s_i+e)+(1-2\pi)s_i+\pi(s_i-e)-\omega_i p & \text{ for } s_i \in [e+g,1] \\  
 \end{cases} \\
\end{split}
\end{equation*}
lead to 
{\small
\begin{equation*}
\begin{split}
 & \Delta u_i^{TTC}(r,p|s_i,\omega_i) = \begin{cases}
  r\Big(\pi(s_i+e-g)+\pi(-s_i+e-g)\Big)-\omega_i p & \text{ for } s_i \in [0,g] \\
  \begin{aligned}
  r\Big(\pi(s_i+e-g)+(1-2\pi)(s_i-g)+\pi(-s_i+e-g)\Big)\\-\omega_i p    
  \end{aligned}
   & \text{ for } s_i \in [g,e-g] \\
  r\Big(\pi(s_i+e-g)+(1-2\pi)(s_i-g)\Big)-\omega_i p & \text{ for } s_i \in [e-g,e+g] \\
  r(s_i-g)-\omega_i p & \text{ for } s_i \in [e+g,1]. \\  
 \end{cases}
\end{split}
\end{equation*}
}

Since $r>0$, $\Delta u_i^{TTC}(r,p|s_i,\omega_i)$ is weakly increasing in $s_i \in [0, 1]$ and strictly increasing in $s_i \in [g,1]$; $\Delta u_i^{TTC}(r,0|g,\omega_i) > 0$ as $\pi e>0$.

\subsection{Proof of Lemma \ref{lm:uniqueness}}

Take any $\varphi \in \{N,DA,TTC\}$.
By Lemma \ref{lm:monotone} and Assumption 2, the best response strategy to $(\hat{r}^{\varphi},p)$ with $p \in [\hat{p}^{\varphi},\bar{p}^{\varphi}]$ has a cutoff signal type $s_{\omega_i}^{\varphi}(\hat{r}^{\varphi},p) \in (g,e-g)$ for any $\omega_i \in \Omega$.
The cutoff signal type $s_{\omega_i}^{\varphi}(\hat{r}^{\varphi},p)$ satisfies $\gamma^{\varphi}(s_{\omega_i}^{\varphi}(\hat{r}^{\varphi},p)) = \omega_i\frac{p}{\hat{r}^{\varphi}}$, where $\gamma^N(x) \coloneqq x-g$, $\gamma^{DA}(x) \coloneqq (1-\pi)(x-g)+\pi e$, and $\gamma^{TTC}(x) \coloneqq (1-2\pi)x+2\pi e-g$, which are obtained by arranging $\Delta u_i^{\varphi}(\hat{r}^{\varphi},p|s_{\omega_i}^{\varphi}(\hat{r}^{\varphi},p),\omega_i)=0$.
Since $\gamma^{\varphi}(\cdot)$ is linear, by taking the weighted average over $\omega_i$, we have 
\begin{equation}\label{eq:lm2_0}
 \gamma^{\varphi}(E[s_{\omega_i}^{\varphi}(\hat{r}^{\varphi},p)]) = \frac{p}{\hat{r}^{\varphi}}.
\end{equation}
When $p=\hat{p}^{\varphi}$, since $(\hat{r}^{\varphi},\hat{p}^{\varphi})$ is the equilibrium values of the reduced problem, equation (\ref{eq:lm2_0}) implies $E[s_{\omega_i}^{\varphi}(\hat{r}^{\varphi},\hat{p}^{\varphi})]=F^{-1}(1-q)$, which is $F(E[s_{\omega_i}^{\varphi}(\hat{r}^{\varphi},\hat{p}^{\varphi})])=1-q$.
By the concavity of $F$, we have 
\begin{equation}\label{eq:lm2_1}
 E[F(s_{\omega_i}^{\varphi}(\hat{r}^{\varphi},\hat{p}^{\varphi}))] \le F(E[s_{\omega_i}^{\varphi}(\hat{r}^{\varphi},\hat{p}^{\varphi})])=1-q.
\end{equation}
When $p=\bar{p}^{\varphi}$, the definition of $\bar{p}^{\varphi}$ and equation (\ref{eq:lm2_0}) imply $E[s_{\omega_i}^{\varphi}(\hat{r}^{\varphi},\bar{p}^{\varphi})]=1-q$.
Since $F(x)\ge x$ for any $x\in[0,1]$, we have 
\begin{equation}\label{eq:lm2_2}
 E[F(s_{\omega_i}^{\varphi}(\hat{r}^{\varphi},\bar{p}^{\varphi}))] \ge E[s_{\omega_i}^{\varphi}(\hat{r}^{\varphi},\bar{p}^{\varphi})]=1-q.
\end{equation}
Since $s_{\omega_i}^{\varphi}(\hat{r}^{\varphi},p)$ is continuous and strictly increasing in $p$ and $F(x)$ is continuous and strictly increasing in $x \in [0,e-g]$, by equations (\ref{eq:lm2_1}) and (\ref{eq:lm2_2}), there must exist a unique $p^{\varphi} \in [\hat{p}^{\varphi},\bar{p}^{\varphi}]$ that satisfies the capacity constraint: $E[F(s_{\omega_i}^{\varphi}(\hat{r}^{\varphi},p^{\varphi}))]=1-q$.
Since the equilibrium rejection probability of the original problem is always equal to $\hat{r}^{\varphi}$ when all cutoff signal types are in $(g,e-g)$ (irrespective of $F(\cdot)$ and the exact values of the cutoffs), the strategy $\sigma^{\varphi}$, characterized by the cutoff signal types $\{s_{\omega_i}^{\varphi}(\hat{r}^{\varphi},p^{\varphi})\}_{\omega_i \in \Omega}$ and $p^{\varphi}$, constitute a symmetric equilibrium.

\subsection{Proof of Lemma \ref{lmA}}
By concavity, $F(\cdot)$ is strictly increasing for any $s_i$ with $F(s_i)<1$.
Then, for a given $d$ in equation (\ref{eq:diff}), the solution $\{s_{\omega_i}(d)\}_{\omega_i \in \Omega} \in [0,1]^{|\Omega|}$ satisfying equations (\ref{eq:cap}) and (\ref{eq:diff}) is unique because of the strict increasingness of $F(\cdot)$ (Assumptions 1--2 guarantee that 
the cutoff solutions in our model always satisfy $F(s_{\omega_i})<1$ for all $\omega_i \in \Omega$).\footnote{We omit $\varphi$ from the notation because the proof is the same for any mechanism $\varphi$.}

Consider $d_1 > d_2 \ge 0$ and their associated solutions $\{s_{\omega_i}(d_1)\}_{\omega_i \in \Omega}$ and $\{s_{\omega_i}(d_2)\}_{\omega_i \in \Omega}$, respectively.
We must have a partition of $\Omega$ into $\overline{\Omega} \neq \emptyset$ and $\underline{\Omega} \neq \emptyset$ such that $\omega_i > \omega_j$ for any $\omega_i \in \overline{\Omega}$ and $\omega_j \in \underline{\Omega}$, and $s_{\omega_i}(d_1) \ge s_{\omega_i}(d_2)$ if and only if $\omega_i \in \overline{\Omega}$.
 This is because otherwise, $s_{\omega_i}-s_{\omega_j} = d(\omega_i-\omega_j)$ would be violated for some $\omega_i, \omega_j \in \Omega$ and $d \in \{d_1,d_2\}$.
 Then, equation (\ref{eq:cap}) implies 
 \begin{equation*}
 \sum_{\omega_i \in \overline{\Omega}}\rho(\omega_i)[F(s_{\omega_i}(d_1))-F(s_{\omega_i}(d_2))] = \sum_{\omega_i \in \underline{\Omega}}\rho(\omega_i)[F(s_{\omega_i}(d_2))-F(s_{\omega_i}(d_1))] > 0.
 \end{equation*}
  By concavity of $F$, we have 
 \begin{equation*}
 \frac{s_{\omega_i}(d_1)-s_{\omega_i}(d_2)}{F(s_{\omega_i}(d_1))-F(s_{\omega_i}(d_2))} \ge \frac{s_{\omega_j}(d_2)-s_{\omega_j}(d_1)}{F(s_{\omega_j}(d_2))-F(s_{\omega_j}(d_1))}
 \end{equation*}
 for any $\omega_i \in \overline{\Omega}$ and $\omega_j \in \underline{\Omega}$.
 Multiplying this for each $\omega_i \in \Omega$, we obtain
 \begin{equation*}
 \sum_{\omega_i \in \overline{\Omega}}\rho(\omega_i)[s_{\omega_i}(d_1)-s_{\omega_i}(d_2)] \ge \sum_{\omega_i \in \underline{\Omega}}\rho(\omega_i)[s_{\omega_i}(d_2)-s_{\omega_i}(d_1)],
 \end{equation*}
 which means $E[s_{\omega_i}(d)]$ is weakly increasing in $d$. 

\subsection{Proof of Lemma \ref{lmB}}
Equations (\ref{eq:expected_cutoff}) determine a linear relationship between $E[s_{\omega_i}^{\varphi}]$ and $d^{\varphi}$. The intercepts of the three lines are $-g$ for N, $-g+\frac{\pi e}{1-\pi}$ for DA, and $\frac{2\pi e-g}{1-2\pi}$ for TTC. It is clear that $-g < -g+\frac{\pi e}{1-\pi}$. Further, 
\begin{equation*}
\begin{split}
& \frac{2\pi e-g}{1-2\pi} - \Big(-g+\frac{\pi e}{1-\pi}\Big) \\
=& \frac{(1-\pi)(2\pi e-g)-(1-2\pi)(\pi(g+e)-g)}{(1-2\pi)(1-\pi)} \\
=& \frac{(1-\pi)2\pi e-(1-2\pi)\pi(g+e)-\pi g}{(1-2\pi)(1-\pi)} \\
=& \frac{(1-\pi)\pi(g+e)+(1-\pi)\pi(e-g)-(1-2\pi)\pi(g+e)-\pi g}{(1-2\pi)(1-\pi)} \\
=& \frac{\pi^2(g+e)+(1-\pi)\pi(e-g)-\pi g}{(1-2\pi)(1-\pi)} \\
=& \frac{\pi(e-2g+2\pi g)}{(1-2\pi)(1-\pi)} \\
>& 0,
\end{split}
\end{equation*}
where the final inequality holds from $e-2g > 0$, implied by Assumption 1. Therefore, as depicted in Figure \ref{fig:lmB}, these three lines are parallel and the intercepts are ordered as $-g < -g+\frac{\pi e}{1-\pi} < \frac{2\pi e-g}{1-2\pi}$. Lemma \ref{lmA}, via equations (\ref{eq:cap}) and (\ref{eq:diff}), implies an additional restriction: $E[s_{\omega_i}^{\varphi}]$ weakly increases in $d^{\varphi}$ and is continuous in $d^{\varphi}$.
Note that for $d^{\varphi}=0$, we have $E[s_{\omega_i}^{\varphi}]=F^{-1}(1-q)$, which is greater than $g$ by Assumption 1.
Furthermore, $E[s_{\omega_i}^{\varphi}] \le E[F(s_{\omega_i}^{\varphi})]=1-q$.
Since there is a unique solution for each $\varphi$ (Lemma \ref{lm:uniqueness}), the three intersections must satisfy $F^{-1}(1-q) \le E[s_{\omega_i}^N] \le E[s_{\omega_i}^{DA}] \le E[s_{\omega_i}^{TTC}] \le 1-q$ and $d^N < d^{DA} < d^{TTC}$.
The strict inequality of the latter comes from the strict difference in the intercepts $-g < -g+\frac{\pi e}{1-\pi} < \frac{2\pi e-g}{1-2\pi}$.

\subsection{Proof of Theorem \ref{thm:n_sorting}}
As proved in Lemma \ref{lmA}, for $d_1 > d_2 > 0$ and their associated solutions $\{s_{\omega_i}(d_1)\}_{\omega_i \in \Omega}$ and $\{s_{\omega_i}(d_2)\}_{\omega_i \in \Omega}$, there must be a partition of $\Omega$ into $\overline{\Omega} \neq \emptyset$ and $\underline{\Omega} \neq \emptyset$ such that $\omega_i > \omega_j$ for any $\omega_i \in \overline{\Omega}$ and $\omega_j \in \underline{\Omega}$, and $s_{\omega_i}(d_1) \ge s_{\omega_i}(d_2)$ if and only if $\omega_i \in \overline{\Omega}$.
This means that the neighborhood population of $\overline{\Omega}$ in $n_1$ (and all other neighborhoods $n_k$ with $k\neq 0$) is smaller under $\{s_{\omega_i}(d_1)\}_{\omega_i \in \Omega}$ than under $\{s_{\omega_i}(d_2)\}_{\omega_i \in \Omega}$, whereas the opposite holds for $\underline{\Omega}$.
Since $n_1$'s average wealth type is lower than $n_0$'s for any $d>0$, neighborhood segregation by wealth increases from cutoffs $\{s_{\omega_i}(d_2)\}_{\omega_i \in \Omega}$ to $\{s_{\omega_i}(d_1)\}_{\omega_i \in \Omega}$.
This and $d^N < d^{DA} < d^{TTC}$ by Lemma \ref{lmB} prove the two statements of Theorem \ref{thm:n_sorting}.

\subsection{Proof of Theorem \ref{thm:s_sorting}}\label{sec:proof_s_sorting}

\noindent
\underline{N and DA.}
Conditional on a wealth type, for a cutoff $s_i \in (g,e-g)$, define $D(s_i)$ to be the \emph{ex post} demand for $c_1$, which is the mass of agents who are not $n_1$ residents and for whom $c_1$ is their first choice. Similarly, for $s_i \in (g,e-g)$, define $S(s_i)$ to be the \emph{ex post} supply of $c_1$, which is the mass of $n_1$ residents for whom $c_1$ is not the first choice. Note that all agents in $S(s_i)$ prefer $c_0$ to $c_1$ and would not apply (resp., point) to $c_1$ under DA (resp., TTC). 
Formally, 
 \begin{equation*}
 \begin{split}
 & D(s_i) \coloneqq (1-\pi)[F(s_i)-F(g)]+\pi[F(g)+F(e-g)], \\
 & S(s_i) \coloneqq \pi[F(e+g)-F(s_i)].
 \end{split}
 \end{equation*}
For any DA cutoff signal types $\{s_{\omega_i}^{DA}\}_{\omega_i \in \Omega}$, we can define the total demand $D$ and supply $S$ of $c_1$ by taking weighted sums of $D(s_{\omega_i}^{DA})$ and $S(s_{\omega_i}^{DA})$: 
 \begin{equation*}
 \begin{split}
 & D \coloneqq \sum_{\omega_i \in \Omega} \rho(\omega_i)D(s_{\omega_i}^{DA}) = (1-\pi)[1-q-F(g)]+\pi[F(g)+F(e-g)], \\
 & S \coloneqq \sum_{\omega_i \in \Omega} \rho(\omega_i)S(s_{\omega_i}^{DA}) = \pi[F(e+g)-(1-q)].
 \end{split}
 \end{equation*}
Note that, for any cutoffs $\{s_{\omega_i}\}_{\omega_i \in \Omega}$ that clear the housing market, $D$ and $S$ do not depend on the cutoffs because of $\sum_{\omega_i \in \Omega}\rho(\omega_i)F(s_{\omega_i}) = 1-q$. By definition, $r^{DA} = \frac{D-S}{D}$.

For calculations below, it is useful the define the following quantity: $\Delta := D(s_i)-S(s_i)-F(s_i)$. Note that $s_i$ enters $D(s_i)$ and $S(s_i)$ as multiples of $F(s_i)$ and they cancel out, so $\Delta$ is not a function of $s_i$. 
This is the crucial observation for the derivation below.

For any $\omega_i \in \Omega$, the (unweighted) population change at $c_1$ from DA to N is
 \begin{equation*}
 \begin{split}
 &(\text{Mass of type-}\omega_i\text{ agents in } c_1\text{ under DA} - \text{Mass of type-}\omega_i\text{ agents in } c_1\text{ under N})/\rho(\omega_i)
 \\
 =& (1-F(s_{\omega_i}^{DA})) - (1-F(s_{\omega_i}^N)) + (1-r^{DA})D(s_{\omega_i}^{DA}) - S(s_{\omega_i}^{DA}) 
 \\
 =& F(s_{\omega_i}^N)-F(s_{\omega_i}^{DA})+D(s_{\omega_i}^{DA}) - S(s_{\omega_i}^{DA}) -r^{DA}D(s_{\omega_i}^{DA}) 
 \\
 =& F(s_{\omega_i}^N)-F(s_{\omega_i}^{DA})+D(s_{\omega_i}^{DA})-S(s_{\omega_i}^{DA})-r^{DA}\Big\{D+D(s_{\omega_i}^{DA})-D\Big\} 
 \\
 =& F(s_{\omega_i}^N)-F(s_{\omega_i}^{DA})+D(s_{\omega_i}^{DA})-S(s_{\omega_i}^{DA})-(D-S)-r^{DA}(1-\pi)[F(s_{\omega_i}^{DA})-(1-q)] 
 \\
 =& F(s_{\omega_i}^N)-F(s_{\omega_i}^{DA})+F(s_{\omega_i}^{DA})+\Delta-(1-q+\Delta)-r^{DA}(1-\pi)[F(s_{\omega_i}^{DA})-(1-q)] 
 \\
 =& F(s_{\omega_i}^N)-(1-q)-r^{DA}(1-\pi)[F(s_{\omega_i}^{DA})-(1-q)],
 \end{split}
 \end{equation*}
where the fourth line follows from $r^{DA}=\frac{D-S}{D}$ and equations for $D(s_{\omega_i}^{DA})$ and $D$ above; and the fifth from the definition of $\Delta$, which, being independent of $s_i$, holds for aggregate demand as well, together with $\sum_{\omega_i \in \Omega} \rho(\omega_i) F(s_{\omega_i}^{DA})=1-q$.

The mass of $\omega_i$ agents in $c_1$ increases under DA if $F(s_{\omega_i}^N) - (1-q)-r^{DA}(1-\pi)[F(s_{\omega_i}^{DA})-(1-q)]$ is positive. 
If $\frac{|F(s_{\omega_i}^{DA}) - (1 - q)|}{|F(s_{\omega_i}^N) - (1 - q)|} > \frac{1}{r^{DA}(1-\pi)}$ holds for $\omega_i \in \tilde{\Omega}(N,DA)$ with both the signs of $F(s_{\omega_i}^N) - (1 - q)$ and $F(s_{\omega_i}^{DA}) - (1 - q)$ being positive (resp., negative), the $\omega_i$-population at $c_1$ decreases (resp., increases).

The condition above applies to a particular wealth index $\omega_i$. To complete the proof, we need to show that the change in our segregation measure, which aggregates the changes across $\omega_i$. We will do so by partitioning our population in (endogenous) subgroups of poor, middle, and rich agents and showing that the effect is ``uniform'' across the group; that is, for example, there are no ``sub-groups'' of poor agents, such that the representation of one subgroup increases under DA relative to N, but the other decreases.

By Lemma \ref{lmA}, there is a partition of $\Omega$ into two non-empty intervals, $\overline{\Omega} \neq \emptyset$ and $\underline{\Omega} \neq \emptyset$ with $\omega_i > \omega_j$ for any $\omega_i \in \overline{\Omega}$ and $\omega_j \in \underline{\Omega}$, such that $s_{\omega_i}^{DA} \ge s_{\omega_i}^N$ if and only if $\omega_i \in \overline{\Omega}$.
Then, there must be a partition of $\Omega$ into three intervals, $\Omega^P \neq \emptyset$, $\Omega^M$, and $\Omega^R \neq \emptyset$, such that for any $(\omega_i,\omega_j,\omega_k) \in \Omega^P \times \Omega^M \times \Omega^R$ (i) $\omega_i > \omega_j > \omega_k$, (ii) $s_{\omega_i}^{DA} \ge s_{\omega_i}^N \ge F^{-1}(1-q)$ and $s_{\omega_k}^{DA} \le s_{\omega_k}^N \le F^{-1}(1-q)$, and (iii) $\Omega^P$ and $\Omega^R$ are largest such intervals. 
That is, by $\frac{|F(s_{\omega_i}^{DA})-(1-q)|}{|F(s_{\omega_i}^N)-(1-q)|} > \frac{1}{r^{DA}(1-\pi)}$ for all $\omega_i \in \tilde{\Omega}(N,DA)$, all wealth types in $\Omega^P$ (resp., $\Omega^R$) decrease (resp., increase) the population share at $c_1$.
Then, by condition (iii) and partitioning of $\Omega$ into $\overline{\Omega}$ and $\underline{\Omega}$, either \big[$F(s_{\omega_i}^N) \le 1-q \le F(s_{\omega_i}^{DA})$ for all $\omega_i \in \Omega^M$\big] or \big[$F(s_{\omega_i}^N) \ge 1-q \ge F(s_{\omega_i}^{DA})$ for all $\omega_i \in \Omega^M$\big] holds, and thus the sign of the population change is the same for all $\omega_i \in \Omega^M$.
Combining the above with earlier observations that $c_1$ is over-demanded---hence, admits exactly $q$ agents---and that all $c_k$, for $k \neq 0$, are symmetric, completes the proof that DA results in a greater school segregation by wealth than N.

The proof is analogous when the neighborhood segregation expansion rate $\frac{|F(s_{\omega_i}^{DA})-(1-q)|}{|F(s_{\omega_i}^N)-(1-q)|}$ is smaller than $\frac{1}{r^{DA}(1-\pi)}$ for all $\omega_i \in \tilde{\Omega}(N,DA)$.

\medskip
\noindent
\underline{N and TTC.}
Define $D(s_i), D, S(s_i)$ and $S$ as before. Conditional on a wealth type, for a cutoff $s_i \in (g,e-g)$, define $X(s_i)$ as the mass of $n_1$ residents for whom some $c_k$ with $k \in \{2,\ldots,m\}$ is the first choice, 
and $X$ as the weighted sum of $X(s_{\omega_i}^{TTC})$: 
 \begin{equation*}
 \begin{split}
  & X(s_i) \coloneqq \pi[F(e-g)-F(s_i)], \\
  & X \coloneqq \sum_{\omega_i \in \Omega} \rho(\omega_i)X(s_{\omega_i}^{TTC}) = \pi[F(e-g)-(1-q)]. \\
 \end{split}
 \end{equation*}
By symmetry and the definition of TTC, $X$ mass of agents 
are assigned to $c_1$ from other neighborhoods $n_k \in N \setminus \{n_0,n_1\}$ without facing competition from other out-of-zone agents. The rejection rate is then $r^{TTC} = \frac{(D-X)-(S-X)}{D-X} = \frac{D-S}{D-X}$.

The change in the (unweighted) mass of agents with wealth type $\omega_i$ attending $c_1$ from TTC to N is
 \begin{equation*}
 \begin{split}
 &(\text{Mass of type-}\omega_i\text{ agents in } c_1\text{ under TTC} - \text{Mass of type-}\omega_i\text{ agents in } c_1\text{ under N})/\rho(\omega_i)\\
 =& (1-F(s_{\omega_i}^{TTC})) - (1-F(s_{\omega_i}^N)) + (1-r^{TTC})(D(s_{\omega_i}^{TTC})-X(s_{\omega_i}^{TTC})) - (S(s_{\omega_i}^{TTC})-X(s_{\omega_i}^{TTC})) \\
 =& F(s_{\omega_i}^N)-F(s_{\omega_i}^{TTC})+D(s_{\omega_i}^{TTC}) - S(s_{\omega_i}^{TTC}) -r^{TTC}(D(s_{\omega_i}^{TTC})-X(s_{\omega_i}^{TTC})) \\
 =& F(s_{\omega_i}^N)-F(s_{\omega_i}^{TTC})+D(s_{\omega_i}^{TTC})-S(s_{\omega_i}^{TTC}) -r^{TTC}\Big\{D-X+D(s_{\omega_i}^{TTC})-X(s_{\omega_i}^{TTC})-(D-X)\Big\} \\
 =& F(s_{\omega_i}^N)-F(s_{\omega_i}^{TTC})+D(s_{\omega_i}^{TTC})-S(s_{\omega_i}^{TTC}) -r^{TTC}\Big\{D(s_{\omega_i}^{TTC})-D-(X(s_{\omega_i}^{TTC})-X)\Big\} \\ 
 =& F(s_{\omega_i}^N)-F(s_{\omega_i}^{TTC})+D(s_{\omega_i}^{TTC})-S(s_{\omega_i}^{TTC})-(D-S) \\
 &-r^{TTC}\Big\{(1-\pi)[F(s_{\omega_i}^{TTC})-(1-q)]-\pi[-F(s_{\omega_i}^{TTC})+(1-q)] \Big\} \\
 =& F(s_{\omega_i}^N)-F(s_{\omega_i}^{TTC})+F(s_{\omega_i}^{TTC})+\Delta-(1-q+\Delta)-r^{TTC}[F(s_{\omega_i}^{TTC})-(1-q)] \\
 =& F(s_{\omega_i}^N)-(1-q)-r^{TTC}[F(s_{\omega_i}^{TTC})-(1-q)] \\
 \end{split}
 \end{equation*}

The equations above relate the expansion rate and inverse rejection probability for a particular wealth type $\omega_i$. The rest of the proof is identical to the DA vs. N case.

\medskip
\noindent
\underline{DA and TTC.}

From the two arguments above, for any $\omega_i \in \Omega$, the change in its (unweighted) mass at $c_1$ from DA to TTC is  
 \begin{equation*}
    \begin{split}
        &(\text{Mass of type-}\omega_i\text{ agents in } c_1\text{ under TTC} - \text{Mass of type-}\omega_i\text{ agents in } c_1\text{ under DA)}/\rho(\omega_i)\\
        =&r^{DA}(1-\pi)[F(s_{\omega_i}^{DA})-(1-q)] - r^{TTC}[F(s_{\omega_i}^{TTC})-(1-q)].         
    \end{split}
 \end{equation*}

If $\frac{|F(s_{\omega_i}^{TTC})-(1-q)|}{|F(s_{\omega_i}^{DA})-(1-q)|} > \frac{r^{DA}(1-\pi)}{r^{TTC}}$ holds for $\omega_i \in \tilde{\Omega}(DA,TTC)$ with both the signs of $F(s_{\omega_i}^N)-(1-q)$ and $F(s_{\omega_i}^{DA})-(1-q)$ being positive (resp., negative), the $\omega_i$-population at $c_1$ decreases (resp., increases). We then complete the proof in the same way as the DA vs. N case.

\subsection{Proof of Lemma \ref{lm:F_max_pop}}

Take arbitrary $(\Pi,s_{\omega^P}^N,s_{\omega^R}^N)$ such that $\mathcal{F}(\Pi,s_{\omega^P}^N,s_{\omega^R}^N) \neq \emptyset$ and fix them.
Let $\Delta \omega := \omega^P-\omega^R$.

\medskip
\noindent
\underline{$\omega^P$'s population change at $c_1$ from N to DA.}

By Theorem \ref{thm:s_sorting}, $\omega^P$'s population change at $c_1$ from N to DA is  
\begin{equation*}
 (F(s_{\omega^P}^N)-(1-q))\Big[1-r^{DA}(1-\pi)\frac{F(s_{\omega^P}^{DA})-(1-q)}{F(s_{\omega^P}^N)-(1-q)}\Big].
\end{equation*}
Note that $r^{DA}=\frac{D-S}{D}$ does not depend on $F$ when $g=0$ and $e=1$ because all the terms in $r^{DA}$ that depend on $F$, i.e., $F(g)$, $F(e-g)$, and $F(e+g)$, take the values of 0 or 1.
Thus, this population change is maximized if $F \in \mathcal{F}(\Pi,s_{\omega^P}^N,s_{\omega^R}^N)$ maximizes $F(s_{\omega^P}^N)-(1-q)$ and minimizes the expansion rate $\frac{F(s_{\omega^P}^{DA})-(1-q)}{F(s_{\omega^P}^N)-(1-q)}$ for $\omega^P$. 
Note that a symmetric argument holds true for $\omega^R$: the population change is equal to $(F(s_{\omega^R}^N)-(1-q))\Big[1-r^{DA}(1-\pi)\frac{F(s_{\omega^R}^{DA})-(1-q)}{F(s_{\omega^R}^N)-(1-q)}\Big]$, which is minimized if $F \in \mathcal{F}(\Pi,s_{\omega^P}^N,s_{\omega^R}^N)$ maximizes $(1-q)-F(s_{\omega^R}^N)$ and minimizes the expansion rate $\frac{F(s_{\omega^R}^{DA})-(1-q)}{F(s_{\omega^R}^N)-(1-q)}$ for $\omega^R$.

For any $F \in \mathcal{F}(\Pi,s_{\omega^P}^N,s_{\omega^R}^N)$, equations (\ref{eq:cap}), (\ref{eq:diff}), and ({\ref{eq:expected_cutoff}}) characterize the cutoff signal types of all three mechanisms.\footnote{A technical condition $1-\rho^P\Delta\omega>0$ holds by Assumption 2 because otherwise the equilibrium would not be characterized by cutoff conditions.}
Under N, the cutoff signal types $(s_{\omega^R},s_{\omega^P})$ satisfy 
\begin{equation}\label{eq:cap_2types}
 \rho^P F(s_{\omega^P})+(1-\rho^P)F(s_{\omega^R}) = 1-q, 
\end{equation}
and 
\begin{equation}\label{eq:Nconst_2types}
 s_{\omega^P}-s_{\omega^R} = \Delta\omega(\rho^P s_{\omega^P}+(1-\rho^P)s_{\omega^R}) \ \Leftrightarrow \ s_{\omega^P} = \frac{1+(1-\rho^P)\Delta\omega}{1-\rho^P\Delta\omega}s_{\omega^R}.
\end{equation}
Similarly, under DA, the cutoff signal types $(s_{\omega^R},s_{\omega^P})$ satisfy equation (\ref{eq:cap_2types}) and 
\begin{equation}\label{eq:DAconst_2types}
 s_{\omega^P}-s_{\omega^R} = \Delta\omega\Big[(\rho^P s_{\omega^P}+(1-\rho^P)s_{\omega^R})+\frac{\pi}{1-\pi}\Big] \ \Leftrightarrow \ s_{\omega^P} = \frac{1+(1-\rho^P)\Delta\omega}{1-\rho^P\Delta\omega}s_{\omega^R} + \frac{\Delta\omega\frac{\pi}{1-\pi}}{1-\rho^P\Delta\omega}.
\end{equation}
Only the values of $F(\cdot)$ at the cutoffs affect these equations; if two signal distributions are identical at the cutoffs, the solutions to these equations would remain the same. Thus, while we consider all $F \in \mathcal{F}(\Pi,s_{\omega^P}^N,s_{\omega^R}^N)$ for which the cutoff solutions are in $(g,e-g)$, it is without loss of generality to focus on piece-wise linear $F$'s that have at most four kinks at $s_i = s_{\omega^R}^N,s_{\omega^P}^N,s_{\omega^R}^{DA},s_{\omega^P}^{DA}$. Note also that equation (\ref{eq:cap_2types}) implies that the expansion rates for the two wealth types coincide with each other.

\begin{figure}[htbp]
    \centering
    \begin{tikzpicture}
        \begin{axis}[
            axis lines=middle,
            xlabel={$s_{\omega^R}$},
            ylabel={$s_{\omega^P}$},
            xmin=0, xmax=11.2,
            ymin=0, ymax=9.5,
            xticklabels={},
            yticklabels={},
            xtick={8.854},
            ytick={9},
            samples=100,
            clip=false,
            xscale=1.2,
            yscale=1.2
        ]
        \addplot[black, domain=0:8] {x} node[right]{};
        \addplot[blue, domain=0:5] {1.5*x} node[right]{N} node[right,yshift=-10pt]{(\ref{eq:Nconst_2types})};
        \addplot[blue, domain=0:4] {1.5*x+3} node[right]{DA} node[right,yshift=-10pt]{(\ref{eq:DAconst_2types})};
        \addplot[cyan, domain=0:2.57] {9-2*x} node[right]{};
        \node[left] at (0,9) {$F^{-1}\left(\frac{1-q}{\rho}\right)$};
        \node[below left] at (0,0) {0};
        \addplot[cyan, domain=2.57:4] {3.857+2.57-x};
        \addplot[cyan, domain=4:8.854] {2.427+2-x/2} node[midway,above]{(\ref{eq:cap_2types})} node[black,below]{$F^{-1}\left(\frac{1-q}{1-\rho}\right)$};
        \fill[cyan] (axis cs:2.57, 3.857) circle (2pt);
        \fill[cyan] (axis cs:1.714, 5.571) circle (2pt);
        \draw[dotted] (2.57, 3.857) -- (2.57, 0) node[below]{$s^N_{\rich}$};
        \draw[dotted] (2.57, 3.857) -- (0, 3.857) node[left]{$s^N_{\poor}$};
        \end{axis}
    \end{tikzpicture}
    \caption{Cutoff signal types under N and DA in the $s_{\omega^R}$-$s_{\omega^P}$ space for some piece-wise linear cdf $F$}    
    \label{fig:cutoff_2types}
    \begin{tabnotes}
        Blue lines labeled N and DA represent the pairs $(s_{\omega^R},s_{\omega^P})$ that solve equations (\ref{eq:Nconst_2types}) and (\ref{eq:DAconst_2types}), respectively. Those lines are parallel. The cyan curve shows the solutions to equation (\ref{eq:cap_2types}) for a piece-wise linear cdf $F$. It is a linearized version of a curve generated by an arbitrary cdf, constrained to pass through the points $(s_{\omega^R}^N,s_{\omega^P}^N)$ and $(s_{\omega^R}^{DA},s_{\omega^P}^{DA})$.
    \end{tabnotes}
\end{figure}
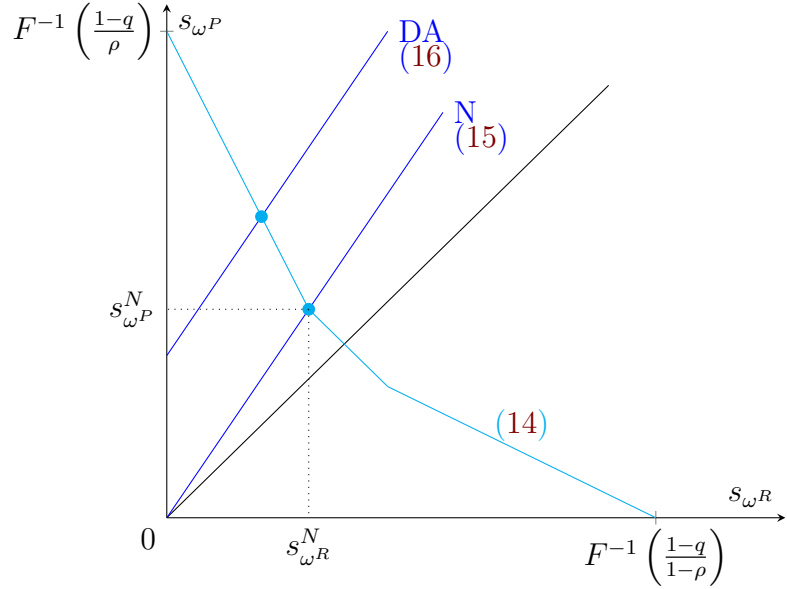

Graphically, the cutoff signal types of N and DA are found in the $s_{\omega^R}$-$s_{\omega^P}$ space.
Although $F(s_i)$ is not defined when $s_i>1$, it is convenient to extend $F$ to such values linearly with the same slope as at $s_i=1$, i.e.,$F(s_i) \coloneqq 1+F'(1)(s_i-1)$ for $s_i>1$. 
In Figure \ref{fig:cutoff_2types}, equations (\ref{eq:Nconst_2types}) and (\ref{eq:DAconst_2types}) are linear lines with the same slope greater than one.
Equation (\ref{eq:cap_2types}) is represented by a convex and decreasing curve connecting $(0,F^{-1}(\frac{1-q}{\rho^P}))$ and $(F^{-1}(\frac{1-q}{1-\rho^P}),0)$.
It must be convex because its slope $-\frac{(1-\rho^P)F'(s_{\omega^R})}{\rho^P F'(s_{\omega^P})}$ is increasing in $s_{\omega^R}$ and decreasing in $s_{\omega^P}$ due to the concavity of $F$.

\begin{figure}[htbp]
    \centering
    \begin{tikzpicture}
        \begin{axis}[
            axis lines=middle,
            xlabel={$s_{\omega^R}$},
            ylabel={$s_{\omega^P}$},
            xmin=0, xmax=6,
            ymin=0, ymax=7,
            xticklabels={},
            yticklabels={},
            xtick=\empty,
            ytick={9},
            samples=100,
            clip=false,
            xscale=0.8,
            yscale=1
        ]
        \draw[red] (1.4,7) -- (2.57, 9-2*2.57);
        \addplot[black, domain=0:6] {x} node[right]{};
        \addplot[blue, domain=0:4.5] {1.5*x} node[right]{N} node[right,yshift=-10pt]{(\ref{eq:Nconst_2types})};
        \addplot[blue, domain=0:2.5] {1.5*x+3} node[right]{DA} node[right,yshift=-10pt]{(\ref{eq:DAconst_2types})};
        \addplot[green, domain=1:2.57] {9-2*x} node[right]{};
        \node[below left] at (0,0) {0};
        \addplot[green, domain=2.57:4] {3.857+2.57-x};
        \addplot[green, domain=4:6] {2.427+2-x/2} node[midway,above]{(\ref{eq:cap_2types})};
        \fill[green] (axis cs:2.57, 3.857) ellipse [x radius=2pt, y radius=1.5pt];
        \fill[green] (axis cs:1.71, 5.56) ellipse [x radius=2pt, y radius=1.5pt];
        \draw[dotted] (2.57, 3.857) -- (2.57, 0) node[below]{$s^N_{\rich}$};
        \draw[dotted] (2.57, 3.857) -- (0, 3.857) node[left]{$s^N_{\poor}$};
        \fill[red] (axis cs:1.855, 5.78) ellipse [x radius=2pt, y radius=1.5pt];
        \fill[red] (axis cs:2.57, 3.857) ellipse [x radius=1pt, y radius=0.75pt];
        \end{axis}
    \end{tikzpicture}%
    \begin{tikzpicture}
        \begin{axis}[
            axis lines=middle,
            xlabel={$s_i$},
            ylabel={cdf},
            xmin=0, xmax=9,
            ymin=0, ymax=10,
            xticklabels={},
            yticklabels={1},
            xtick={2.5,5,7.1,7.5},
            ytick={9},
            samples=100,
            clip=false,
            xscale=1.2,
            yscale=1
        ]
        \addplot[green, domain=0:5] {1.5*x} node[right]{};
        \draw[green] (5,1.5*5) -- (9,9) node[midway,above,xshift=-12pt]{$G$};
        \addplot[red, domain=0:2] {2*x};
        \addplot[red, domain=2:5] {4+0.9*(x-2)};
        \draw[red] (5,4+0.9*3) -- (9,9) node[midway,below,xshift=-7pt]{$F$};
        \draw[black] (9,5.5) -- (0,5.5) node[left] {$1-q$};
        \draw[dotted] (2.5,9) -- (2.5,0) node[below]{$s^N_{\rich}$};
        \draw[dotted] (5,9) -- (5,0) node[below]{$s^N_{\poor}$};
        \draw[dotted] (7.1,9) -- (7.1,0) node[above, xshift=-18pt]{$s^{DA}_{\poor}(G)$};
        \draw[dotted] (7.5,9) -- (7.5,0) node[below, xshift=14pt]{$s^{DA}_{\poor}(F)$};
        \fill[green] (axis cs:1.8, 2.7) ellipse [x radius=1.8pt, y radius=2pt];
        \fill[red] (axis cs:2, 4) ellipse [x radius=1.8pt, y radius=2pt];
        \fill[green] (axis cs:2.5, 2.5*1.5) ellipse [x radius=1.8pt, y radius=2pt];
        \fill[red] (axis cs:2.5, 4+0.9*0.5) ellipse [x radius=1.8pt, y radius=2pt];        
        \fill[green] (axis cs:5, 5*1.5) ellipse [x radius=1.8pt, y radius=2pt];
        \fill[red] (axis cs:5, 4+0.9*3) ellipse [x radius=1.8pt, y radius=2pt];
        \fill[green] (axis cs:7.1, 8.29) ellipse [x radius=1.8pt, y radius=2pt];
        \fill[red] (axis cs:7.5, 8.11) ellipse [x radius=1.8pt, y radius=2pt];        
        \end{axis}
    \end{tikzpicture}    
    \caption{Cutoff signal types under N and DA in the $s_{\omega^R}$-$s_{\omega^P}$ space: the case where $s_{\omega_i}^{DA}(F) > s_{\omega_i}^{DA}(G)$ for both $\omega_i\in\{\omega^P, \omega^R\}$}    
    \label{fig:lm5_case2}
\end{figure}
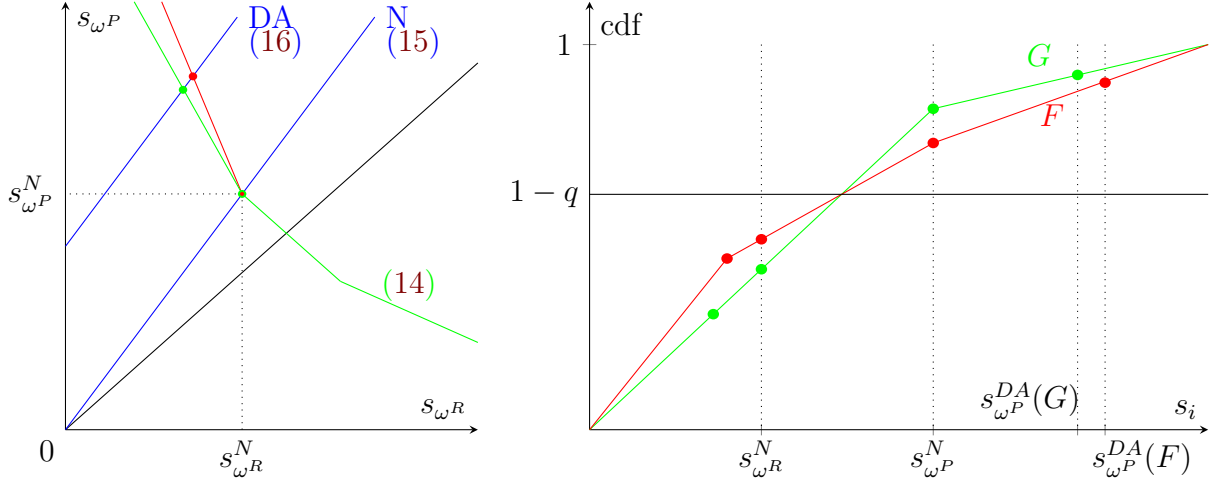

Since we focus on piece-wise linear $F$'s between four cutoffs, any such $F \in \mathcal{F}(\Pi,s_{\omega^P}^N,s_{\omega^R}^N)$ satisfies $F((1-\rho^P)s_{\omega^R}^N+\rho^P s_{\omega^P}^N)=1-q$.
Let $G$ be the cdf of $s_i$ with only one kink at $s_{\omega^P}^N$ satisfying $G((1-\rho^P)s_{\omega^R}^N+\rho^P s_{\omega^P}^N)=1-q$ (the green line in Figures \ref{fig:lm5_case1}--\ref{fig:lm5_case2}). 
We will show that $G$ maximizes $\omega^P$'s population change in the class of piece-wise linear cdfs between four cutoffs in $\mathcal{F}(\Pi,s_{\omega^P}^N,s_{\omega^R}^N)$.
Since $G$ clearly maximizes $G(s_{\omega^P}^N)-(1-q)$ and $(1-q)-G(s_{\omega^R}^N)$, it suffices to prove that $G$ minimizes the expansion rate (for $\omega^P$ or $\omega^R$).

Since $G'(s_i)$ only changes at $s_{\omega^P}^N$, the slope of the line (\ref{eq:cap_2types}) for $G$ in the $s_{\omega^R}$-$s_{\omega^P}$ space is constant between $(0,G^{-1}(\frac{1-q}{\rho^P}))$ and $(s_{\omega^R}^N,s_{\omega^P}^N)$.
Take any other $F$ (the red lines in Figures \ref{fig:lm5_case1}--\ref{fig:lm5_case2}). 
For each $\omega_i\in\{\omega^P, \omega^R\}$, let $s_{\omega_i}^{DA}(G)$ and $s_{\omega_i}^{DA}(F)$ be the cutoff types of DA under $G$ and $F$, respectively. 
Since the DA cutoffs must be on line (\ref{eq:DAconst_2types}), we have either [1] $s_{\omega_i}^{DA}(F) \le s_{\omega_i}^{DA}(G)$ for both $\omega_i\in\{\omega^P, \omega^R\}$, or [2] $s_{\omega_i}^{DA}(F) > s_{\omega_i}^{DA}(G)$ for both $\omega_i\in\{\omega^P, \omega^R\}$.

\medskip
\noindent
[1] When $s_{\omega_i}^{DA}(F) \le s_{\omega_i}^{DA}(G)$ for both $\omega_i\in\{\omega^P, \omega^R\}$ (Figure \ref{fig:lm5_case1}): 
\begin{figure}[htbp]
    \centering
    \begin{tikzpicture}
        \begin{axis}[
            axis lines=middle,
            xlabel={$s_{\omega^R}$},
            ylabel={$s_{\omega^P}$},
            xmin=0, xmax=6,
            ymin=0, ymax=7,
            xticklabels={},
            yticklabels={},
            xtick=\empty,
            ytick={9},
            samples=100,
            clip=false,
            xscale=0.8,
            yscale=1
        ]
        \draw[red] (1.34,5) -- (2.57, 9-2*2.57);
        \draw[red] (0.75,6.5) -- (1.34,5);
        \addplot[black, domain=0:6] {x} node[right]{};
        \addplot[blue, domain=0:4] {1.5*x} node[right]{N} node[right,yshift=-10pt]{(\ref{eq:Nconst_2types})};
        \addplot[blue, domain=0:2.5] {1.5*x+3} node[right]{DA} node[right,yshift=-10pt]{(\ref{eq:DAconst_2types})};
        \addplot[green, domain=1:2.57] {9-2*x} node[right]{};
        \node[below left] at (0,0) {0};
        \addplot[green, domain=2.57:4] {3.857+2.57-x};
        \addplot[green, domain=4:6] {2.427+2-x/2} node[midway,above]{(\ref{eq:cap_2types})};
        \fill[green] (axis cs:2.57, 3.857) ellipse [x radius=2pt, y radius=1.5pt];
        \fill[green] (axis cs:1.71, 5.56) ellipse [x radius=2pt, y radius=1.5pt];
        \draw[dotted] (2.57, 3.857) -- (2.57, 0) node[below]{$s^N_{\rich}$};
        \draw[dotted] (2.57, 3.857) -- (0, 3.857) node[left]{$s^N_{\poor}$};
        \fill[red] (axis cs:1.34,5) ellipse [x radius=2pt, y radius=1.5pt];
        \fill[red] (axis cs:2.57, 3.857) ellipse [x radius=2pt, y radius=1.5pt];
        \end{axis}
    \end{tikzpicture}%
    \begin{tikzpicture}
        \begin{axis}[
            axis lines=middle,
            xlabel={$s_i$},
            ylabel={cdf},
            xmin=0, xmax=9,
            ymin=0, ymax=10,
            xticklabels={},
            yticklabels={1},
            xtick={2.5,5,6.5,7.5},
            ytick={9},
            samples=100,
            clip=false,
            xscale=1.1,
            yscale=1
        ]
        \addplot[green, domain=0:5] {1.5*x} node[right]{};
        \draw[green] (5,1.5*5) -- (9,9) node[midway,above]{$G$};
        \addplot[red, domain=0:0.7] {3*x};
        \draw[red] (0.7,3*0.7) -- (2.5,4.75);
        %
        \draw[red] (2.5,4.75) -- (9,9) node[midway,below]{$F$};
        \draw[black] (9,5.5) -- (0,5.5) node[left] {$1-q$};
        \draw[dotted] (2.5,9) -- (2.5,0) node[below]{$s^N_{\rich}$};
        \draw[dotted] (5,9) -- (5,0) node[below]{$s^N_{\poor}$};
        \draw[dotted] (1.5,9) -- (1.5,0) node[above]{$s^{DA}_{\rich}(G)$};
        \draw[dotted] (0.7,9) -- (0.7,0) node[below]{$s^{DA}_{\rich}(F)$};
        \fill[green] (axis cs:1.5, 1.5*1.5) circle (2pt);
        \fill[red] (axis cs:0.7,3*0.7) circle (2pt);
        \fill[green] (axis cs:2.5, 2.5*1.5) circle (2pt);
        \fill[red] (axis cs:2.5, 4.75) circle (2pt);        
        \fill[green] (axis cs:5, 5*1.5) circle (2pt);
        \fill[red] (axis cs:5, 6.4) circle (2pt);
        \fill[green] (axis cs:7, 8.24) circle (2pt);
        \fill[red] (axis cs:6.5, 7.379) circle (2pt);        
        \end{axis}
    \end{tikzpicture}    
    \caption{Cutoff signal types under N and DA in the $s_{\omega^R}$-$s_{\omega^P}$ space: the case where $s_{\omega_i}^{DA}(F) \le s_{\omega_i}^{DA}(G)$ for both $\omega_i\in\{\omega^P, \omega^R\}$}    
    \label{fig:lm5_case1}
\end{figure}

 Consider the expansion rate for $\omega^R$.
 For $s_i \in [0, (1-\rho^P)s_{\omega^R}^N+\rho^P s_{\omega^P}^N]$, $G(s_i)$ is linear whereas $F(s_i)$ is weakly concave.
 Combined with $s_{\omega^R}^{DA}(F) \le s_{\omega^R}^{DA}(G)$, we obtain 
 \begin{equation*}
 \begin{split}
 & \frac{(1-q)-G(s_{\omega^R}^{DA}(G))}{(1-q)-G(s_{\omega^R}^N)} = \frac{((1-\rho^P)s_{\omega^R}^N+\rho^P s_{\omega^P}^N)-s_{\omega^R}^{DA}(G)}{((1-\rho^P)s_{\omega^R}^N+\rho^P s_{\omega^P}^N)-s_{\omega^R}^N} \\
 & \le \frac{((1-\rho^P)s_{\omega^R}^N+\rho^P s_{\omega^P}^N)-s_{\omega^R}^{DA}(F)}{((1-\rho^P)s_{\omega^R}^N+\rho^P s_{\omega^P}^N)-s_{\omega^R}^N}
 \le \frac{(1-q)-F(s_{\omega^R}^{DA}(F))}{(1-q)-F(s_{\omega^R}^N)},
 \end{split}
 \end{equation*}
where the first equality follows from the linearity of $G$ on that interval; the second from  assumption $s_{\omega_i}^{DA}(F) \le s_{\omega_i}^{DA}(G)$; and the third from weak concavity of $F$. Overall, the inequality means that the expansion rate under $G$ is lower than that under $F$. 
 
\medskip
\noindent
[2] When $s_{\omega_i}^{DA}(F) > s_{\omega_i}^{DA}(G)$ for both $\omega_i\in\{\omega^P, \omega^R\}$ (Figure \ref{fig:lm5_case2}):\footnote{Although not directly relevant to the proof, we can show that $1-q \le \rho^P$ must hold in this case. Suppose, by contradiction, that $1-q>\rho^P$. Then, in Figure \ref{fig:cutoff_2types}, the vertical intercepts of the lines (\ref{eq:cap_2types}) under both cdf's ($F^{-1}(\frac{1-q}{\rho^P})$ and $G^{-1}(\frac{1-q}{\rho^P})$) are greater than one.
 Let $s_{\omega^R}(F)$ and $s_{\omega^R}(G)$ be the values of $s_{\omega^R}$ that satisfy the capacity constraint (\ref{eq:cap_2types}) under each cdf when $s_{\omega^P}=1$. The convexity of the line (\ref{eq:cap_2types}) for $F$ and the linearity of the line (\ref{eq:cap_2types}) for $G$ imply that $0<s_{\omega^R}(G)<s_{\omega^R}(F)$. On the other hand, the capacity constraint (\ref{eq:cap_2types}) implies $F(s_{\omega^R}(F))=G(s_{\omega^R}(G))=\frac{1-q-\rho^P}{1-\rho^P}$, which further implies $s_{\omega^R}(G) \ge s_{\omega^R}(F)$ by the concavity of $F(s_i)$ for $s_i \in [0, (1-\rho^P)s_{\omega^R}^N+\rho^P s_{\omega^P}^N]$. These inequalities contradict each other.}

 Consider the expansion rate for $\omega^P$.
 By $G(s_{\omega^P}^N) \ge F(s_{\omega^P}^N)$, $F'(s_i)\ge G'(s_i)$ must hold for $s_i \in [s_{\omega^P}^N, \min\{s_{\omega^P}^{DA}(G),s_{\omega^P}^{DA}(F)\}] = [s_{\omega^P}^N, s_{\omega^P}^{DA}(G)]$ because otherwise $F(1)=1$ cannot hold for a concave $F$.
 Then, 
 \begin{equation*}
 \begin{split}
 & \frac{G(s_{\omega^P}^{DA}(G))-(1-q)}{G(s_{\omega^P}^N)-(1-q)} \le \frac{F(s_{\omega^P}^{DA}(G))-(1-q)}{F(s_{\omega^P}^N)-(1-q)}
 < \frac{F(s_{\omega^P}^{DA}(F))-(1-q)}{F(s_{\omega^P}^N)-(1-q)},
 \end{split}
 \end{equation*}
where the first inequality holds by the established relation on the slopes, and the second by $s_{\omega_i}^{DA}(F) > s_{\omega_i}^{DA}(G)$. Overall, the inequality means that the expansion rate under $G$ is lower than that under $F$.

\medskip
\noindent
\underline{$\omega^P$'s population change at $c_1$ from N to TTC.}

Analogously, $\omega^P$'s population change at $c_1$ from N to TTC is  
\begin{equation*}
 (F(s_{\omega^P}^N)-(1-q))\Big[1-r^{TTC}\frac{F(s_{\omega^P}^{DA})-(1-q)}{F(s_{\omega^P}^N)-(1-q)}\Big].
\end{equation*}
Since $r^{TTC}$ does not depend on $F(\cdot)$ under $g=0$ and $e=1$, this population change is maximized if $F \in \mathcal{F}(\Pi,s_{\omega^P}^N,s_{\omega^R}^N)$ maximizes $F(s_{\omega^P}^N)-(1-q)$ and minimizes the expansion rate $\frac{F(s_{\omega^P}^{TTC})-(1-q)}{F(s_{\omega^P}^N)-(1-q)}$ for $\omega^P$.

Under TTC, the cutoff signal types $(s_{\omega^R},s_{\omega^P})$ satisfy equation (\ref{eq:cap_2types}) and 
\begin{equation}\label{eq:TTCconst_2types}
 s_{\omega^P}-s_{\omega^R} = \Delta\omega\Big[(\rho^P s_{\omega^P}+(1-\rho^P)s_{\omega^R})+\frac{2\pi}{1-2\pi}\Big] \ \Leftrightarrow \ s_{\omega^P} = \frac{1+(1-\rho^P)\Delta\omega}{1-\rho^P\Delta\omega}s_{\omega^R} + \frac{\Delta\omega\frac{2\pi}{1-2\pi}}{1-\rho^P\Delta\omega}.
\end{equation}
The only difference between DA and TTC is the intercepts of the equations (\ref{eq:DAconst_2types}) and (\ref{eq:TTCconst_2types}).
Therefore, the proof is analogous to the change from N to DA.

\subsection{Proof of Proposition \ref{prop:s_sorting_2types}}

\medskip
\noindent
\underline{N and DA.}
By Lemma \ref{lm:F_max_pop}, it suffices to show that for any $(\Pi,s_{\omega^P}^N,s_{\omega^R}^N)$ with $\mathcal{F}(\Pi,s_{\omega^P}^N,s_{\omega^R}^N)\neq \emptyset$, $\omega^P$'s population change from N to DA at $c_1$ under the ``optimal" cdf $G \in \mathcal{F}(\Pi,s_{\omega^P}^N,s_{\omega^R}^N)$ (i.e., the piece-wise linear cdf that has one kink at $s_i=s_{\omega^P}^N$) is negative if and only if $1-q < \rho^P$.

Take $(\Pi,s_{\omega^P}^N,s_{\omega^R}^N)$ with $\mathcal{F}(\Pi,s_{\omega^P}^N,s_{\omega^R}^N)\neq \emptyset$ and fix them.
Let $\alpha \coloneqq \frac{1+(1-\rho^P)\Delta\omega}{1-\rho^P\Delta\omega}$, $\beta \coloneqq \frac{\Delta\omega\frac{\pi}{1-\pi}}{1-\rho^P\Delta\omega}$, and $x \coloneqq s_{\omega^R}^N$.
Then by equation (\ref{eq:Nconst_2types}), $s_{\omega^P}^N = \alpha x$.
The optimal cdf $G \in \mathcal{F}(\Pi,s_{\omega^P}^N,s_{\omega^R}^N)$ is $G(s_i)=\frac{1-q}{(\rho^P\alpha+(1-\rho^P))x}s_i$ for $s_i \in [0,\alpha x]$ because it is linear for $s_i \in [0,\alpha x]$ and satisfies $G((1-\rho^P)s_{\omega^R}^N+\rho^P s_{\omega^P}^N)=1-q$.
Then, we obtain $G(s_{\omega^R}^N)=\frac{1-q}{\rho^P\alpha+1-\rho^P}$.

We can then solve explicitly for the portion of the line (\ref{eq:cap_2types}) in Figure \ref{fig:cutoff_2types} for $s_{\omega^R} \in [0, x]$:
\begin{equation*}
 s_{\omega^P} = G^{-1}\Big(\frac{1-q}{\rho^P}\Big)-\frac{G^{-1}(\frac{1-q}{\rho^P})-\alpha x}{x}s_{\omega^R}.
\end{equation*}
Together with equation (\ref{eq:DAconst_2types}), we obtain $s_{\omega^R}^{DA}=(1-\frac{\beta}{G^{-1}(\frac{1-q}{\rho^P})})x$.
The expansion rate under $G$ is 
\begin{equation*}
\begin{split}
 & \frac{|G(s_{\omega^R}^{DA})-(1-q)|}{|G(s_{\omega^R}^N)-(1-q)|} = \frac{(1-q)-\frac{1-q}{\rho^P\alpha+1-\rho^P}(1-\frac{\beta}{G^{-1}(\frac{1-q}{\rho^P})})}{(1-q)-\frac{1-q}{\rho^P\alpha+1-\rho^P}} = 1 + \frac{\beta}{\rho^P(\alpha-1)}\frac{1}{G^{-1}(\frac{1-q}{\rho^P})} \\
 &= 1+\frac{\frac{\Delta\omega\frac{\pi}{1-\pi}}{1-\rho^P\Delta\omega}}{\rho^P(\frac{1+(1-\rho^P)\Delta\omega}{1-\rho^P\Delta\omega}-1)}\frac{1}{G^{-1}(\frac{1-q}{\rho^P})} = 1+\frac{1}{\rho^P G^{-1}(\frac{1-q}{\rho^P})}\frac{\pi}{1-\pi}.
\end{split}
\end{equation*}
On the other hand, 
\begin{equation*}
\frac{1}{r^{DA}(1-\pi)} = \frac{1}{\frac{1-q}{(1-\pi)(1-q)+\pi}(1-\pi)} = 1+\frac{1}{1-q}\frac{\pi}{1-\pi}
\end{equation*}
because $r^{DA}=1-\frac{S}{D}=1-\frac{\pi q}{(1-\pi)(1-q)+\pi}=\frac{1-q}{(1-\pi)(1-q)+\pi}$ when $e=1$ and $g=0$.
By Theorem \ref{thm:s_sorting}, $\omega^P$'s population change is negative when the expansion rate is greater than $\frac{1}{r^{DA}(1-\pi)}$.
Under $G$, this happens if and only if $\rho^P G^{-1}(\frac{1-q}{\rho^P}) < 1-q$, which is $\frac{1-q}{\rho^P} < G(\frac{1-q}{\rho^P})$.
Since $G$ is concave and $G(1)=1$, this condition is equivalent to $\frac{1-q}{\rho^P} < 1$.

\medskip
\noindent
\underline{N and TTC.}
Under $g=0$ and $e=1$, 
\begin{equation*}
\frac{1}{r^{TTC}} = \frac{D-X}{D-S} = 1
\end{equation*}
because of $S=X$.
By Theorem \ref{thm:n_sorting}, the expansion rate from N to TTC is always greater than one for both $\omega^P$ and $\omega^R$.
Then, Theorem \ref{thm:s_sorting} implies that TTC results in greater school segregation by wealth than N for all $F \in \mathcal{F}(\Pi)$.

\medskip
\noindent
\underline{DA and TTC.}
By $|\Omega|=2$, Theorem \ref{thm:n_sorting}, and the capacity constraint, we have $s_{\omega^R}^{TTC} < s_{\omega^R}^{DA} < F^{-1}(1-q) < s_{\omega^P}^{DA} < s_{\omega^P}^{TTC}$ and $\tilde{\Omega}(DA,TTC)=\Omega$. 
These imply that the sufficient condition of statement 3 of Theorem \ref{thm:s_sorting} is satisfied for any $F \in \mathcal{F}(\Pi)$.

\subsection{Proof of Theorem \ref{thm:price}}
\noindent
\underline{$p^N \le p^{DA}$.}
Recall that $r^{DA} = \frac{D-S}{D}$ and 
 \begin{equation*}
 r^{DA}_{\mathrm{uniform}} = \frac{1-q-g}{(1-\pi)(1-q-g)+\pi e}.
 \end{equation*}
When $r^{DA} \ge r^{DA}_{\mathrm{uniform}}$ holds, 
 \begin{equation*}
 \begin{split}
 p^{DA} &= r^{DA}(1-\pi)\Big\{E[s_{\omega_i}^{DA}]-g+\frac{\pi e}{1-\pi}\Big\} \\
 &\ge r^{DA}_{\mathrm{uniform}}(1-\pi)\Big\{E[s_{\omega_i}^N]-g+\frac{\pi e}{1-\pi}\Big\} \\
 &= \frac{1-q-g}{1-q-g+\frac{\pi e}{1-\pi}}\frac{E[s_{\omega_i}^N]-g+\frac{\pi e}{1-\pi}}{E[s_{\omega_i}^N]-g}\Big\{E[s_{\omega_i}^N]-g\Big\} \\
 &\ge E[s_{\omega_i}^N]-g \\
 &= p^N,
 \end{split}
 \end{equation*}
where the first inequality follows from $r^{DA} \ge r^{DA}_{\mathrm{uniform}}$ and Lemma \ref{lmB}, and the last inequality follows from $\frac{1-q-g}{1-q-g+\frac{\pi e}{1-\pi}} \ge \frac{E[s_{\omega_i}^N]-g}{E[s_{\omega_i}^N]-g+\frac{\pi e}{1-\pi}} > 0$.

\medskip
\noindent
\underline{$p^{DA} < p^{TTC}$.}
First, we establish 
 \begin{equation}\label{eq:p_DA_TTC}
 \begin{split}
   \frac{r^{DA}}{r^{TTC}} &= \frac{D-X}{D} \\
 &= \frac{1-q-(1-2\pi)F(g)}{(1-\pi)(1-q)-(1-2\pi)F(g)+\pi F(e-g)} \\
 &= \frac{(1-2\pi)(1-q)+2\pi(1-q)-(1-2\pi)F(g)}{(1-\pi)(1-q)+\pi F(e-g)-(1-2\pi)F(g)} \\
 &\le \frac{(1-2\pi)(1-q)+2\pi(1-q)-(1-2\pi)g}{(1-\pi)(1-q)+\pi F(e-g)-(1-2\pi)g} \\
 &< \frac{(1-2\pi)(1-q)+2\pi (e-g)-(1-2\pi)g}{(1-\pi)(1-q)+\pi(e-g)-(1-2\pi)g} \\
 &= \frac{1-2\pi}{1-\pi}\frac{1-q-g+\frac{2\pi (e-g)}{1-2\pi}}{1-q-g+\frac{\pi e}{1-\pi}}
 \end{split}
 \end{equation}
 because $\frac{r^{DA}}{r^{TTC}}<1$, $F(g) \ge g$, $e-g>1-q$, and $F(e-g)\ge e-g$.
 Then by equation (\ref{eq:expected_cutoff}), 
 \begin{equation*}
 \begin{split}
 \frac{p^{DA}}{p^{TTC}} &= \frac{(1-\pi)r^{DA}}{(1-2\pi)r^{TTC}}\frac{E[s_{\omega_i}^{DA}]-g+\frac{\pi e}{1-\pi}}{E[s_{\omega_i}^{TTC}]+\frac{2\pi e-g}{1-2\pi}} \\
 &\le \frac{(1-\pi)r^{DA}}{(1-2\pi)r^{TTC}}\frac{E[s_{\omega_i}^{DA}]-g+\frac{\pi e}{1-\pi}+(1-q-E[s_{\omega_i}^{DA}])}{E[s_{\omega_i}^{TTC}]+\frac{2\pi e-g}{1-2\pi}+(1-q-E[s_{\omega_i}^{TTC}])} \\
 &= \frac{(1-\pi)r^{DA}}{(1-2\pi)r^{TTC}}\frac{1-q-g+\frac{\pi e}{1-\pi}}{1-q+\frac{2\pi e-g}{1-2\pi}} \\
 &= \frac{(1-\pi)r^{DA}}{(1-2\pi)r^{TTC}}\frac{1-q-g+\frac{\pi e}{1-\pi}}{1-q-g+\frac{2\pi(e-g)}{1-2\pi}} \\
 &< 1,
 \end{split}
 \end{equation*}
 where the first inequality holds because $\frac{E[s_{\omega_i}^{DA}]-g+\frac{\pi e}{1-\pi}}{E[s_{\omega_i}^{TTC}]+\frac{2\pi e-g}{1-2\pi}}<1$ and $E[s_{\omega_i}^{DA}] \le E[s_{\omega_i}^{TTC}] \le 1-q$ by Lemma \ref{lmB}, and the last inequality follows from equation (\ref{eq:p_DA_TTC}).

\end{document}

%% file: ex_pic_definitions3.tex
        \def\xMax{10}
        \def\xq{5}
        \def\xpoor{5}
        \def\xrich{1}
        \def\shiftN{1}
        \def\clrN{olive}
        \def\shiftDA{2}
        \def\clrDA{teal}
        \def\shiftTTC{4}    
        \def\clrTTC{red}
        \def\w{\omega}
        \coordinate (O) at (0,0);

%% file: ex_pic_slopes.tex
        \draw[-,\clrN,thick](\xMax-\xq-\shiftN,\xrich) -- (\xMax-\xq+\shiftN,\xpoor) node[above] {N};
        \draw[-,\clrDA,thick](\xMax-\xq-\shiftDA,\xrich) -- (\xMax-\xq+\shiftDA,\xpoor) node[above right] {DA};      
        \draw[-,\clrTTC,thick](\xMax-\xq-\shiftTTC,\xrich) -- (\xMax-\xq+\shiftTTC,\xpoor) node[above right] {TTC};
        \node[left,align=right] at (\xMax,{(\xrich+\xpoor)/3.5}) {\itshape Agents\\\itshape living in $n_1$};
        \draw[->,dotted] (\xMax-0.9,\xpoor-2.5)--(\xMax-0.5,\xpoor-0.01);
        \draw[->,dotted] (\xMax-3,\xrich+0.4)--(\xMax/2+1,\xrich+0.01);
        \draw[<-,loosely dashed] (\xMax-\xq+0.1,{(\xpoor+\xrich)/2}) -- (\xMax-\xq/2-1,{(\xpoor+\xrich)/2});
        \node at (\xMax-\xq/2,{(\xpoor+\xrich)/2}) {$\frac{n_1\text{ capacity}}{2}$};
        \draw[->,loosely dashed] (\xMax-\xq/2+1,{(\xpoor+\xrich)/2}) -- (\xMax,{(\xpoor+\xrich)/2});

%% file: ex_pic_mainbox3.tex
        \draw[->] (\xLeft,0) -- (\xMax+2,0);
        \node[align=left,below right] at ({\xLeft},0) {\itshape Exp.value \itshape of $c_1$};
        \draw[->] (\xLeft,0) -- (\xLeft,\xpoor+1.3) node[right] {$\omega_i$, \itshape wealth index};
        \draw[dashed,-] (\xLeft,\xpoor) -- (\xMax,\xpoor) ;
        \node[below right] at (\xLeft,{\xpoor}) {\itshape poor};
        \draw[dashed,-] (\xLeft,\xrich) -- (\xMax,\xrich); 
        \node[above right] at (\xLeft,{\xrich})  {\itshape rich};
        \draw[-](0,0) node[below] {$s_i=0$}-- (0,\xpoor);
        \draw[-](\xMax,0) node[below] {$s_i=1$}-- (\xMax,\xpoor);

%% file: ex_pic_TTC_reshufle3.tex
        \draw[-,\clrTTC](\xMax-\xq-\shiftTTC,\xrich) -- (\xMax-\xq+\shiftTTC,\xpoor);
        \draw[-,\clrTTC](-\xMax+\xq+\shiftTTC,\xrich) -- (-\xMax+\xq-\shiftTTC,\xpoor);
        \def\boxWidthN1{0.333}
        fill=\clrTTC, opacity=0.5
        \filldraw[fill=red,pattern=north east lines,pattern color=red,draw=\clrTTC, opacity=0.4] (\xq+\shiftTTC,\xpoor) rectangle (\xMax,\xpoor+2*\boxWidthN1) node[midway] {$=,\rightarrow$};
        \draw[draw=\clrDA,fill=green,pattern=vertical lines,pattern color=green, opacity=0.4] (\xq+\shiftTTC,\xpoor+2*\boxWidthN1) rectangle (\xMax,\xpoor+3*\boxWidthN1) node[midway]{$\leftarrow$};
        \filldraw[fill=red,pattern=north east lines,pattern color=red,draw=\clrTTC, opacity=0.4] (\xq-\shiftTTC,\xrich) rectangle (\xMax,\xrich-2*\boxWidthN1) node[midway] {$=, \rightarrow$};
        \draw[draw=\clrDA,fill=green,pattern=vertical lines,pattern color=green, opacity=0.4] (\xq-\shiftTTC,\xrich-2*\boxWidthN1) rectangle (\xMax,\xrich-3*\boxWidthN1) node[midway]{$\leftarrow$};
        \filldraw[fill=red,pattern=north east lines,pattern color=red,draw=\clrTTC, opacity=0.4] (-\xq-\shiftTTC,\xpoor) rectangle (-\xMax,\xpoor+2*\boxWidthN1) node[midway] {$=, \leftarrow$};
        \draw[draw=\clrDA,fill=green,pattern=vertical lines,pattern color=green, opacity=0.4] (-\xq-\shiftTTC,\xpoor+2*\boxWidthN1) rectangle (-\xMax,\xpoor+3*\boxWidthN1) node[midway]{$\rightarrow$};
        \filldraw[fill=red,pattern=north east lines,pattern color=red,draw=\clrTTC, opacity=0.4] (-\xq+\shiftTTC,\xrich) rectangle (-\xMax,\xrich-2*\boxWidthN1) node[midway] {$=, \leftarrow$};
        \draw[draw=\clrDA,fill=green,pattern=vertical lines,pattern color=green, opacity=0.4] (-\xq+\shiftTTC,\xrich-2*\boxWidthN1) rectangle (-\xMax,\xrich-3*\boxWidthN1) node[midway]{$\rightarrow$};
        \def\boxWidthNoN1{0.45*\boxWidthN1}
        \node[above,fill=white,text=black] at ({(-\xMax-\xq-\shiftTTC+\xMax+\xq+\shiftTTC)/2},\xrich-9*\boxWidthN1) {\itshape Students in $n_{-1}$ and $n_1$ exchange their seats};
        %
        \draw ({(-\xMax+\xq-\shiftTTC)/2},\xrich-2.7*\boxWidthN1) edge[<->, \clrTTC, bend right=23] ({(\xMax-\xq+\shiftTTC)/2},\xrich-2.7*\boxWidthN1);
        \node[above,\clrTTC,fill=white] at (0,{\xrich}){TTC};

%% file: ex_pic_DAreshufle3.tex
        \draw[-,\clrDA,thick](\xMax-\xq-\shiftDA,\xrich) -- (\xMax-\xq+\shiftDA,\xpoor) node[below right] {DA}; 
        \def\boxWidthN1{0.3333}
        \filldraw[fill=red,pattern=north east lines,pattern color=red, draw=\clrDA, opacity=0.4] (\xq+\shiftDA,\xpoor) rectangle (\xMax,\xpoor+2*\boxWidthN1) node[midway] {$=, \rightarrow$};
        \filldraw[draw=\clrDA,fill=green,pattern=vertical lines,pattern color=green, opacity=0.4] (\xq+\shiftDA,\xpoor+2*\boxWidthN1) rectangle (\xMax,\xpoor+3*\boxWidthN1) node[midway] {$\leftarrow$};
        \filldraw[fill=red,pattern=north east lines,pattern color=red,draw=\clrDA, opacity=0.4] (\xq-\shiftDA,\xrich) rectangle (\xMax,\xrich-2*\boxWidthN1) node[midway] {$=, \rightarrow$};
        \filldraw[draw=\clrDA,fill=green,pattern=vertical lines,pattern color=green, opacity=0.4] (\xq-\shiftDA,\xrich-2*\boxWidthN1) rectangle (\xMax,\xrich-3*\boxWidthN1) node[midway] {$\leftarrow$};
        \def\boxWidthNoN1{3*\boxWidthN1/4}
        \filldraw[fill=red,pattern=crosshatch,pattern color=blue,draw=\clrDA, opacity=0.4] (\xLeft,\xpoor) rectangle (\xq+\shiftDA,\xpoor+\boxWidthNoN1) node[midway] {$ \rightarrow$};
        \filldraw[fill=red,pattern=crosshatch,pattern color=blue,draw=\clrDA, opacity=0.4] (\xLeft,\xrich) rectangle (\xq-\shiftDA,\xrich-\boxWidthNoN1) node[midway] {$\rightarrow$};
        \draw[<->,dotted](\xLeft/2,{\xrich-\boxWidthN1/2.5}) -- (\xLeft/2,{\xpoor+\boxWidthN1/2.5}) node[midway,fill=white,align=left]{\itshape Out-of-zone applicants\\\itshape assigned to $c_1$};
        \draw[<->,dotted](\xMax-0.3,{\xrich-5*\boxWidthN1/2}) -- (\xMax-0.3,{\xpoor+5*\boxWidthN1/2}) node[midway,fill=white,align=left]{\itshape In-zone\\\itshape applicants\\\itshape leaving $c_1$};
        \draw[<->,dotted] ({\xMax-(\xq-\shiftDA)/2},\xrich-0.2) -- (1,{(\xrich+\xpoor)/2}) node[fill=white,align=left]{\itshape In-zone\\\itshape applicants\\\itshape assigned to $c_1$} -- ({\xMax-(\xq-\shiftDA)/2},\xpoor+0.2);

%% file: merged_refs.bib
@STRING{AEJ  = "{A}merican {E}conomic {J}ournal: Microeconomics" }

@STRING{JPE  = "{J}ournal of {P}olitical {E}conomy" }

@incollection{AbdulkadirogluAndersson2023SchoolChoiceReview,
  author       = {Atila Abdulkadiroğlu and Tommy Andersson},
  title        = {School Choice},
  booktitle    = {Handbook of the Economics of Education},
  editor       = {Eric A. Hanushek and Steven G. Rivkin},
  volume       = {6},
  pages        = {135--185},
  year         = {2023},
  publisher    = {North-Holland},
  address      = {Amsterdam},
  doi          = {10.1016/bs.hesed.2023.07.003}
}

@article{BodohHickman2018MatchingContest,
  author       = {Aaron L. Bodoh-Creed and Brent R. Hickman},
  title        = {College Assignment as a Large Contest},
  journal      = {Journal of Economic Theory},
  volume       = {175},
  pages        = {88--126},
  year         = {2018},
  doi          = {10.1016/j.jet.2018.01.006}
}

@article{HarrisLarsen2023WhatFamiliesWant,
  author       = {Harris, Douglas N. and Larsen, Matthew F.},
  title        = {What Schools Do Families Want (and Why)? Evidence on Revealed Preferences from New Orleans},
  journal      = {Educational Evaluation and Policy Analysis},
  year         = {2023},
  volume       = {45},
  number       = {3},
  pages        = {496--519},
  doi          = {10.3102/01623737221134528},
}

@article{GlazermanDotter2017NewOrleanPreferences,
  author       = {Glazerman, Steven and Dotter, Dallas},
  title        = {Market Signals: Evidence on the Determinants and Consequences of School Choice from a Citywide Lottery},
  journal      = {Educational Evaluation and Policy Analysis},
  year         = {2017},
  volume       = {39},
  number       = {4},
  pages        = {593--619},
  doi          = {10.3102/0162373717702964}
}

@article{Pathak2011Review,
  author       = {Parag A. Pathak},
  title        = {The Mechanism Design Approach to Student Assignment},
  journal      = {Annual Review of Economics},
  volume       = {3},
  number       = {1},
  pages        = {513--536},
  year         = {2011},
  doi          = {10.1146/annurev-economics-111809-125054},
  url          = {https://doi.org/10.1146/annurev-economics-111809-125054}
}

@article{EllisonPathak2021AA,
  author       = {Ellison, Glenn and Pathak, Parag A.},
  title        = {The Efficiency of Race-Neutral Alternatives to Race-Based Affirmative Action: Evidence from Chicago's Exam Schools},
  journal      = {American Economic Review},
  volume       = {111},
  number       = {3},
  pages        = {943--975},
  year         = {2021},
  doi          = {10.1257/aer.20161290},
}

@article{PathakReesJonesSonmez2020ReversingReserves,
  author       = {Pathak, Parag A. and Rees-Jones, Alex and S\"{o}nmez, Tayfun},
  title        = {Reversing Reserves},
  journal      = {Management Science},
  volume       = {69},
  number       = {11},
  pages        = {6940--6953},
  year         = {2023},
  doi          = {10.1287/mnsc.2023.4836},
}

@article{Bleemer2023AA,
  author       = {Bleemer, Zachary},
  title        = {Affirmative Action and Its Race-Neutral Alternatives},
  journal      = {Journal of Public Economics},
  volume       = {220},
  pages        = {104839},
  year         = {2023},
  doi          = {10.1016/j.jpubeco.2023.104839},
}

@article{DurKominersPathakSonmez2018ReserveDesign,
  author       = {Dur, Umut and Kominers, Scott Duke and Pathak, Parag A. and S\"onmez, Tayfun},
  title        = {Reserve Design: Unintended Consequences and the Demise of Boston's Walk Zones},
  journal      = {Journal of Political Economy},
  volume       = {126},
  number       = {6},
  pages        = {2457--2479},
  year         = {2018},
  doi          = {10.1086/699977},
}

@article{CarlsonEtAl2020AA,
  author       = {Carlson, Deven and Bell, Elizabeth and Cowen, Joshua M. and Cowan, James and Witte, John F.},
  title        = {Socioeconomic-Based School Assignment Policy and Racial Segregation Levels: Evidence from the Wake County Public School System},
  journal      = {American Educational Research Journal},
  volume       = {57},
  number       = {1},
  pages        = {258--304},
  year         = {2020},
  doi          = {10.3102/0002831219851729},
}

@techreport{Musset2012SchoolChoiceEquity,
  author       = {Musset, Pauline},
  title        = {School Choice and Equity: Current Policies in OECD Countries and a Literature Review},
  type         = {{OECD Education Working Paper}},
  number       = {66},
  year         = {2012},
  month        = {January},
  doi          = {10.1787/5k9fq23507vc-en},
  url          = {https://doi.org/10.1787/5k9fq23507vc-en}
}

@article{AveryPathak2021PublicSchool,
  title        = {The Distributional Consequences of Public School Choice},
  author       = {Avery, Christopher and Pathak, Parag A.},
  journal      = {American Economic Review},
  volume       = {111},
  number       = {1},
  pages        = {129--152},
  year         = {2021},
  doi          = {10.1257/aer.20151147},
}

@article{BarseghyanClarkCoate2019PeerPublicSchool,
  title        = {Peer Preferences, School Competition, and the Effects of Public School Choice},
  author       = {Barseghyan, Levon and Clark, Damon and Coate, Stephen},
  journal      = {American Economic Journal: Economic Policy},
  volume       = {11},
  number       = {4},
  pages        = {124--158},
  year         = {2019},
  doi          = {10.1257/pol.20170484},
}

@article{DeFrajaMartinezMora2014DesegragatinTracking,
  title        = {The desegregating effect of school tracking},
  author       = {De Fraja, Gianni and Mart{\'i}nez-Mora, Francisco},
  journal      = {Journal of Urban Economics},
  volume       = {80},
  pages        = {164--177},
  year         = {2014},
  doi          = {10.1016/j.jue.2014.05.002},
}

@article{PathakSonmez2008SincereSophisticated,
  author  = {Pathak, Parag A. and S\"onmez, Tayfun},
  title   = {Leveling the Playing Field: Sincere and Sophisticated Players in the Boston Mechanism},
  journal = {American Economic Review},
  volume  = {98},
  number  = {4},
  pages   = {1636--1652},
  year    = {2008}
}

@techreport{Gonczarowski2024SeveralFields,
  author       = {Yannai A. Gonczarowski and Michael Yin and Shirley Zhang},
  title        = {Multi-District School Choice: Playing on Several Fields},
  institution  = {arXiv},
  type         = {Preprint},
  number       = {arXiv:2403.04530},
  year         = {2024},
  url          = {https://arxiv.org/abs/2403.04530},
}

@techreport{OpenEnrollment2018Jeong,
  author       = {Byeong-Hyeon Jeong},
  title        = {School Choice in Context: Can Open Enrollment Cure Segregation?},
  year         = {2022},
  type         = {{SSRN Working Paper}},
  doi          = {10.2139/ssrn.4031432},
  url          = {https://papers.ssrn.com/sol3/papers.cfm?abstract_id=4031432}
}

@techreport{Grigoryan2021,
  author       = {Aram Grigoryan},
  title        = {School Choice and the Housing Market},
  type         = {{SSRN Working Paper}},
  year         = {2021},
  url          = {https://econ.duke.edu/sites/econ.duke.edu/files/documents/Abstract%20_Aram%20Grigoryan_10-18-21_0.pdf}
}

@article{Caetano2019,
  author = {Gregorio S. Caetano},
  title = {Neighborhood Sorting and the Value of Public School Quality},
  journal = {Journal of Urban Economics},
  volume = {114},
  pages = {103193},
  year = {2019},
  doi = {10.1016/j.jue.2019.103193}
}

@techreport{Agostinelli2024,
  author = {Francesco Agostinelli and Margaux Luflade and Paolo Martellini},
  title = {On the Spatial Determinants of Educational Access},
  year = {2024},
  number = {32246},
  type = {{NBER Working Paper}},
  url = {https://www.nber.org/papers/w32246}
}

@inproceedings{LocationSchool2023Park,
  author = {Minseon Park and Dong Woo Hahm},
  title = {Location Choice, Commuting, and School Choice},
  booktitle = {Proceedings of the 24th ACM Conference on Economics and Computation (EC '23)},
  year = {2023},
  pages = {850--860},
  type = {{Proceedings of the 24th ACM Conference on Economics and Computation (EC '23)}},
  url = {https://researchr.org/publication/ParkH23-1}
}

@techreport{Pietrabissa2024,
  author = {Giorgio Pietrabissa},
  title = {School Access and City Structure},
  institution = {CEMFI},
  type = {Working Paper},
  year = {2024},
  url = {https://giorgiopietrabissa.github.io/files/school_sorting.pdf}
}

@article{LocalExpend1956Tiebout,
 ISSN = {00223808, 1537534X},
 URL = {http://www.jstor.org/stable/1826343},
 author = {Charles M. Tiebout},
 journal = {Journal of Political Economy},
 number = {5},
 pages = {416--424},
 publisher = {University of Chicago Press},
 title = {A Pure Theory of Local Expenditures},
 urldate = {2024-03-05},
 volume = {64},
 year = {1956}
}

@article{MobilityPrivateVouchers2000Nechyba,
Author = {Nechyba, Thomas J.},
Title = {Mobility, Targeting, and Private-School Vouchers},
Journal = {American Economic Review},
Volume = {90},
Number = {1},
Year = {2000},
Month = {March},
Pages = {130-146},
DOI = {10.1257/aer.90.1.130},
URL = {https://www.aeaweb.org/articles?id=10.1257/aer.90.1.130}}

@techreport{AmenityHousingPrices2018Moon,
  title={Access to local amenity and housing prices},
  author={Moon, Terry S},
  year={2018},
  type={{Working Paper, Princeton University}}
}

@incollection{NeigborhoodChoiceSchools2003Epple,
  title={Neighborhood schools, choice, and the distribution of educational benefits},
  author={Epple, Dennis N and Romano, Richard},
  booktitle={The Economics of School Choice},
  pages={227--286},
  year={2003},
  publisher={University of Chicago Press}
}

@techreport{PricingNeighborhoods2023Eshaghnia,
  title={Pricing Neighborhoods},
  author={Eshaghnia, Sadegh and Heckman, James J and Razavi, Goya},
  year={2023},
  type = {{NBER Working Paper}}
}

@article{RiskSegregation2021Calsamiglia,
  title={School choice design, risk aversion and cardinal segregation},
  author={Calsamiglia, Caterina and Mart{\'\i}nez-Mora, Francisco and Miralles, Antonio},
  journal={The Economic Journal},
  volume={131},
  number={635},
  pages={1081--1104},
  year={2021},
  publisher={Oxford University Press}
}

@article{Stratification2023Calsamiglia,
  title={Catchment Areas, Stratification, And Access To Better Schools},
  author={Calsamiglia, Caterina and Miralles, Antonio},
  journal={International Economic Review},
  volume={64},
  number={4},
  pages={1469--1492},
  year={2023},
  publisher={Wiley Online Library}
}

@techreport{SchoolSorting2024Greaves,
  title={School Choice and Neighborhood Sorting: Equilibrium Consequences of Geographic School Admissions},
  author={Greaves, Ellen and Turon, H{\'e}l{\`e}ne},
  year={2024},
  type={{IZA Discussion Paper}}
}

@article{abdulkadirouglu2003school,
  title={School Choice: A Mechanism Design Approach},
  author={Abdulkadiro{\u{g}}lu, Atila and S{\"o}nmez, Tayfun},
  journal={American Economic Review},
  volume={93},
  number={3},
  pages={729--747},
  year={2003}
}

@article{SchoolHousingPrices:Black:1999,
  title={Do better schools matter? Parental valuation of elementary education},
  author={Black, Sandra E},
  journal={The Quarterly Journal of Economics},
  volume={114},
  number={2},
  pages={577--599},
  year={1999},
  publisher={MIT Press}
}

@article{ReviewSchoolHousingPrices:NHY:2011,
title = {The capitalization of school quality into house values: A review},
journal = {Journal of Housing Economics},
volume = {20},
number = {1},
pages = {30-48},
year = {2011},
issn = {1051-1377},
doi = {https://doi.org/10.1016/j.jhe.2011.02.001},
url = {https://www.sciencedirect.com/science/article/pii/S1051137711000027},
author = {Phuong Nguyen-Hoang and John Yinger}
}

@article{SchoolHousingPricesBoston:La:2015,
title = {Capitalization of school quality into housing prices: Evidence from Boston Public School district walk zones},
journal = {Economics Letters},
volume = {134},
pages = {102-106},
year = {2015},
issn = {0165-1765},
doi = {https://doi.org/10.1016/j.econlet.2015.07.001},
url = {https://www.sciencedirect.com/science/article/pii/S0165176515002748},
author = {Vincent La}
}

@article{azevedo/leshno:16,
  title={A supply and demand framework for two-sided matching markets},
  author={Azevedo, Eduardo M and Leshno, Jacob D},
  journal=JPE,
   volume  = 124,
   pages   = {1235-1268},
   year    = 2016,
}

@Article{acy:15,
  author  = {Atila Abdulkadiro{\u{g}}lu and Yeon-Koo Che and Yosuke Yasuda},
  title   = {Expanding `Choice' in School Choice},
   journal = AEJ,
  volume  = 7,
  pages   = {1-42},
  year    = {2015},
}

@article{leshno2021cutoff,
  title={The cutoff structure of top trading cycles in school choice},
  author={Leshno, Jacob D and Lo, Irene},
  journal={The Review of Economic Studies},
  volume={88},
  number={4},
  pages={1582--1623},
  year={2021},
  publisher={Oxford University Press}
}
